%% file: 0-main.tex
\documentclass[acmtog,nonacm,balance=True]{acmart}
\renewcommand\footnotetextcopyrightpermission[1]{}
\renewcommand\footnotetextauthorsaddresses[1]{}

\usepackage{booktabs} 

\usepackage{bm}
\usepackage{xspace}
\usepackage{enumitem}

\usepackage{multirow}
\AtEndPreamble{
    \usepackage[capitalize]{cleveref}
    \crefname{section}{Sec.}{Secs.}
    \Crefname{section}{Section}{Sections}
    \Crefname{table}{Table}{Tables}
    \crefname{table}{Tab.}{Tabs.}
}

\newcommand{\fst}[1]{\textbf{#1}}

\newcommand{\pndf}{$\mathcal{P}$-NDF\xspace}
\definecolor{myblue}{HTML}{4C72b0}

\def\eg{e.g.~}

\usepackage[ruled]{algorithm2e} 

\begin{document}
\title{Position-Normal Manifold for Efficient Glint Rendering on High-Resolution Normal Maps}

\author{Liwen Wu}
\orcid{0009-0007-2773-2032}
\affiliation{%
  \institution{University of California San Diego}
  \country{USA}}
\email{liw026@ucsd.edu}

\author{Fujun Luan}
\orcid{0000-0001-5926-6266}
\affiliation{%
  \institution{Adobe Research}
  \country{USA}}
\email{fluan@adobe.com}

\author{Miloš Hašan}
\orcid{0000-0003-3808-6092}
\affiliation{%
  \institution{Adobe Research}
  \country{USA}}
\email{mihasan@adobe.com}

\author{Ravi Ramamoorthi}
\orcid{0000-0003-3993-5789}
\affiliation{%
  \institution{University of California San Diego}
  \country{USA}}
\email{ravir@cs.ucsd.edu}


\begin{abstract}
Detailed microstructures on specular objects often exhibit intriguing glinty patterns under high-frequency lighting,
which is challenging to render using a conventional normal-mapped BRDF.
In this paper, we present a manifold-based formulation of the glint normal distribution functions (NDF) that precisely captures the surface normal distributions over queried footprints.
The manifold-based formulation transfers the integration for the glint NDF construction to a problem of mesh intersections.
Compared to previous works that rely on complex numerical approximations,
our integral solution is exact and much simpler to compute,
which also allows an easy adaptation of a mesh clustering hierarchy to accelerate the NDF evaluation of large footprints.
Our performance and quality analysis shows that our NDF formulation achieves similar glinty appearance compared to the baselines but is an order of magnitude faster.
Within this framework, we further present a novel derivation of analytical shadow-masking for normal-mapped diffuse surfaces---a component that is often ignored in previous works.
\end{abstract}

%
%

%
%

\input{figures/teaser}
\maketitle

\input{1-introduction}
\input{2-relatedwork}

\input{3-method}

\input{4-experiments}
\input{5-conclusion}

\bibliographystyle{ACM-Reference-Format}
\bibliography{bibliography}
\newpage

\clearpage
\newpage
\begin{appendix}
    \input{6-supplemental.tex}
\end{appendix}
\end{document}

%% file: figures/teaser.tex
\begin{teaserfigure}
\centering
    \setlength\tabcolsep{0.0pt}
    \includegraphics[width=0.99\linewidth]{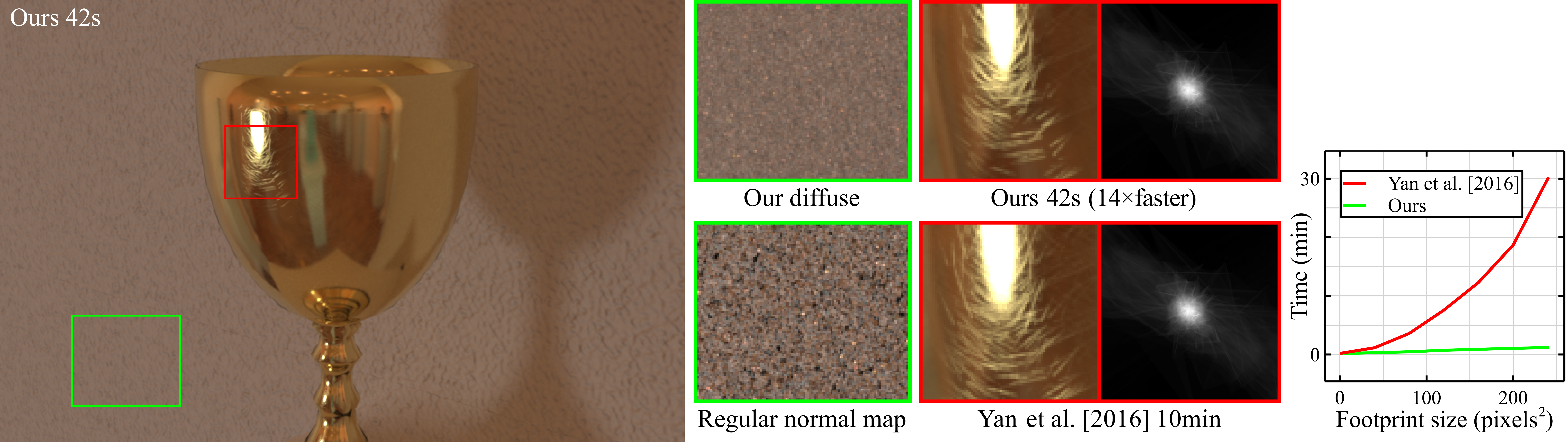}
    \caption{
\textbf{Our manifold-based \pndf{} vs the baselines.}
\citet{yan2014rendering,yan2016position} construct the normal distribution function of a footprint query (\pndf{}) and convolve it with a tiny amount of Gaussian roughness, which has no closed-form solution and requires slow numerical approximations.
Instead, we show the convolution can be avoided. Our representation converts the \pndf{} evaluation to simple manifold intersections, which is an exact solution with similar quality (insets with red outlines) but is much faster for large footprint size queries (bottom right plot) that benefit from a cluster hierarchy.
Furthermore, an analytical projected-area integration can be derived using our approach, which can be used for improved rendering of normal-mapped diffuse surfaces,
whereas the standard approach shows aliasing artifacts (insets with green outlines; rendered at 4 SPP).
}
    \label{fig:teaser}
\end{teaserfigure}

%% file: 1-introduction.tex
\section{Introduction}
\label{sec:introduction}
Many glossy objects in the real world show complex specular glinty appearance such as scratches and metallic flakes, 
which cannot be faithfully reproduced by conventional BRDF models with a smooth normal distribution function (NDF).
The works of \citet{yan2014rendering,yan2016position} capture the accurate distribution of normals defined over a high-resolution normal map,
which enable the detailed modeling of material glints, but are inefficient.
In this paper, we present a mesh-based framework of normal distribution evaluation that is accurate and efficient to compute.

\citet{yan2014rendering} treats the normal map of a specular surface as a delta position-normal distribution.
The normal distribution function of a surface patch (\pndf{}) is then formulated as the distribution of normals randomly taken from the patch, convolved by a tiny amount of Gaussian roughness.
This cannot be solved in closed-form and requires an expensive numerical integration using a piecewise polynomial approximation. 
Alternatively, the \pndf evaluation can be accelerated by approximating the graph of the normal map function with a 4D mixture of Gaussians~\cite{yan2016position}. 
However, to preserve the details, 
the resulting Gaussian mixture is very large and cannot be easily simplified. 
To achieve further speed-up, we turn to a different approach.

We show that, with some care, the Gaussian roughness convolution can be avoided, allowing us to treat the normal map function graph as a 2D manifold in the 4D position-normal space. 
In this formulation, \pndf{} evaluation is equivalent to projecting the 4D manifold onto the 2D normal plane (\cref{subsec:normal-mesh}).
This results in a simple mesh-intersection algorithm when representing the manifold by a mesh,
which is computationally much more efficient than the previous numerical integration.

Within a mesh-based framework,
we further introduce a mesh simplification approach to hierarchically cluster similar triangles to coarser grids (\cref{subsec:clustering}).
The \pndf calculation only needs to be performed on the simplified mesh for large footprint queries,
reducing the number of mesh intersection tests required.

Figure~\ref{fig:teaser} demonstrates that our manifold-based \pndf{} achieves similar results compared to the baselines,
while our evaluation speed is $\sim$14$\times$ faster than~\citet{yan2016position}; we achieve even more speedup for large footprint queries (\cref{subsec:performance}).
Our formulation further allows an analytical shadowing-masking solution for a piecewise constant footprint kernel,
which we show is a low frequency function for specular surfaces (\cref{subsec:shadow-masking}) but can be important for diffuse appearance modeling (\cref{subsec:shadow-masking-result}). 
In summary, our contributions include:
\begin{enumerate}
    \item a novel manifold-based \pndf formulation that is efficient to compute as a projection of a 4D mesh into 2D,
    \item a cluster hierarchy that further accelerates the above mesh projection for large footprint size queries, and
    \item an analytical shadowing-masking derivation that can also be used for anti-aliasing normal-mapped diffuse reflections.
\end{enumerate}

%% file: 2-relatedwork.tex
\section{Related work}
\label{sec:related-work}

\paragraph{Normal map filtering.}
Detailed surface reflection patterns are commonly modeled by the normal mapping technique,
which can produce aliasing when the pixel sampling rate is below the normal map frequency.
To properly filter the normal map,
conventional methods work on mip-mapping its statistics over texture patches and interpolate over the query footprint to reconstruct the underlying normal distribution function (NDF).
The NDF can be approximated by a Gaussian~\cite{toksvig2005mipmapping,olano2010lean,dupuy2013linear} that additionally incorporates normal variance as an extrinsic roughness to the normal-mapped BRDF;
\citet{chermain2020procedural,chermain2021importance,chermain2021real}, \citet{wu2019accurate}, and \citet{zhao2016downsampling} further construct a mixture of Gaussians.
Alternatively, the filtering can be performed in the frequency domain by mip-mapping the spherical harmonic coefficients of the NDF~\cite{han2007frequency}.
In recent years, learning-based approaches have been proposed to measure the anti-aliased BRDF parameters through neural networks~\cite{gauthier2022mipnet},
and neural networks can also be optimized to explicitly parameterize surface displacement and BRDFs with mip-mapping support~\cite{kuznetsov2021neumip,zeltner2023real}.
All of these methods successfully capture the meso-level normal variations (\eg small footprint queries) but over-smooth the high-frequency NDF details from the surface microstructures.
As demonstrated in Fig.~4 of \citet{yan2014rendering},
large footprint queries still lead to interesting NDFs that cannot be smoothly approximated for accurate renderings (\cref{fig:other-baselines}).

\paragraph{Glint NDF construction.}
The glinty appearance can be accurately modeled by constructing the footprint's exact normal distribution (\pndf{}) through a discrete or continuous formulation.
The discrete formulation~\cite{jakob2014discrete} treats the normal map texels as a set of facets and counts the number of reflecting facets within the footprint to obtain the reflection response.
which can be sped up using histograms~\cite{wang2018fast,atanasov2021multiscale}
or even brought to real time with improved facet counting strategies~\cite{zirr2016real,deliot2023real}.
Owing to its discontinuity, however, the discrete glint produces spiky highlights and cannot model curved surfaces with continuously changing normals.
Instead, the continuous glint model~\cite{yan2014rendering} explicitly integrates the normal distributions on the interpolated normal map to construct the \pndf{},
which gives smoother glinty effects but is difficult to accelerate.
While using a mixture of Gaussians approximation of the normal map graph~\cite{yan2016position} results in a faster \pndf computation than~\citet{yan2014rendering}’s numerical integration,
this method is still inefficient for large footprint queries.
\citet{deng2022constant} uses tensor decomposition to store a pre-computed \pndf{} over sampled locations,
which can be efficiently queried but is inaccurate owing to the spatial domain discretization.
Our method also works on the continuous normal map with acceleration structures that preserves the accurate normal distributions on the spatial domain yet is efficient for large footprint queries as shown in \cref{fig:teaser}'s plot.
Besides the \pndf construction, shadow-masking terms have only been studied in the discrete glint formulation~\cite{atanasov2021multiscale,chermain2020microfacet} with \citet{chermain2019glint} taking multiple scattering into account.
We show an accurate shadow-masking derivation of our continuous formulation that also extends the microstructure modeling to diffuse surfaces.

\paragraph{Other glinty appearance modeling methods.}
A comprehensive review can be found in the work of \citet{zhu2022recent}.
The pre-baking idea is also used by~\citet{raymond2016multi} to calculate the BRDF for repeated scratch patterns with multiple scattering support.
\citet{wang2020example}, \citet{tan2022real}, and \citet{zhu2019stationary} study normal map synthesis algorithms for glint rendering with low storage cost.
\citet{shah2024neural} use neural networks to properly interpolate $\mathcal{P}$-NDFs.
Rather than calculating the NDF, 
manifold exploration~\cite{jakob2012manifold} can also be applied to find the glinty specular paths~\cite{zeltner2020specular,wang2020path,fan2024specular},
and \citet{fan2022efficient} use differentiable renderings to evolve complex glint patterns from a simpler setup.
These methods, however, cannot be easily integrated into a standard path tracing pipeline.

%% file: 3-method.tex
\input{figures/manifold}
\section{Preliminaries}
\label{sec:background}
Our model, like previous works, is an extension of the Cook-Torrance BRDF model~\cite{cook1982reflectance,walter2007microfacet} with notations specified in \cref{tab:notation}:
\input{equations/brdf}%
Here $f$ is defined in the local shading frame, so the cosine terms $(\bm{\omega}_i)_z,(\bm{\omega}_o)_z$ are just the $z$ components. 
While $D$ is a statistical approximation for the entire surface in traditional microfacet theory,
it becomes the \pndf, an explicit distribution of normals on a specific footprint,
in \citet{yan2014rendering}'s glint rendering framework:
\input{equations/gndf}%
Here, $\mathbf{n}$ and $\mathbf{m}$ are on the projected hemisphere space that has only $xy$ components 
(\eg $\tilde{\mathbf{n}}_{xy}=\mathbf{n},\tilde{\mathbf{n}}_z=\sqrt{1\!-\!\Vert\mathbf{n}\Vert^2}$),
and the kernel $k_\mathbf{r}$ is selected to cover the footprint size (\eg measured by the ray differentials).
\cref{eq:gndf} is thus a pdf on the projected hemisphere domain. It may have singularities corresponding to zero Jacobian determinants,
which are avoided by \citeauthor{yan2014rendering} using Gaussian convolution with an intrinsic roughness.
Instead of convolution, we show in \cref{subsec:normal-mesh} that an alternative strategy by clamping is just as effective. Therefore, we can focus on directly accelerating eq. (\ref{eq:gndf}).

\paragraph{Sampling and evaluation}
For Monte Carlo rendering, 
it is necessary to be able to both sample and evaluate the \pndf.
$\mathbf{m}$ can be easily sampled from $D$ by first sampling the footprint kernel $\mathbf{u}\!\sim\!k_\mathbf{r}(\mathbf{u}\!-\!\mathbf{x})$ (\eg a 2D Gaussian of mean $\mathbf{x}$ and variance $\mathbf{r}^2$) then querying the normal $\mathbf{m}\!=\!\mathbf{n}(\mathbf{u})$;
and the \pndf evaluation is discussed in the following section.

\input{tables/notation}
\input{figures/normal_map}
\input{figures/ndf_comparison}
\section{Position-Normal Manifold}
\label{sec:method}
We treat the normal map $\mathbf{n}(\mathbf{u})$ as a manifold on the position-normal space,
for which evaluating \cref{eq:gndf} is finding a finite number of projections $\mathbf{u}_i$ to the normal plane $\mathbf{m}$ then taking a sum (\cref{fig:manifold}).
This is in closed form as long as $\mathbf{n}(\mathbf{u}_i)\!=\!\mathbf{m}$ is solvable.
For that purpose, we take a mesh-based manifold representation (\cref{subsec:normal-mesh}) that allows easy $\mathbf{u}_i$ finding accelerated by mesh clustering (\cref{subsec:clustering}).
Benefiting from the simplicity of our formulation, we also show a novel \pndf{} shadow-masking in \cref{subsec:shadow-masking} with detailed derivations in the supplementary.

\input{figures/acceleration}
\subsection{Mesh-based manifold representation}
\label{subsec:normal-mesh}
We connect adjacent normal map pixels, with their locations and normals, 
into a triangle mesh (\cref{fig:normal_map}).
This produces a 4D manifold parameterization $(\mathbf{u,n(u)})$ with normal query given by barycentric interpolation:
\input{equations/interpolation}%
Its Jacobian determinant is twice the triangle area projected to the normal space, denoted as normal triangles $\mathbf{n}(\triangle_\mathbf{u_0u_1u_2}),\mathbf{n}(\triangle_\mathbf{u_3u_2u_1})$:
\input{equations/jacobian}%
With this mesh-based representation,
the projection search is equivalent to finding every normal triangle $\mathbf{n}(\triangle_\mathbf{abc})$ that intersects $\mathbf{m}$ by checking the barycentric coordinate $(\lambda_0,\lambda_1,\lambda_2)\!=\!\text{bary}(\mathbf{m},\mathbf{n}(\triangle_\mathbf{abc}))$.
The \pndf evaluation then sums up the kernel contribution divided by the Jacobian determinant over all the intersections:
\input{equations/ndf_eval}%
$\mathbf{1}_{\mathbf{n}(\triangle_\mathbf{abc})}(\mathbf{m})$ checks $\forall\lambda_i\!\in\![0,1]$ to indicate the intersection, 
and $\mathbf{a}\lambda_0\!+\!\mathbf{b}\lambda_1\!+\!\mathbf{c}\lambda_2$ is the intersection's $\mathbf{u}$-coordinate.

Note that \citet{yan2014rendering} also triangulates a normal map but with a very different purpose:
they seek a numerical approximation of a convolved \pndf,
while \cref{eq:ndf_eval} gives an exact solution of \cref{eq:gndf}.
Both methods show very similar results as the intrinsic Gaussian roughness is generally small (\cref{fig:ndf_comparison}).
However, our approach is as simple as point-triangle intersections with kernel evaluations, which is much more efficient arithmetically.
Unlike \citeauthor{yan2014rendering} that restricts to Gaussian filters,
our evaluation can also apply arbitrary $k$, such as a cheaper disk filter, for further speedup (\cref{subsec:ablation}).

\paragraph{Preventing Jacobian singularity.}
Normal triangles can have zero area (\eg mirror reflection with identical vertex normals). Worse, triangles can be arbitrarily close to zero area, which causes unpleasant spiky highlights in renderings; this was the primary reason for introducing the convolution by Yan et al.
Instead, we simply replace the offending triangles with Jacobian smaller than $\epsilon\!=\!10^{-6}$ by equilateral ones with Jacobian exactly $\epsilon$. 
This amounts to clamping the Jacobian to a small value $\max(2\Vert\mathbf{n}(\triangle_{\mathbf{abc}})\Vert,\epsilon)$ and modifying the \pndf sampling 
to match the clamped pdf:
\input{equations/debias}%
$\text{EqTri}(\cdot,\cdot)$ warps $\mathbf{u}\!-\!\lfloor\mathbf{u}\rfloor$ in the unit square to an equilateral triangle centered at the origin of area $\epsilon/2$ which has Jacobian $\epsilon$. We could also use any other shape (e.g. a disk).

\subsection{Acceleration by mesh clustering}
\label{subsec:clustering}

The \pndf evaluation by point-triangle intersection is similar to ray tracing,
which can be accelerated by a bounding volume hierarchy.
Like~\citet{yan2014rendering} and~\citet{jakob2014discrete},
a min-max hierarchy of the normal triangle $\mathbf{n}(\triangle)$ is used to efficiently skip triangles that never intersect the normal query (\cref{fig:acceleration}a).
However, this gets slow as the query footprint size increases,
because there are too many intersection candidates to check.
Inspired by Nanite~\cite{karis2021nanite} and Lightcuts~\cite{walter2005lightcuts},
we build another hierarchy to group normal triangles into bigger clusters and check intersections over the cluster instead when it gives a good approximation.

Figure~\ref{fig:acceleration} shows a 1D-normal-1D-position example of the intersection computation with our cluster hierarchy.
Extending to the full 4D case,
each cluster at level $l$ simplifies triangles from a $2^l$$\times$$2^l$ sub-grid to a 1$\times$1 grid of two triangles with vertex normals $\mathbf{n}^l_0\cdots\mathbf{n}^l_3$,
yielding a barycentric interpolation $\mathbf{n}^l(\mathbf{u}/2^l)$ to approximate $\mathbf{n}(\mathbf{u})$.
These normals are selected by minimizing the L2 distance between $\mathbf{n}^l(\mathbf{u}/2^l)$ and $\mathbf{n}(\mathbf{u})$,
which is weighted by the inverse Jacobian to match the triangle's contribution to \cref{eq:ndf_eval}:
\input{equations/normal_optimization}%
Because $\mathbf{n}^l(\mathbf{u}/2^l)$ for $\mathbf{n}_i^l$ is linear, 
the normals' gradient in \cref{eq:normal_optimization} is in the form
$\mathbf{A}(\mathbf{n}^l_0,\mathbf{n}^l_1,\mathbf{n}^l_2,\mathbf{n}^l_3)\!^\top\!+\!\mathbf{B}$ (see derivation of $\mathbf{A,B}$ in supplementary),
so the cluster normals can be estimated by least squares with the residual $e^l$ indicating the approximation error:
\input{equations/normal_optimization2}%
To get a cut of the cluster tree that best approximates the \pndf,
we use a heuristic that the residual should satisfy $e^l\!\leq\!\mathbf{r}_x\mathbf{r}_y\tau$ for a pre-defined threshold $\tau$ and the kernel footprint size $\mathbf{r}$.
Given a \pndf query, we therefore traverse from top to bottom (pruned by the min-max hierarchy) until the criterion is met,
at which point $\mathbf{n}^l(\mathbf{u}/2^l)$ is directly used for the \pndf calculation.
The \pndf sampling (\cref{eq:debias}) should also query $\mathbf{n}^l(\mathbf{u}/2^l)$ rather than the original normal map $\mathbf{n}(\mathbf{u})$ to ensure consistent sampling and evaluation pdfs.

In practice, our heuristic with $\tau\!=\!10^{-3}\!\sim\!10^{-4}$ gives reasonable \pndf reconstruction without hurting the rendering over a range of different footprint sizes (\cref{fig:cluster_comparison}),
but it noticeably reduces the number of intersection tests.
Meanwhile, both our min-max and cluster hierarchy are perfectly balanced quad trees,
which can be compactly stored as mip-maps with traversal as efficient as texture fetching.
All these acceleration strategies allow our \pndf evaluation to stay fast especially for the large footprint size query (\cref{subsec:performance}).

\input{figures/cluster_comparison}
\subsection{Shadow-masking}
\label{subsec:shadow-masking}
For the actual rendering with a \pndf, 
it is also necessary to know the shadow-masking term~\cite{smith1967geometrical,ashikmin2000microfacet} $G(\bm{\omega}_i,\bm{\omega}_o,\mathbf{m})\!=\!\tfrac{H(\tilde{\mathbf{m}}\!^\top\!\bm{\omega}_i)H(\tilde{\mathbf{m}}\!^\top\!\bm{\omega}_o)}{(1+\Lambda(\bm{\omega}_i))(1+\Lambda(\bm{\omega}_o))}$ or its height-correlated version \cite{ross2005detailed} $\tfrac{H(\tilde{\mathbf{m}}\!^\top\!\bm{\omega}_i)H(\tilde{\mathbf{m}}\!^\top\!\bm{\omega}_o)}{1+\Lambda(\bm{\omega}_i)+\Lambda(\bm{\omega}_o)}$. 
$\Lambda(\bm{\omega})\!=\!\tfrac{P(\!\bm{\omega}\!)}{\bm{\omega}_z}\!-\!1$ depends on the projected area $P(\bm{\omega})$, so the key is to solve the integral:
\input{equations/lambda}%
To that end, we approximate the footprint kernel to be piecewise constant for each triangle of the normal map and show the integral is tractable in this situation.

\input{figures/integral_domain}
\paragraph{Analytical projected area.}
Given a query $\bm{\omega}$, 
we first rotate the xy-plane by the angle $-\!\arctan{\tfrac{\bm{\omega}_y}{\bm{\omega}_x}}$ to let $\bm{\omega}_y=0$.
Under this canonical setting,
$\max(\cdot)$ in \cref{eq:lambda} clamps the integral domain to the region with $\tilde{\mathbf{m}}\!^\top\!\bm{\omega}\geq 0$, 
a semi-circle and a semi-ellipse $(\tfrac{\mathbf{m}_x}{\bm{\omega}_z})^2+\mathbf{m}_y^2=1$ on the projected hemisphere (proof in supplementary);
and \cref{eq:ndf_eval} restricts the integral domain to $\mathbf{n}(\triangle)$ for each triangle $\triangle$.
Their intersection gives the final integral domain $\mathbf{n}^+(\triangle)$,
whose boundary consists of $M$ segments with endpoints $\mathbf{n}^i,\mathbf{n}^{i+1}$ ($\mathbf{n}^{M}\!=\!\mathbf{n}^0$) that are either lines or ellipse arcs (\cref{fig:integral_domain}).
The endpoints of the ellipse arcs are obtained by solving the semi-ellipse's intersection with the triangle edges,
which is a quadratic equation (details in supplementary). 
Let $k_\mathbf{r}(\triangle)$ denote the constant kernel value for each triangle.
$P(\bm{\omega})$ is then a weighted sum of area integrals over $\mathbf{n}^+(\triangle)$:
\input{equations/project}%
By applying Stokes' theorem, each area integral can be converted to line integrals of the boundary line segments and the ellipse arcs,
both of which have closed-form solutions (derivations in supplementary):
\input{equations/line-integral}%
As shown in \cref{fig:projected_area} left,
the equations above give the exact projected-area integral for a box filter,
and the picewise constant approximation in general is close to the Monte Carlo reference.

\paragraph{Approximation by a smooth \pndf}
The \pndf's projected area is a low-frequency function because the specular surface normal has small variation,
and $\max(\tilde{\mathbf{m}}\!^\top\!\bm{\omega},0)$ is low-pass filtering the $D$.
Therefore, it is reasonable to use a smooth GGX~\cite{walter2007microfacet} projected area $P'(\bm{\omega})$ for efficient approximation by fitting its roughness $\bm{\alpha}$ and tangent frame $\mathbf{Q}$ parameters:
\input{equations/ggx_projected}%
As suggested by~\citet{heitz2015sggx},
$P'(\bm{\omega})^2=\bm{\omega}^{\smash{\top}}_{\smash{xy}}\bm{\Omega}\bm{\omega}_{\smash{xy}}$ for $\bm{\omega}_z\!=\!0$ is a linear function of $\bm{\Omega}$.
We thus can solve $\bm{\Omega}$ to match $P(\bm{\omega})^2$ sampled in grazing angles by least squares then apply eigenvalue decomposition to get $\bm{\alpha}$ and $\mathbf{Q}$.
These are calculated at every grid center and mip-level of the normal map then interpolated for inference.
As shown in \cref{fig:projected_area} right,
the approximation mainly differs with the ground truth in grazing angles, 
yet the error is very minimal.
So, we only need to use the approximation in our experiments when dealing with specular reflections.
For a diffuse surface with large normal variations,
accurate projected area still exhibits spatial details that requires our analytical formulation as described below.

\input{figures/projected_area}
\paragraph{Application to diffuse surfaces.}
Similar to specular reflections, 
aliasing issues with the normal-mapped diffuse reflections can be improved by explicitly integrating the normal-mapped BRDF response within the queried footprint: 
\input{equations/glint_diffuse}%
$(\tilde{\mathbf{n}}(\mathbf{u}))_z$ is the projection factor~\cite{dupuy2013linear};
and dividing the result above by the incident cosine term,
we obtain an aggregated diffuse BRDF $f_d=\tfrac{P(\bm{\omega}_i)}{\pi(\bm{\omega}_i)_z}$ that directly relates to the projected area integral.
By applying our analytical projected area,
$f_d$ can thus anti-alias while preserving the diffuse appearances when one pixel covers too many normal map details (\cref{subsec:shadow-masking-result}).

%% file: figures/manifold.tex
\begin{figure*}[t]
    \centering
    \setlength\tabcolsep{1.0pt}
    \includegraphics[width=0.99\linewidth]{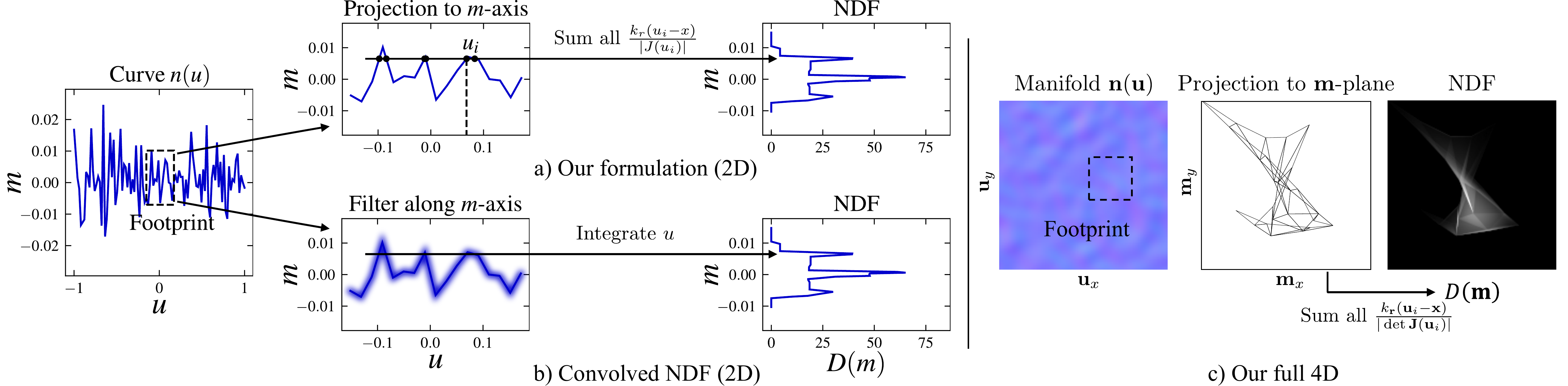}
    \caption{
    \textbf{Our position-normal manifold formulation}
    a) converts the \pndf{} integration to finding the manifold projections $\mathbf{u}_i$ followed by accumulating a finite number of $\tfrac{k_\mathbf{r}(\mathbf{u}_i-\mathbf{x})}{\vert\det\mathbf{J}(\mathbf{u}_i)\vert}$.
In contrast, b) \citet{yan2014rendering}’s convolved formulation requires computing a complex integral to reason about the NDF.
The left images show toy examples of 1D normal and 1D position, and c) shows the full 4D case.
    }
    \label{fig:manifold}
\end{figure*}

%% file: equations/brdf.tex
\begin{equation}
    f(\bm{\omega}_i,\bm{\omega}_o,\mathbf{x})=\frac{F(\bm{\omega}_o,\mathbf{m})G(\bm{\omega}_i,\bm{\omega}_o,\mathbf{m})D(\mathbf{m},\mathbf{x})}{4(\bm{\omega}_i)_z(\bm{\omega}_o)_z}.
    \label{eq:brdf}
\end{equation}

%% file: equations/gndf.tex
\begin{equation}
\small
    D(\mathbf{m},\mathbf{x})\!=\!\int k_\mathbf{r}(\mathbf{u}\!-\!\mathbf{x})\delta(\mathbf{n}(\mathbf{u})\!-\!\mathbf{m})\mathrm{d}\mathbf{u}\!=\!\sum_{\forall \mathbf{n}(\mathbf{u}_i)=\mathbf{m}} \frac{k_\mathbf{r}(\mathbf{u}_i\!-\!\mathbf{x})}{\vert\det\mathbf{J}(\mathbf{u}_i)\vert}.
    \label{eq:gndf}
\end{equation}

%% file: tables/notation.tex
\begin{table}[t]
    \centering
    \caption{\textbf{Notations.}}
    \resizebox{0.99\linewidth}{!}{
    \begin{tabular}{ll}
    \toprule
    Symbol & Definition\\
    \midrule
         $\square_{x},\square_{xy}$ & vector swizzle operation\\
         $\delta(\cdot),H(\cdot)$ & Dirac delta and Heaviside function\\
         $\mathbf{u}$ & unnormalized texture coordinate\\
         $\mathbf{x}$ & ray intersection in texture space\\
         $\bm{\omega}_i,\bm{\omega}_o$ & incident (light), outgoing (viewing) directions\\
         $\tilde{\mathbf{m}},\tilde{\mathbf{n}}(\mathbf{u})$ & micro-normal (half vector): $\tfrac{\bm{\omega}_i+\bm{\omega}_o}{\Vert\bm{\omega}_i+\bm{\omega}_o\Vert_2}$, normal map\\
         $\mathbf{m},\mathbf{n}(\mathbf{u})$ & micro-normal, normal map on projected hemisphere\\
         $\mathbf{J}(\mathbf{u})$ & Jacobian of $\mathbf{n}(\mathbf{u})$\\
         $\mathbf{u}_{0},\mathbf{u}_1,\mathbf{u}_2,\mathbf{u}_3$ & 
         $(\lfloor\!\mathbf{u}_x\!\rfloor,\lfloor\!\mathbf{u}_y\!\rfloor),
         (\lceil\!\mathbf{u}_x\!\rceil,\lfloor\!\mathbf{u}_y\!\rfloor),
         (\lfloor\!\mathbf{u}_x\!\rfloor,\lceil\!\mathbf{u}_y\!\rceil),
         (\lceil\!\mathbf{u}_x\!\rceil,\lceil\!\mathbf{u}_y\!\rceil)$\\
         $\triangle_{\mathbf{abc}},|\triangle_{\mathbf{abc}}|$ &
         triangle with vertices $\mathbf{a,b,c}$ and its (signed) area\\
         $\mathbf{n}(\triangle_\mathbf{abc})$ & (normal) triangle with vertices $\mathbf{n(a),n(b),n(c)}$\\
         $\text{bary}(\mathbf{x},\triangle_\textbf{abc})$ &
         barycentric coordinates: $
         \tfrac{|\triangle_\textbf{xbc}|}{|\triangle_\textbf{abc}|},
         \tfrac{|\triangle_\textbf{axc}|}{|\triangle_\textbf{abc}|},
         \tfrac{|\triangle_\textbf{abx}|}{|\triangle_\textbf{abc}|}
         $\\
         $k_\mathbf{r}(\mathbf{u}-\mathbf{x})$ & footprint kernel of size $2\mathbf{r}_x\!\times\!2\mathbf{r}_y$ ($k\!=\!0\;\forall \vert\mathbf{x\!-\!u}\vert\!>\!\mathbf{r}$)\\
         $\triangle \in k_\mathbf{r}$ & triangle within $[\mathbf{x}_x\!-\!\mathbf{r}_x,\mathbf{x}_x\!+\!\mathbf{r}_x]\!\times\![\mathbf{x}_y\!-\!\mathbf{r}_y,\mathbf{x}_y\!+\!\mathbf{r}_y]$\\
         $\mathbf{1}_{\triangle}\!(\mathbf{x})$ & indicator function (if point $\mathbf{x}$ intersects triangle $\triangle$)\\
         $D(\mathbf{m},\mathbf{x})$ & normal distribution function of a patch (\pndf{})\\
         $G(\bm{\omega}_i,\bm{\omega}_o,\mathbf{m})$ & shadow-masking term 
         \\
         $F(\bm{\omega}_o,\mathbf{m})$ & Fresnel term\\
    \bottomrule
    \end{tabular}
    }
    \label{tab:notation}
\end{table}

%% file: figures/normal_map.tex
\begin{figure}[t]
    \centering
    \setlength\tabcolsep{1.0pt}
    \includegraphics[width=0.99\linewidth]{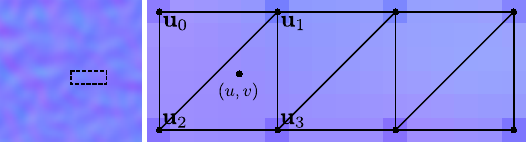}
    \caption{
    \textbf{Normal map texels are placed on a triangle mesh grid} ($\mathbf{u}_0\cdots\mathbf{u}_3$),
    and barycentric interpolation is used to create the continuous $\mathbf{n}(\mathbf{u})$.
    The right image shows the zoom-in of the dotted region.
    }
    \label{fig:normal_map}
\end{figure}

%% file: figures/ndf_comparison.tex
\begin{figure}[t]
    \centering
    \setlength\tabcolsep{1.0pt}
    \resizebox{0.8\linewidth}{!}{
    \begin{tabular}{ccc}
    \includegraphics[trim={2 8 6 0},clip,width=0.3\linewidth]{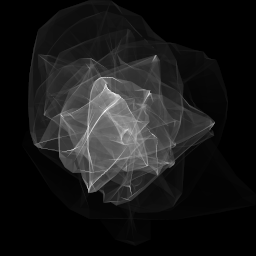}&
    \includegraphics[trim={2 8 6 0},clip,width=0.3\linewidth]{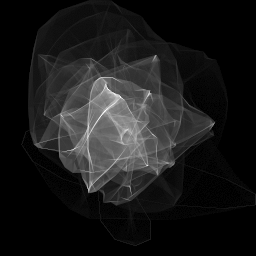}&
    \includegraphics[trim={2 8 6 0},clip,width=0.3\linewidth]{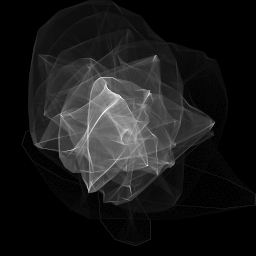}\\
    Our evaluation & Binning & With convolution
    \end{tabular}
    }
    \caption{
    \textbf{Comparison of \pndf{} evaluation.} 
    Our analytical evaluation matches the reference given by the binning approach~\cite{yan2014rendering}.
    It is also close to \citeauthor{yan2014rendering}'s convolved formulation with small intrinsic roughness ($10^{-4}$ here).
    }
    \label{fig:ndf_comparison}
\end{figure}

%% file: figures/acceleration.tex
\begin{figure*}[t]
    \centering
    \setlength\tabcolsep{1.0pt}
    \includegraphics[width=0.99\linewidth]{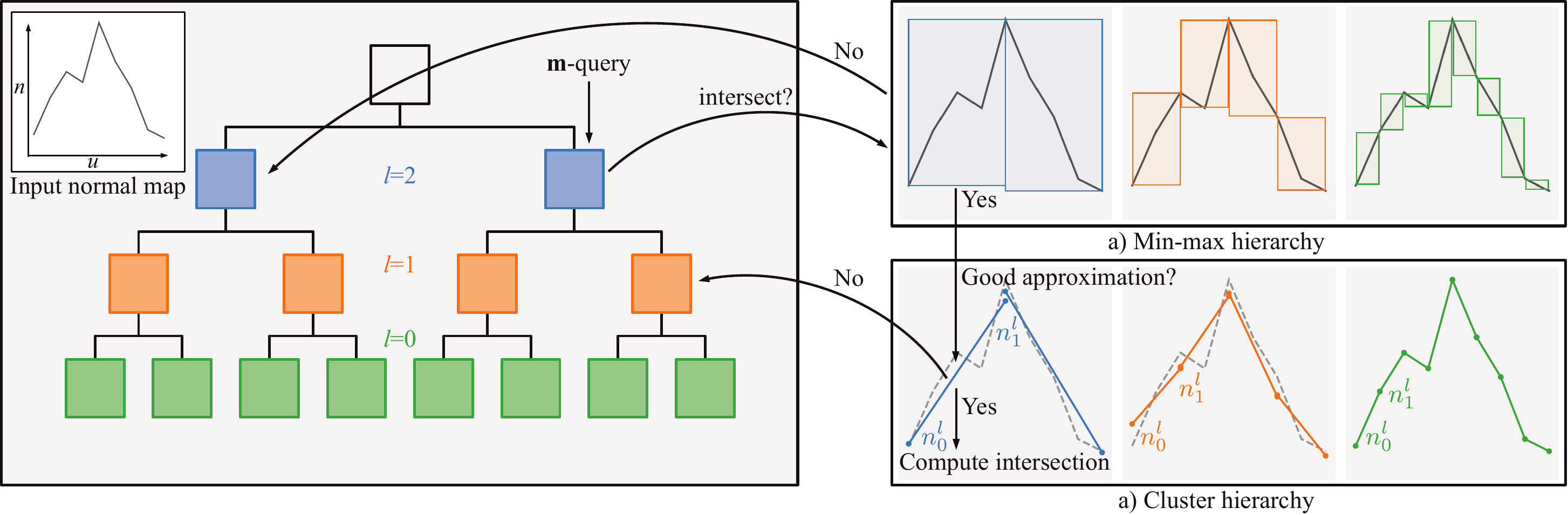}
    \vspace{-4pt}
    \caption{
    \textbf{Acceleration structures used by our method,} shown as a 2D toy example. 
    a)~a min-max hierarchy records the normal triangles' bounding box for every $2^l\!\times\!2^l$ spatial region (shown as $2^l$ here),
    which helps prune out the never-intersected triangles.
    b)~a cluster hierarchy simplifies the normal map into coarser grids (clusters) and is partitioned by a cut according to the error criteria;
    the intersection only needs to be checked against the clusters on the cut.
    }
    \label{fig:acceleration}
\end{figure*}

%% file: equations/interpolation.tex
\begin{equation}
\begin{aligned}
    \mathbf{n}(\mathbf{u})=
    \begin{cases}
        \mathbf{n}_0(1-u-v)+
        \mathbf{n}_1u+\mathbf{n}_2v & u+v\!<\!1\\
        \mathbf{n}_3(u+v-1)+
        \mathbf{n}_2(1\!-\!u)+\mathbf{n}_1(1\!-\!v)
         & \text{otherwise}
    \end{cases}\\
\text{where } (u,v)=\mathbf{u}\!-\!\lfloor\mathbf{u}\rfloor, \mathbf{n}_i=\mathbf{n}(\mathbf{u}_i).
\end{aligned}
    \label{eq:interpolation}
\end{equation}

%% file: equations/jacobian.tex
\begin{equation}
\small
|\det\mathbf{J}(\mathbf{u})|\!=\!
\begin{cases}
    |(\mathbf{n}_2\!-\!\mathbf{n}_0)\!\times\!(\mathbf{n}_1\!-\!\mathbf{n}_0)| 
    = 2\Vert\mathbf{n}(\triangle_\mathbf{u_0u_1u_2})\Vert& 
    u+v\!<\!1\\
    |(\mathbf{n}_2\!-\!\mathbf{n}_3)\!\times\!(\mathbf{n}_1\!-\!\mathbf{n}_3)|
    = 2\Vert\mathbf{n}(\triangle_\mathbf{u_3u_2u_1})\Vert& 
    \text{otherwise}
\label{eq:jacobian}
\end{cases}\!.
\end{equation}

%% file: equations/ndf_eval.tex
\begin{equation}
D(\mathbf{m},\mathbf{x})
=\sum_{\forall \triangle_\mathbf{abc}\in k_\mathbf{r}}
\frac{
k_\mathbf{r}(
\mathbf{a}\lambda_0\!+\!\mathbf{b}\lambda_1\!+\!\mathbf{c}\lambda_2\!-\!\mathbf{x})
\mathbf{1}_{\mathbf{n}(\triangle_\mathbf{abc})}\!(\mathbf{m})}
{2\Vert\mathbf{n}(\triangle_{\mathbf{abc}})\Vert}.
    \label{eq:ndf_eval}
\end{equation}

%% file: equations/debias.tex
\begin{equation}
\begin{aligned}
    \mathbf{m}\!=\!
    \begin{cases}
        \mathbf{n}(\mathbf{u}) & \vert\det\mathbf{J}(\mathbf{u})\vert\geq \epsilon\\        
        \mathbf{n}(\lfloor\mathbf{u}\rfloor\!+\!\tfrac{1}{2})\!+\!\text{EqTri}(\mathbf{u}\!-\!\lfloor\mathbf{u}\rfloor,\tfrac{\epsilon}{2}) & \text{otherwise}
    \end{cases}\\
    \text{where }  \mathbf{u}\!\sim\!k_\mathbf{r}(\mathbf{u}\!-\!\mathbf{x}). 
\end{aligned}
    \label{eq:debias}
\end{equation}

%% file: equations/normal_optimization.tex
\begin{equation}
(\mathbf{n}^l_0,\mathbf{n}^l_1,\mathbf{n}^l_2,\mathbf{n}^l_3)=
\underset{(\mathbf{n}^l_0,\mathbf{n}^l_1,\mathbf{n}^l_2,\mathbf{n}^l_3)}{\text{argmin}}\int 
    \frac{\Vert \mathbf{n}^l(\mathbf{u}/2^l)\!-\!\mathbf{n}(\mathbf{u})\Vert^2}{\vert\det\mathbf{J}(\mathbf{u})\vert} \mathrm{d}\mathbf{u}.
\label{eq:normal_optimization}
\end{equation}

%% file: equations/normal_optimization2.tex
\begin{equation}
    (\mathbf{n}^l_0,\mathbf{n}^l_1,\mathbf{n}^l_2,\mathbf{n}^l_3)\!^\top
    \!=\!-(\mathbf{A}\!^\top\mathbf{A})\!^{-1}\mathbf{A}^\top\mathbf{B},
    e^l\!=\!\Vert\mathbf{A}
    (\mathbf{n}^l_0,\mathbf{n}^l_1,\mathbf{n}^l_2,\mathbf{n}^l_3)\!^\top
    \!+\mathbf{B}\Vert_2.
    \label{eq:normal_optmization2}
\end{equation}

%% file: figures/cluster_comparison.tex
\begin{figure}[t]
    \centering
    \setlength\tabcolsep{1.0pt}
    \resizebox{0.99\linewidth}{!}{
    \begin{tabular}{cccc}
    \multicolumn{4}{c}{With mesh cluster hierarchy}\\
    \includegraphics[width=0.3\linewidth]{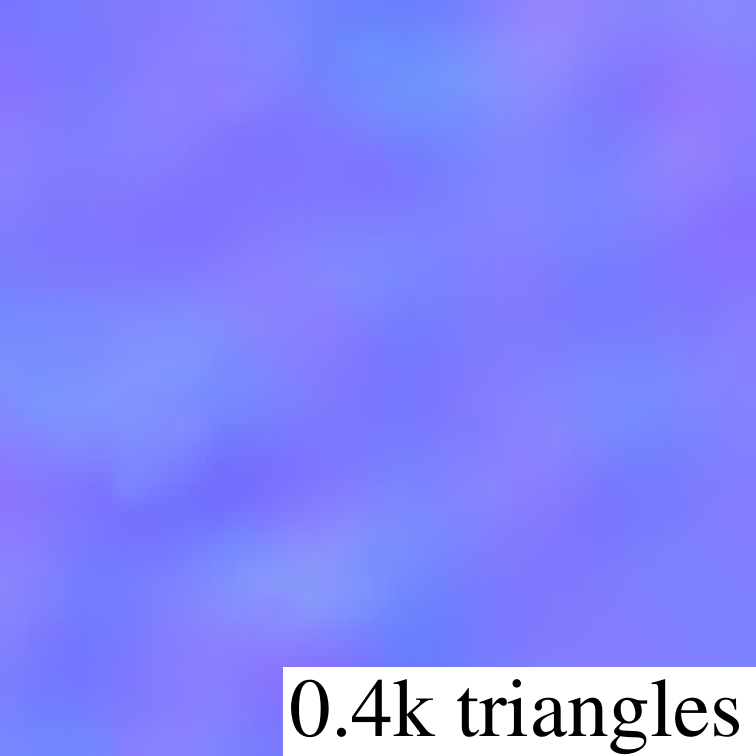}&
    \includegraphics[width=0.3\linewidth]{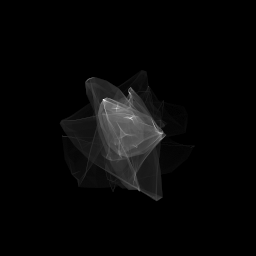}&
    \includegraphics[width=0.3\linewidth]{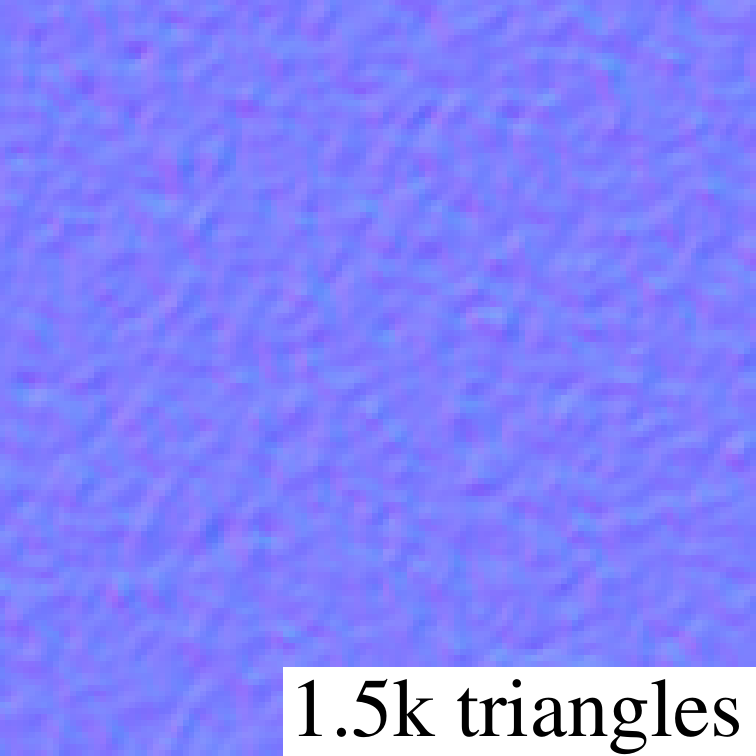}&
    \includegraphics[width=0.3\linewidth]{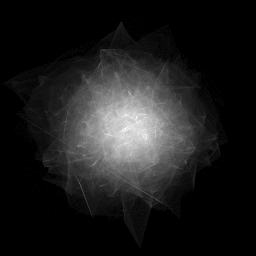}\\
    \multicolumn{4}{c}{Ground truth}\\
    \includegraphics[width=0.3\linewidth]{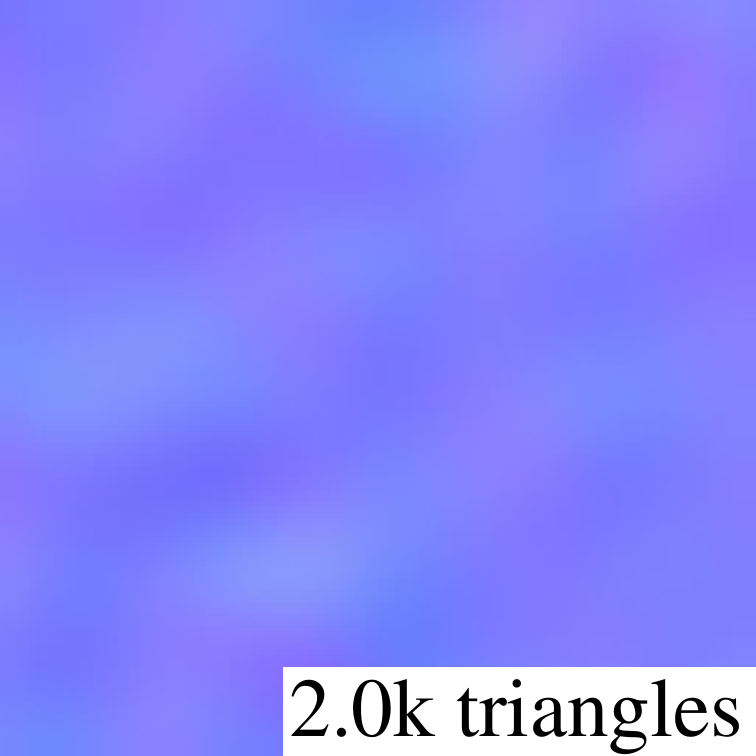}&
    \includegraphics[width=0.3\linewidth]{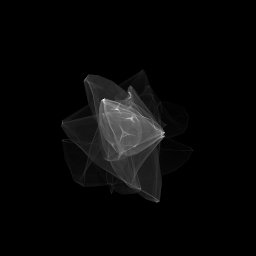}&
    \includegraphics[width=0.3\linewidth]{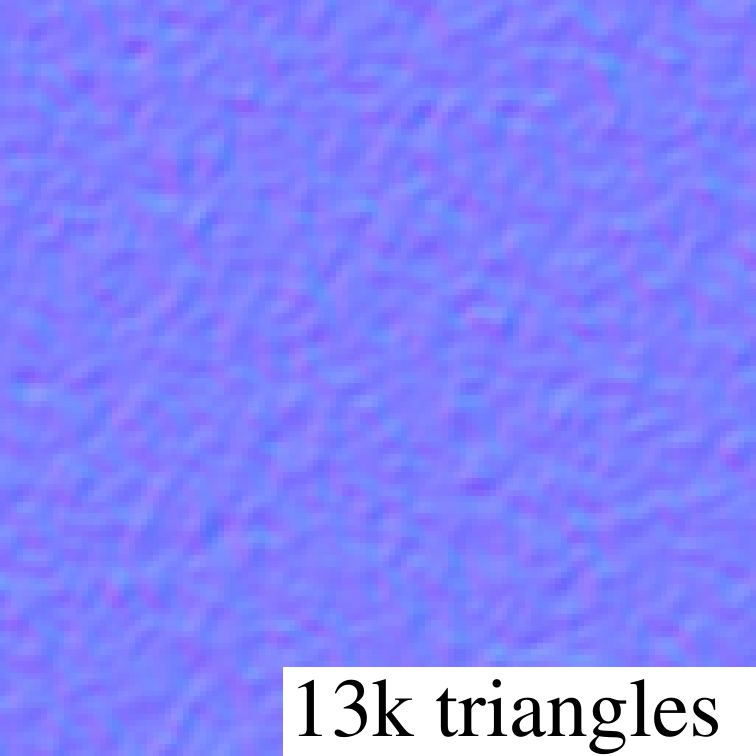}&
    \includegraphics[width=0.3\linewidth]{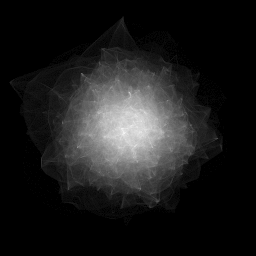}\\
    
    \multicolumn{2}{c}{$32\times32$ footprint} & 
    \multicolumn{2}{c}{$256\times256$ footprint}\\
    
    \end{tabular}
    }
    \caption{
    \textbf{Mesh cluster hierarchy} successfully uses fewer triangles to represent the normal map (column 1,3).
    This works for \pndf{} evaluations of both small (column 2) and large (column 4) footprint.
    }
    \label{fig:cluster_comparison}
\end{figure}

%% file: equations/lambda.tex
\begin{equation}
    P(\bm{\omega})=\int D(\mathbf{m},\mathbf{x})\max{(\tilde{\mathbf{m}}\!^\top\!\bm{\omega},0)}\mathrm{d}\tilde{\mathbf{m}}.
    \label{eq:lambda}
\end{equation}

%% file: figures/integral_domain.tex
\begin{figure}[t]
    \centering
    \setlength\tabcolsep{1.0pt}
    \includegraphics[width=0.95\linewidth]{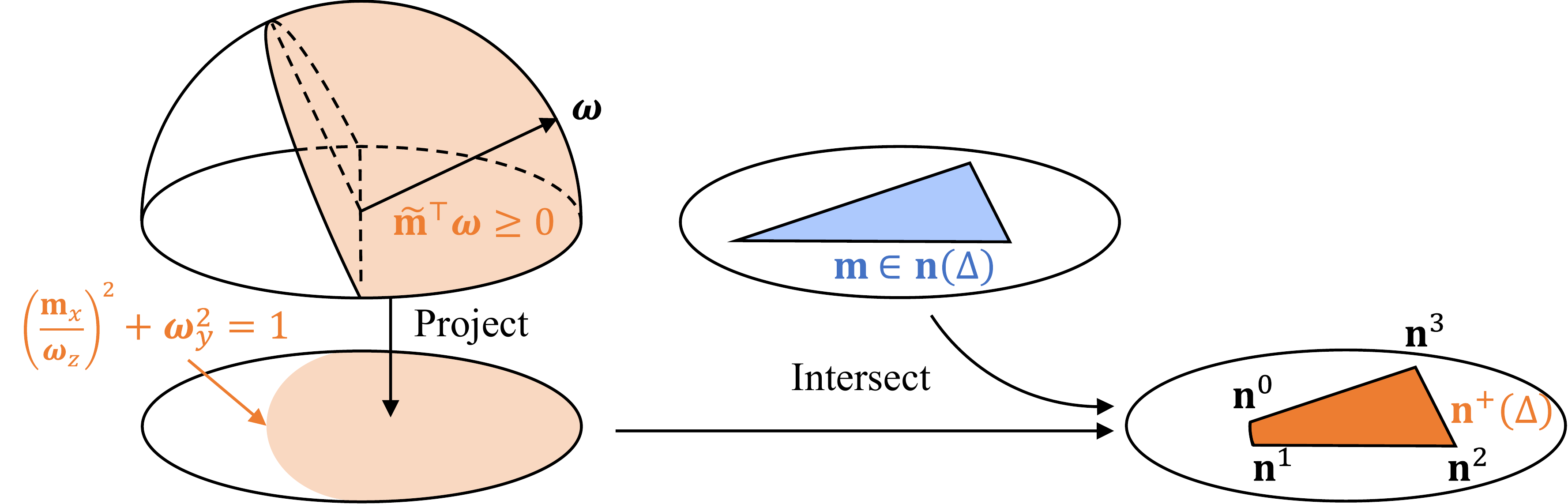}
    \caption{
    \textbf{Projected area integral domain for each triangle}
    is the intersection of the normal triangle (middle) and $\bm{\omega}$'s visible normals (top left) on the projected hemisphere (bottom left).
    Its boundary (right) consists of lines (\eg $\mathbf{n}^1\mathbf{n}^2$) and ellipse arcs (\eg $\mathbf{n}^0\mathbf{n}^1$).
    }
    \label{fig:integral_domain}
\end{figure}

%% file: equations/project.tex
\begin{equation}
\begin{aligned}
P(\bm{\omega})
=\int \sum_{\forall \triangle\in k_\mathbf{r}}
\tfrac{
k_\mathbf{r}(\triangle)
\mathbf{1}_{\mathbf{n}(\triangle_\mathbf{abc})}\!(\mathbf{m})
}
{2\Vert\mathbf{n}(\triangle)\Vert}
\max{(\tilde{\mathbf{m}}\!^\top\!\bm{\omega},0)}\mathrm{d}\tilde{\mathbf{m}}
\\
=\!  
\sum_{\forall \triangle \in k_\mathbf{r}}
\tfrac{k_\mathbf{r}(\triangle)}{2\Vert\mathbf{n}(\triangle)\Vert}
 \int_{\mathbf{n}(\triangle)}
 \!\tfrac{\max(\tilde{\mathbf{m}}\!^\top\!\bm{\omega},0)}{\tilde{\mathbf{m}}_z}\mathrm{d}\mathbf{m}\\
=\sum_{\forall \triangle \in k_\mathbf{r}}
\tfrac{k_\mathbf{r}(\triangle)}{2\Vert\mathbf{n}(\triangle)\Vert}
 \!\int_{\mathbf{n}^+(\triangle)}\!(
\bm{\omega}_x\tfrac{\tilde{\mathbf{m}}_x}{\tilde{\mathbf{m}}_z}+\bm{\omega}_z)\mathrm{d}\mathbf{m}.
\end{aligned}
\label{eq:area-integral}
\end{equation}

%% file: equations/line-integral.tex
\begin{align}
\int_{\mathbf{n}^+(\triangle)}\!(
\bm{\omega}_x\tfrac{\tilde{\mathbf{m}}_x}{\tilde{\mathbf{m}}_z}+\bm{\omega}_z)\mathrm{d}\mathbf{m}
\!=\!\sum_{i=0}^{M-1} \oint_{\mathbf{n}^i}^{\mathbf{n}^{i+1}}\!
(\bm{\omega}_z\tilde{\mathbf{m}}_x-\bm{\omega}_x\tilde{\mathbf{m}}_z)\mathrm{d}\tilde{\mathbf{m}}_y\\
\begin{aligned}
=\sum_{i=0}^{M-1}
\begin{cases}
\small
\tfrac{\bm{\omega}_z}{2}(\mathbf{n}^{i+1}_y-\mathbf{n}^{i}_y)(\mathbf{n}^{i+1}_x\!+\!\mathbf{n}^i_x) &\\[-1ex]
    + \tfrac{\bm{\omega}_x}{2}r^2\mathbf{d}_y\left[\arcsin{p}+p\sqrt{1\!-\!p^2}\right]^{\mathbf{d}\!^\top\!\mathbf{n}^{i+1}/r}_{\mathbf{d}\!^\top\!\mathbf{n}^i/r} & \text{for a line}\\[3ex]
\tfrac{1}{2}\left[\arcsin{p}+p\sqrt{1-p^2}\right]_{\mathbf{n}^{i+1}_y}^{\mathbf{n}^{i}_y} &
\text{for an arc}\\
\end{cases}\\
\mathbf{d}=(\mathbf{n}^{i+1}\!-\!\mathbf{n}^{i})/\Vert\mathbf{n}^{i+1}\!-\!\mathbf{n}^{i}\Vert_2,
\quad
r = \sqrt{1-(\mathbf{d}_{\smash{y}}\mathbf{n}_{\smash{x}}^{\smash{i}}\!-\!\mathbf{d}_{\smash{x}}\mathbf{n}_{\smash{y}}^{\smash{i}})^{\smash{2}}}.
\end{aligned}
\label{eq:line-integral}
\end{align}

%% file: equations/ggx_projected.tex
\begin{equation}
    P'(\bm{\omega})=
\tfrac{1}{2}\bm{\omega}_{z}+\sqrt{\bm{\omega}_{\smash{z}}^{\smash{2}}+\bm{\omega}^{\smash{\top}}_{\smash{xy}}\bm{\Omega}\bm{\omega}_{\smash{xy}}},
\quad
\bm{\Omega}=\bm{Q}^\top\text{diag}(\bm{\alpha^2})\bm{Q}.
\label{eq:our-ggx}
\end{equation}

%% file: figures/projected_area.tex
\begin{figure}[t]
    \centering
    \setlength\tabcolsep{0.2pt}
    \resizebox{0.99\linewidth}{!}{
    \begin{tabular}{c@{\hskip 2pt}|@{\hskip 2pt}c}
         \begin{tabular}{ccc}
    \multicolumn{3}{c}{Box filter}\\
    \includegraphics[width=0.3\linewidth]{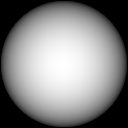}&
    \includegraphics[width=0.3\linewidth]{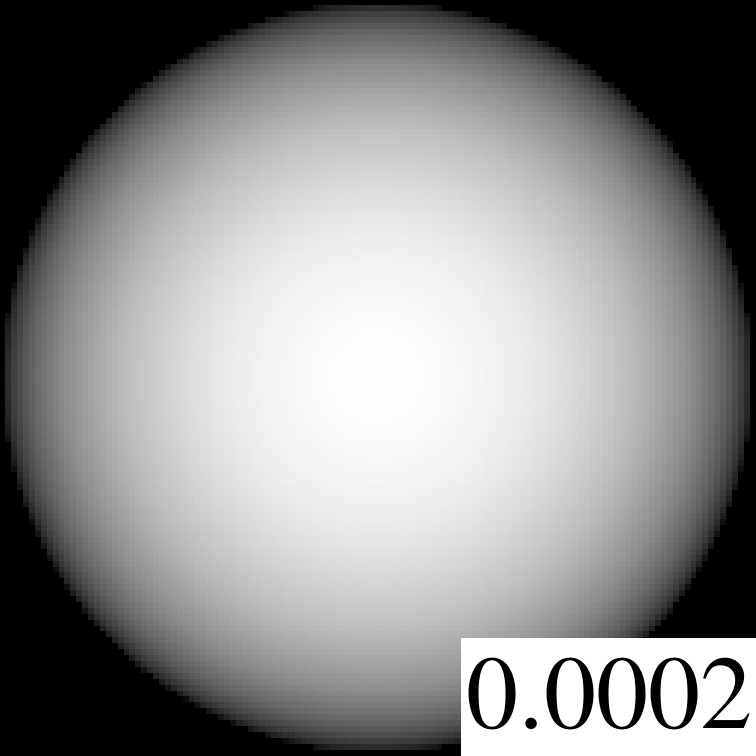}&
    \includegraphics[width=0.3\linewidth]{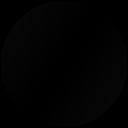}\\

    \multicolumn{3}{c}{Gaussian filter}\\
    \includegraphics[width=0.3\linewidth]{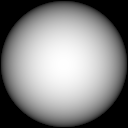}&
    \includegraphics[width=0.3\linewidth]{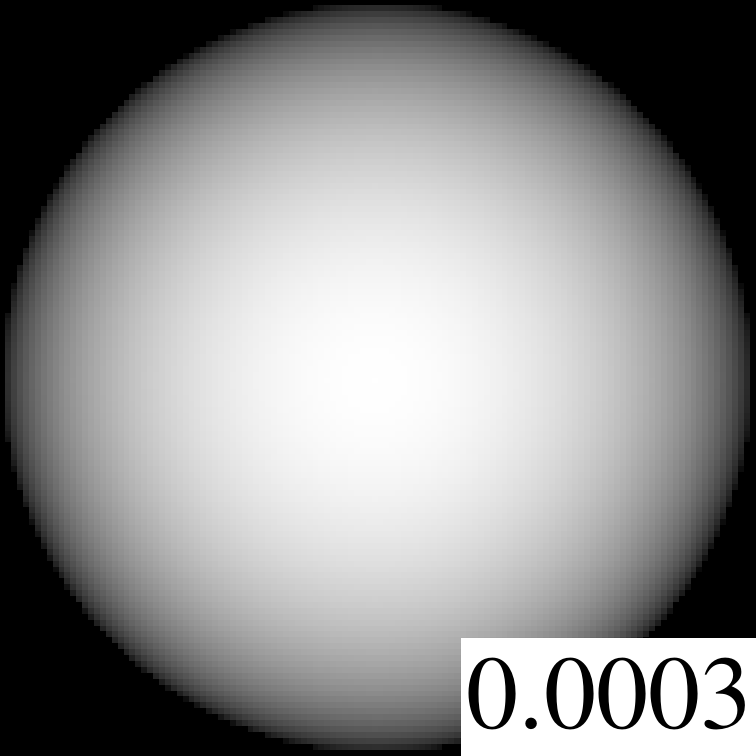}&
    \includegraphics[width=0.3\linewidth]{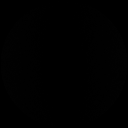}\\

    Monte Carlo & Our analytical & Difference$10\times$\\
    \end{tabular}&
    \begin{tabular}{c}
        Our GGX\\
         \includegraphics[width=0.3\linewidth]{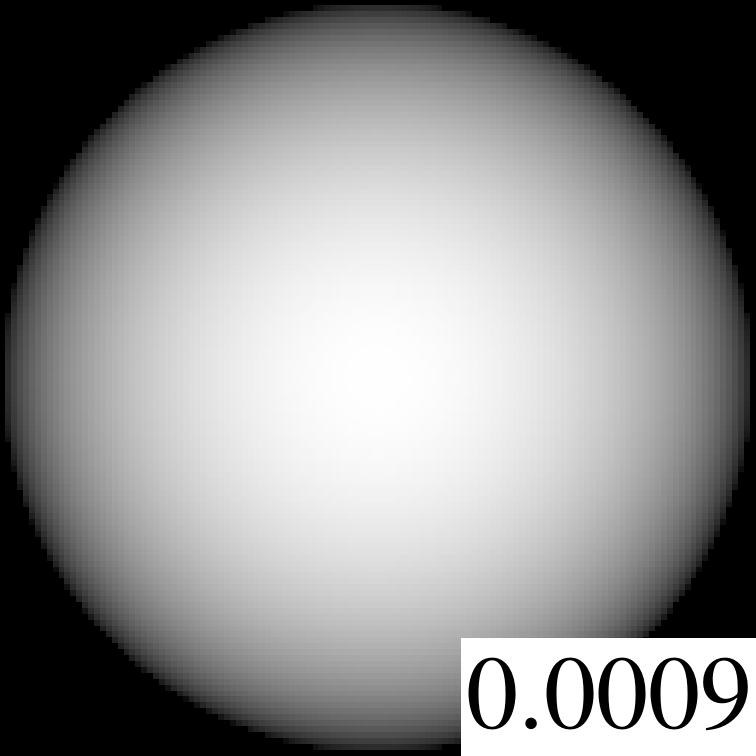}\\
         \;\\
         \includegraphics[width=0.3\linewidth]{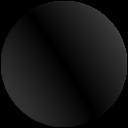}\\
          Difference$10\times$
    \end{tabular}
    \end{tabular}
    }
    \caption{
    \textbf{Our analytical projected-area integral in \cref{eq:line-integral} matches the ground truth} given by Monte Carlo estimation (column 1-3).
    It is a low-frequency function so can be reasonably approximated with a GGX projected area function in \cref{eq:our-ggx} (column 4).
    The numbers show the root mean square error (RMSE).
    }
    \label{fig:projected_area}
\end{figure}

%% file: equations/glint_diffuse.tex
\begin{equation}
\begin{aligned}
\int \frac{\max(\tilde{\mathbf{n}}(\mathbf{u})\!^\top\!\bm{\omega}_i,0)}{\pi(\tilde{\mathbf{n}}(\mathbf{u}))_z}
k_\mathbf{r}(\mathbf{u\!-\!x})\mathrm{d}\mathbf{u}
\\
= \iint\frac{\max(\tilde{\mathbf{m}}\!^\top\!\bm{\omega}_i,0)}{\pi\tilde{\mathbf{m}}_z}\delta(\mathbf{n}(\mathbf{u})\!-\!\mathbf{m})k_\mathbf{r}(\mathbf{u\!-\!x})\mathrm{d}\mathbf{m}\mathrm{d}\mathbf{u}\\
    = \int\frac{\max(\tilde{\mathbf{m}}\!^\top\!\bm{\omega}_i,0)}{\pi}D(\mathbf{m})\mathrm{d}\tilde{\mathbf{m}}
    = \frac{1}{\pi}P(\bm{\omega}_i).
\end{aligned}
\end{equation}

%% file: 4-experiments.tex
\input{figures/other_baseline}
\section{Results}
\label{sec:experiments}

We use Mitsuba 0.6~\cite{mitsuba} for code implementation and path tracing with multiple importance sampling of emitters and BRDF.
The renderings use three $1024^2$ normal maps (\cref{fig:performance}) of 8MB each,
and each normal map takes 34MB to store its acceleration hierarchies (generated within a minute).
Figure~\ref{fig:other-baselines} shows only continuous glint models can capture microstructure details over large footprints, 
so we mainly compare with \citet{yan2014rendering, yan2016position} in terms of equal SPP rendering speed and equal speed rendering quality (\cref{subsec:performance}).
Since the \pndf model for each method is different,
we only compare rendering results qualitatively but provide the quantitative error measure over our method variants in \cref{subsec:shadow-masking-result,subsec:ablation}.
All the experiments are run on a Ryzen 9900X 12-Core CPU.
The code is available at: \url{https://github.com/lwwu2/glint24}.

\input{figures/performance2}
\subsection{Performance comparison}
\label{subsec:performance}
We measure the rendering time of scenes in \cref{fig:performance} with 256 SPP and $800\times800$ resolution for different footprint scales (the number of texels covered by a unit footprint),
and \cref{tab:performance} shows the results for our model (ours), our method without the clustering (ours no cluster), and the two baselines from \citet{yan2014rendering, yan2016position}.
We choose a Gaussian filter for $k_\mathbf{r}$ with a clustering threshold $\tau\!=\!10^{-3}$ except for `scratch' that uses $\tau\!=\!10^{-4}$ for more accuracy (\cref{sec:conclusion}).
For the balance of both performance and quality,
we use $1.0$ sampling rate in \citet{yan2016position}.
It can be seen that ours and \citeauthor{yan2014rendering}’s glint appearance are very similar,
except that their intrinsic roughness tends to produce longer specular tails for the brush and scratch surfaces.
However, our computation cost is noticeably smaller owing to the simplicity of our \pndf{} formulation:
the rendering time is around half of \citet{yan2016position} even without the clustering hierarchy,
and the full model brings down the time for large footprint size rendering ($256^2$) to minutes compared to the baselines that take more than half an hour.
Such an efficiency allows more samples to be allocated without hurting the performance.
As demonstrated in \cref{fig:performance2}, 
our method scales up well to more complex scenes and shows similar equal SPP renderings as the baseline but less Monte Carlo variance in equal rendering time.
\input{figures/performance}

\input{tables/performance}

\subsection{Shadow-masking analysis}
\label{subsec:shadow-masking-result}
For simplicity, we implement a brute-force projected area computation, but the acceleration structures can be similarly applied. 
The effect of the shadow-masking is demonstrated in \cref{fig:shadow-masking}.
It can be seen that the difference between our analytical shadow-masking and the GGX approximation is small,
which matches our discussion in \cref{subsec:shadow-masking} that the shadow-masking/projected area for a specular surface is a smooth function.
\citet{yan2014rendering,yan2016position} do not have a shadow-masking derivation and simply take the Beckmann shadow-masking using fixed roughness,
which can be inaccurate near grazing angles.
Because the shadow-masking takes the inverse of the projected area integral,
it is difficult to get the ground truth rendering using Monte Carlo estimation,
but \cref{fig:projected_area} suggests our analytical result should be very close to the true reference.
For the application of diffuse BRDF aggregation,
Fig.~\ref{fig:shadow-masking-diffuse} shows the shading of $f_d$ gets flat when the surface roughness (normal variation) increases.
This is very similar to an Oren-Nayar BRDF~\cite{oren1994generalization}, 
except our model looks darker owing to the lack of the interreflection term.
However, with the analytical projected area,
the rendering of our diffuse model is able to model the microstructure details that are ignored by the smooth diffuse models.
While the standard normal mapping technique can achieve similar effects by tracing an extensive number of samples,
our results have little aliasing even for a small SPP (\cref{fig:diffuse}).
This is because our model explicitly considers the reflections from all the microfacets within the pixel footprint.
\input{figures/shadow-masking}
\input{figures/shadow-masking-diffuse}
\input{figures/diffuse}

\subsection{Ablation study}
\label{subsec:ablation}

\input{figures/performance-bias}
\paragraph{Performance-error trade-off.}
In \cref{fig:performance-bias}, we study the impact of different clustering thresholds $\tau$ on the scratch normal map under $64^2$ footprint scale.
Increasing $\tau$ results in early termination of the tree traversal, which directly affects the number of intersection tests to speed up the inference (\cref{tab:performance-bias}).
However, the normal approximation at higher tree levels is also less accurate, giving inconsistent rendering compared to the reference without the clustering.

\input{tables/performance-bias}
\paragraph{Different footprint kernel.}
A Gaussian footprint kernel is used in \cref{subsec:performance} to match \citeauthor{yan2016position}’s glint NDF formulation,
but our method can use arbitrary footprint kernels.
Figure~\ref{fig:kernel} shows the renderings on the isotropic normal map of a Gaussian, disk, and box filter,
where the footprint scale is $128^2$ for the Gaussian and $64^2$ for the others.
Unlike the Gaussian that has a long tail, the disk/box filter can use smaller footprint size to achieve similar glint appearance.
As a result, their rendering speeds (numbers in \cref{fig:kernel}) are $2\times$ faster benefiting from the fewer intersection tests.
For a Gaussian with its $(6\sigma)^2$ as the footprint size, 
we found  a disk/box filter with $(3\sigma)^2$ footprint size produces a similar NDF (\cref{fig:kernel} insets; also see supplementary video).
\input{figures/kernel}

%% file: figures/other_baseline.tex
\begin{figure}[t]
    \centering
    \setlength\tabcolsep{0.0pt}
    \resizebox{0.99\linewidth}{!}{
    \begin{tabular}{cc}
    $2\times2$ footprint & $128\times128$ footprint\\
    \includegraphics[width=0.55\linewidth]{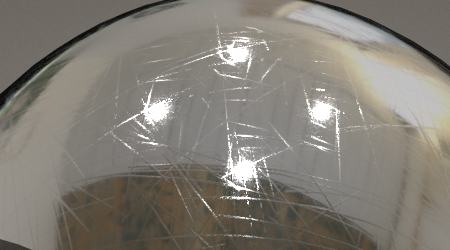}&
    \includegraphics[width=0.55\linewidth]{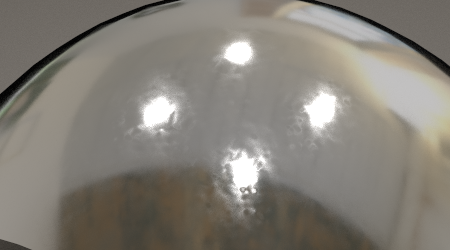}\\[-2.5pt]
    \multicolumn{2}{c}{\citet{chermain2021real}}\\
    \includegraphics[width=0.55\linewidth]{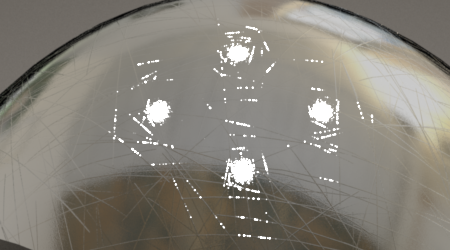}&
    \includegraphics[width=0.55\linewidth]{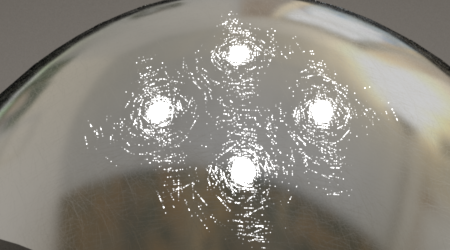}\\[-2.5pt]
    \multicolumn{2}{c}{\citet{atanasov2021multiscale}}\\
    \includegraphics[width=0.55\linewidth]{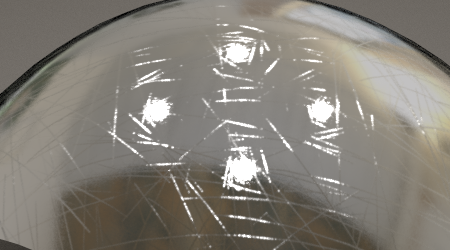}&
    \includegraphics[width=0.55\linewidth]{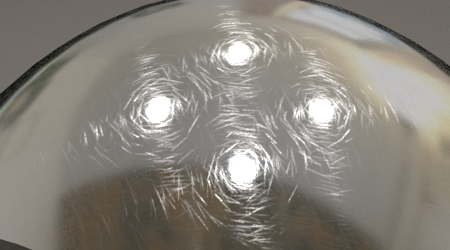}\\[-2.5pt]
    \multicolumn{2}{c}{\citet{yan2016position}}\\
    \includegraphics[width=0.55\linewidth]{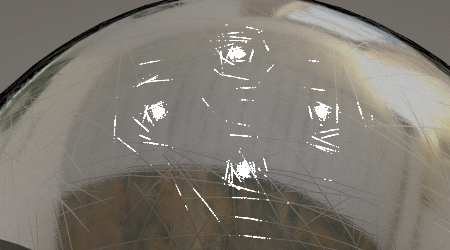}&
    \includegraphics[width=0.55\linewidth]{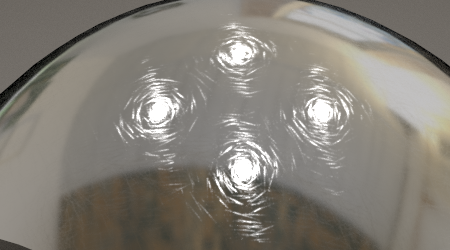}\\[-2.5pt]
    \multicolumn{2}{c}{Ours}
    \end{tabular}
    }
    \caption{\textbf{Continuous glint model vs. other formulations.}
    \citeauthor{chermain2021real} approximate the \pndf using averaged statistics as in LEADR mapping~\cite{dupuy2013linear},
    and \citeauthor{atanasov2021multiscale} utilize the discrete glint formulation~\cite{jakob2014discrete}.
    Both methods are designed for modeling the \pndf over small texture patches (1st column) but produce blurry (1st row) or discontinuous highlights (2nd row) when the texel numbers within a footprint are large (2nd column).
    Therefore, we only compare with the continuous model (\citeauthor{yan2016position}) in this paper.}
    \label{fig:other-baselines}
\end{figure}

%% file: figures/performance2.tex
\begin{figure*}[p]
    \centering
    \setlength\tabcolsep{0.5pt}
    \resizebox{0.99\linewidth}{!}{
    \begin{tabular}{cc}
    \citet{yan2016position}&Ours\\
    \includegraphics[width=0.5\linewidth]{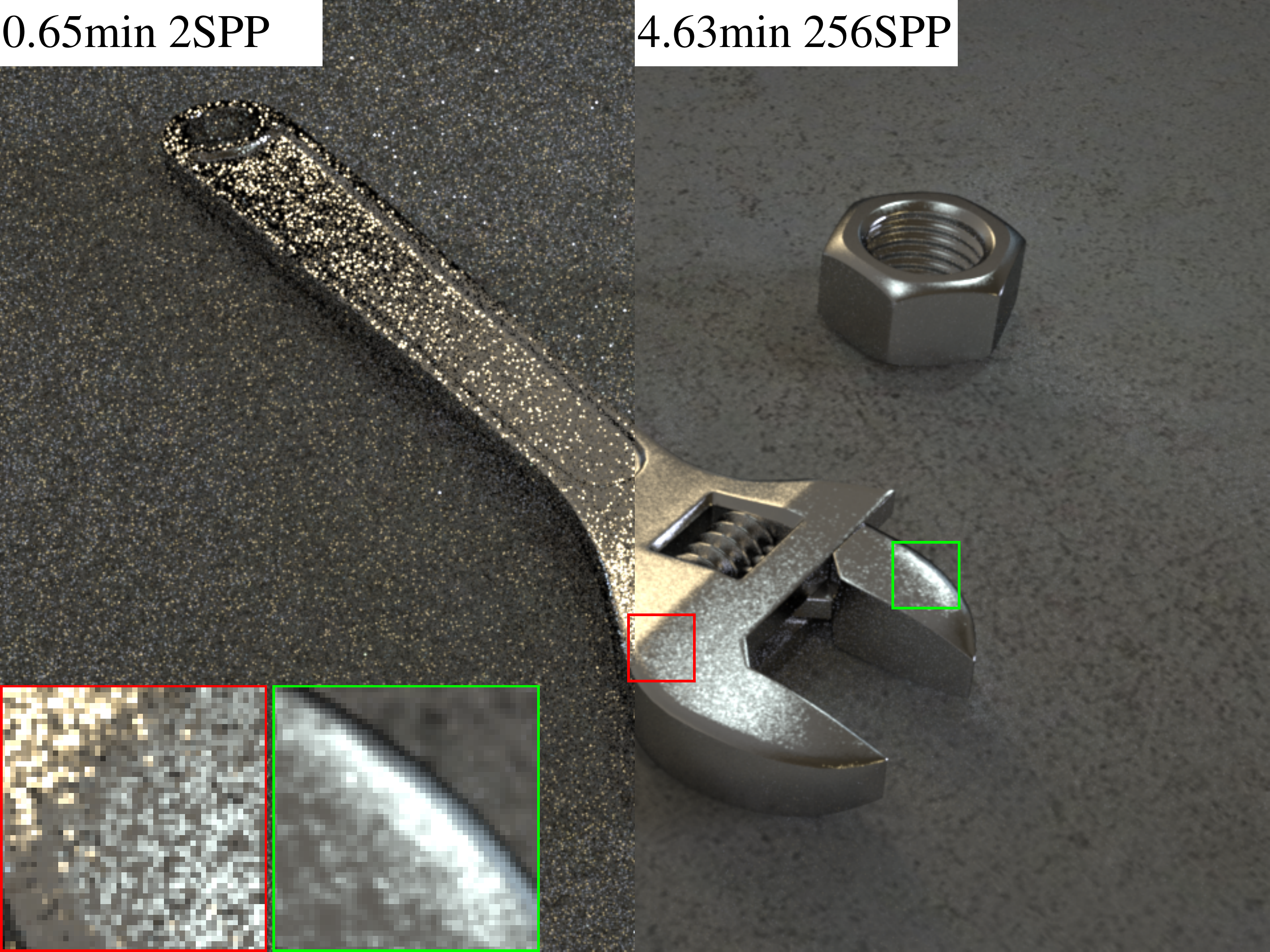}&
    \includegraphics[width=0.5\linewidth]{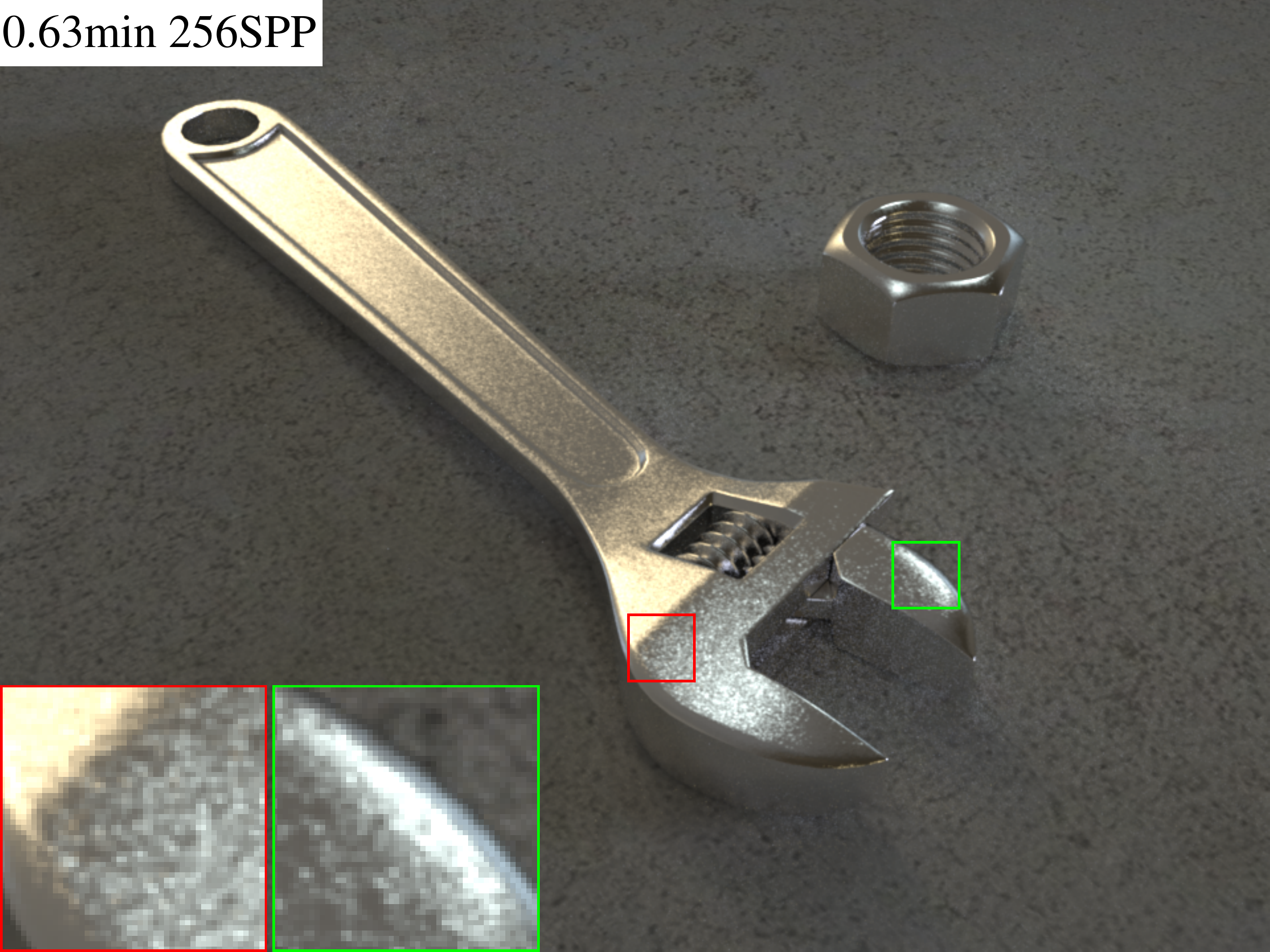}\\[-2pt]
    \includegraphics[width=0.5\linewidth]{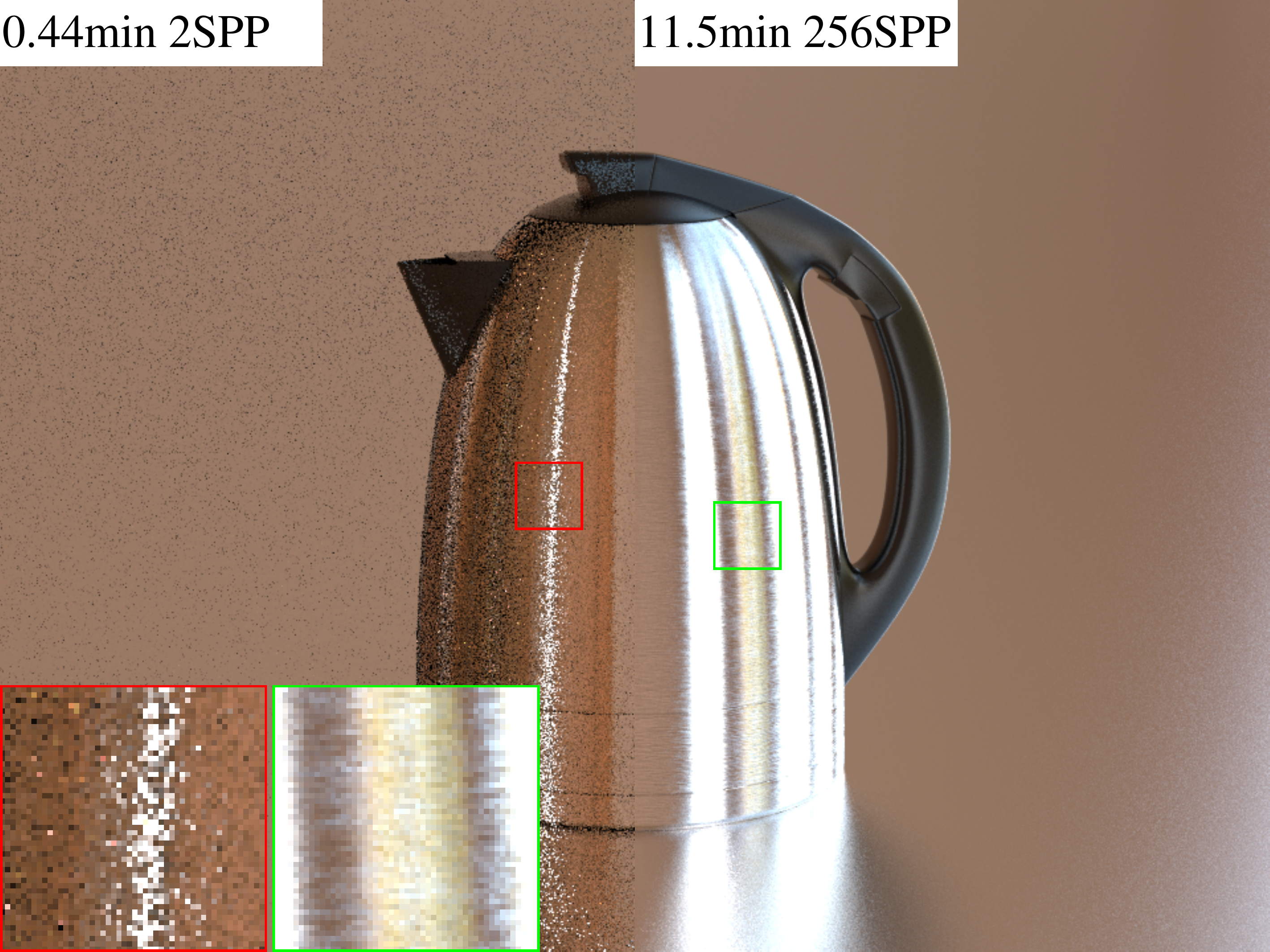}&
    \includegraphics[width=0.5\linewidth]{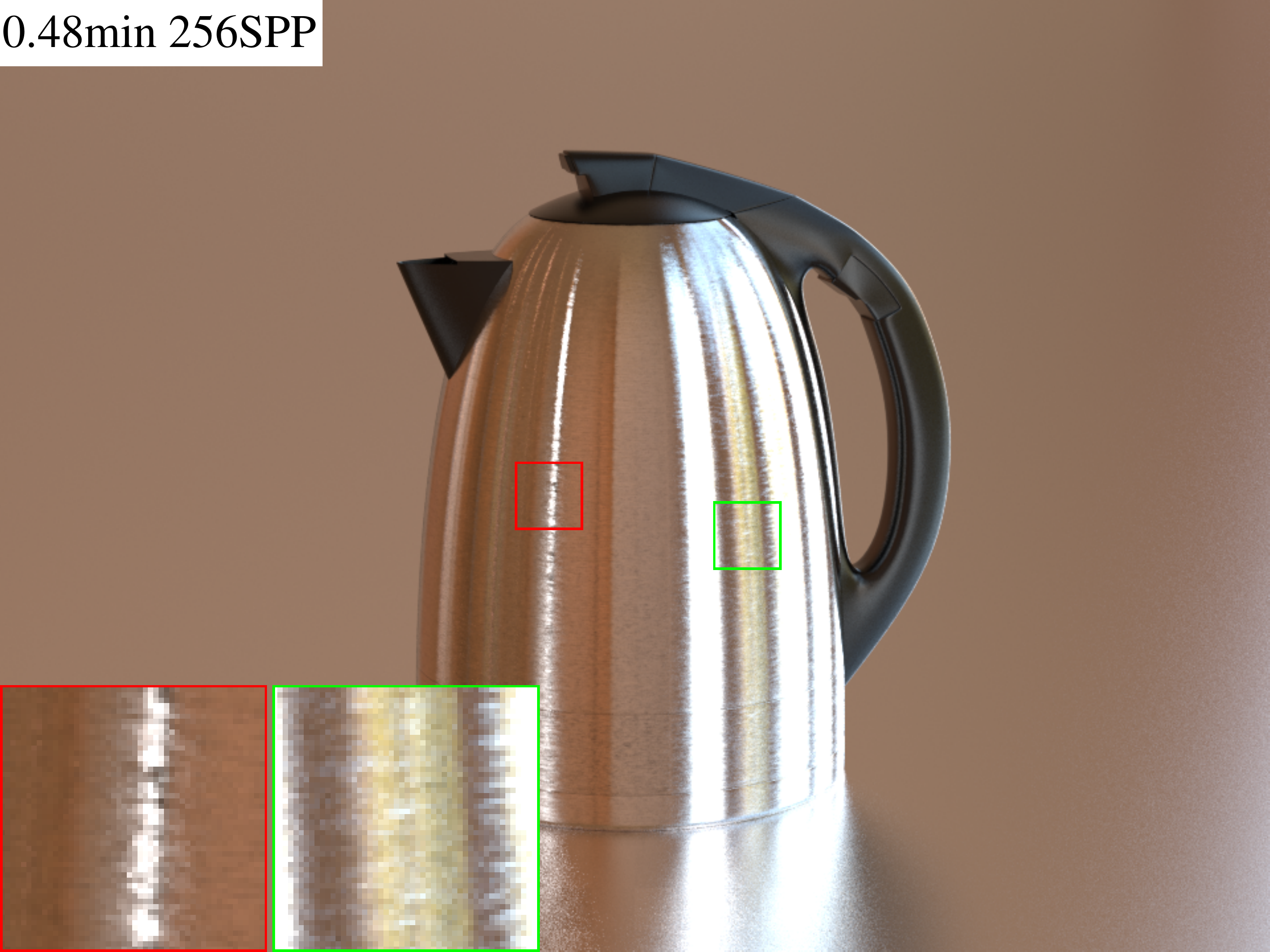}\\[-2pt]
    \includegraphics[width=0.5\linewidth]{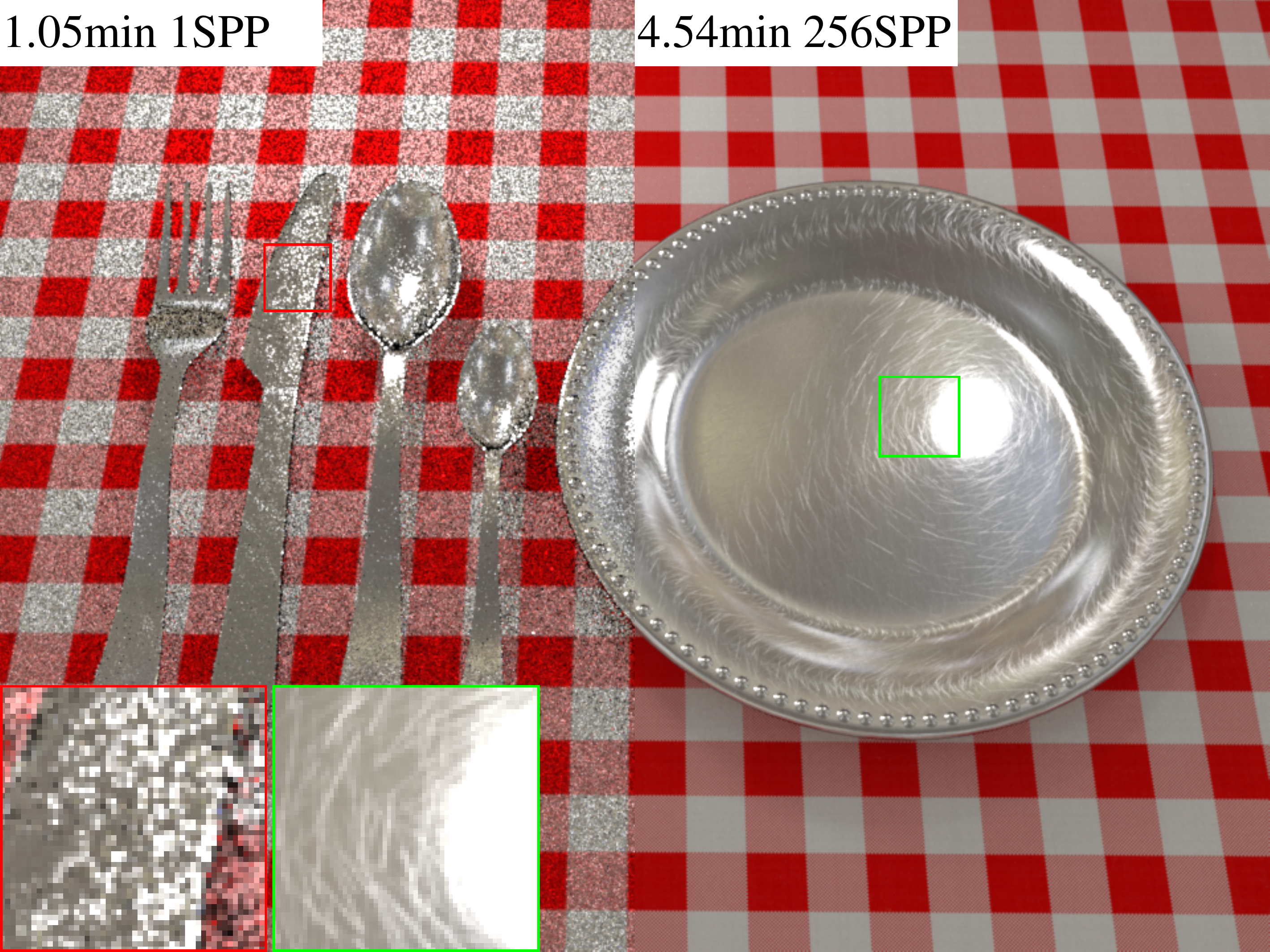}&
    \includegraphics[width=0.5\linewidth]{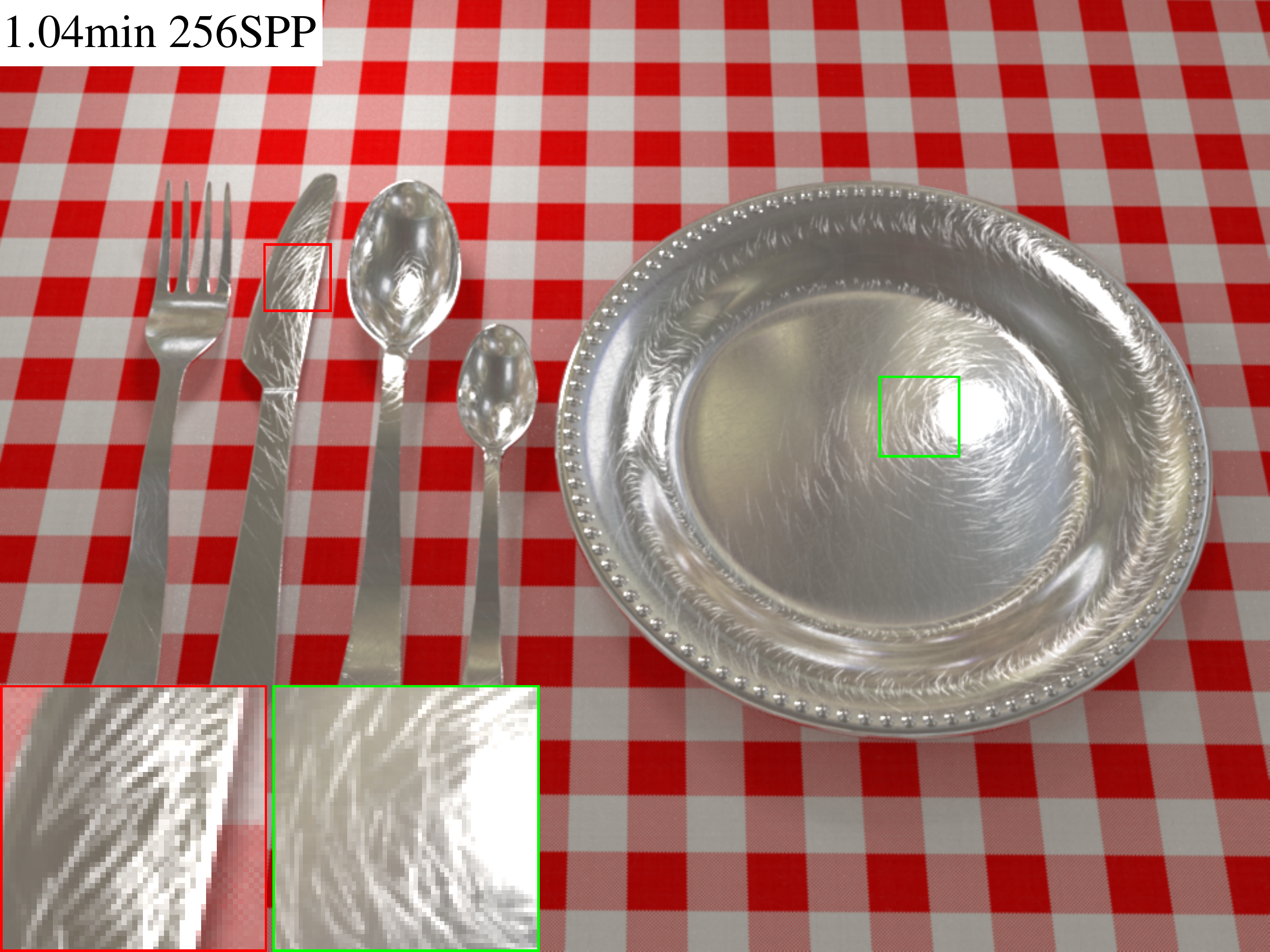}\\[-2pt]
    \end{tabular}
    }
    \caption{
    \textbf{Equal time rendering comparison} shows our method (right) is able to use more samples to reduce variance given similar time budget.
    In contrast, it takes more time for the baseline (left) to obtain less noisy images.
    The intrinsic roughness smooths the NDF response, so the left images may have slightly darker (top) or longer highlights (bottom).
    The red and green insets are rendered in equal time and equal SPP respectively.
    From top to bottom, we use the isotropic, brush, and scratch normal map,
    with all images rendered in $960\!\times\!720$ resolution.
    }
    \label{fig:performance2}
\end{figure*}

%% file: figures/performance.tex
\begin{figure}[t]
    \centering
    \setlength\tabcolsep{0.5pt}
    \resizebox{0.99\linewidth}{!}{
    \begin{tabular}{cc ccc}
    & & $64^2$ & $128^2$ & $256^2$\\[-2pt]
    \includegraphics[width=0.32\linewidth]{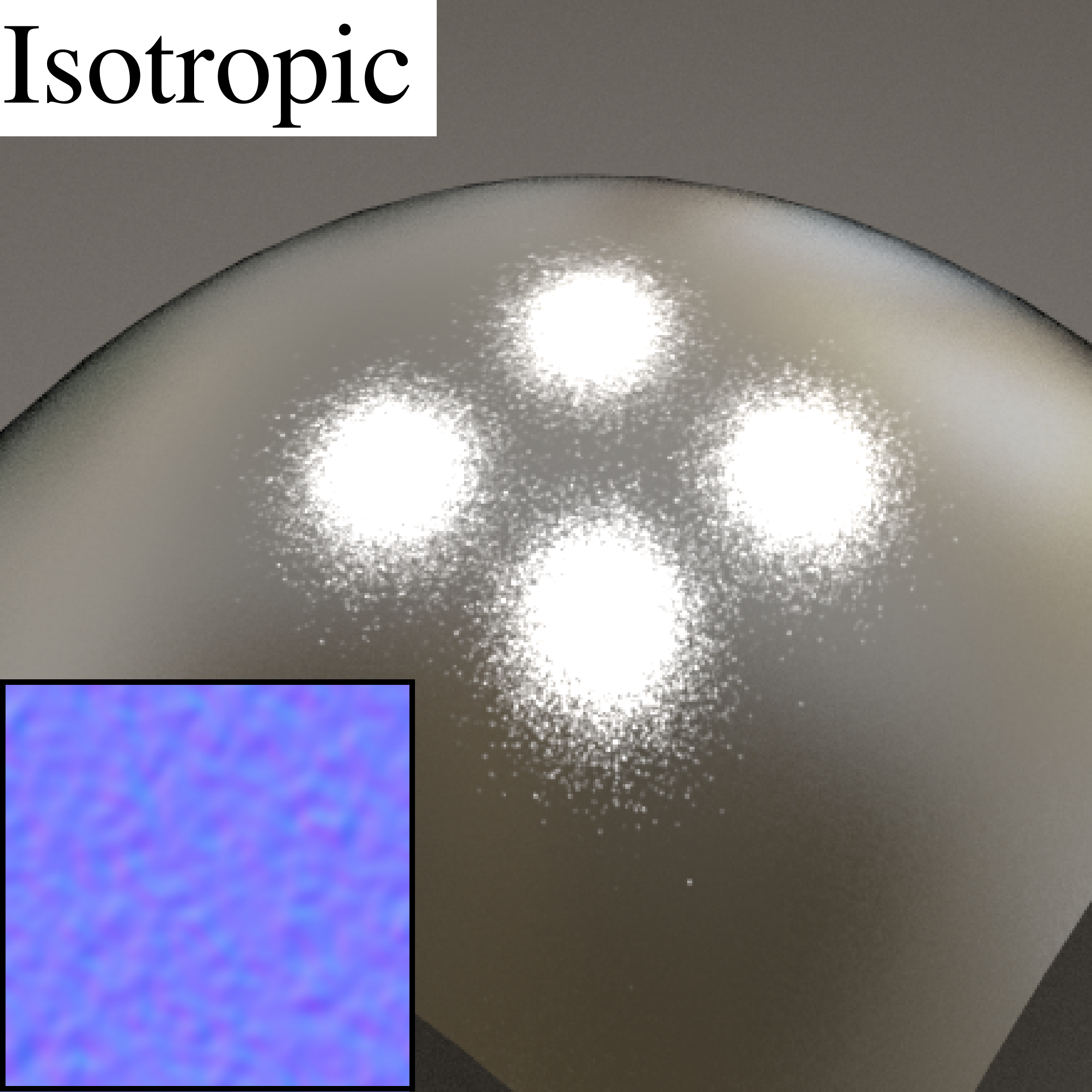}&
    \includegraphics[width=0.32\linewidth]{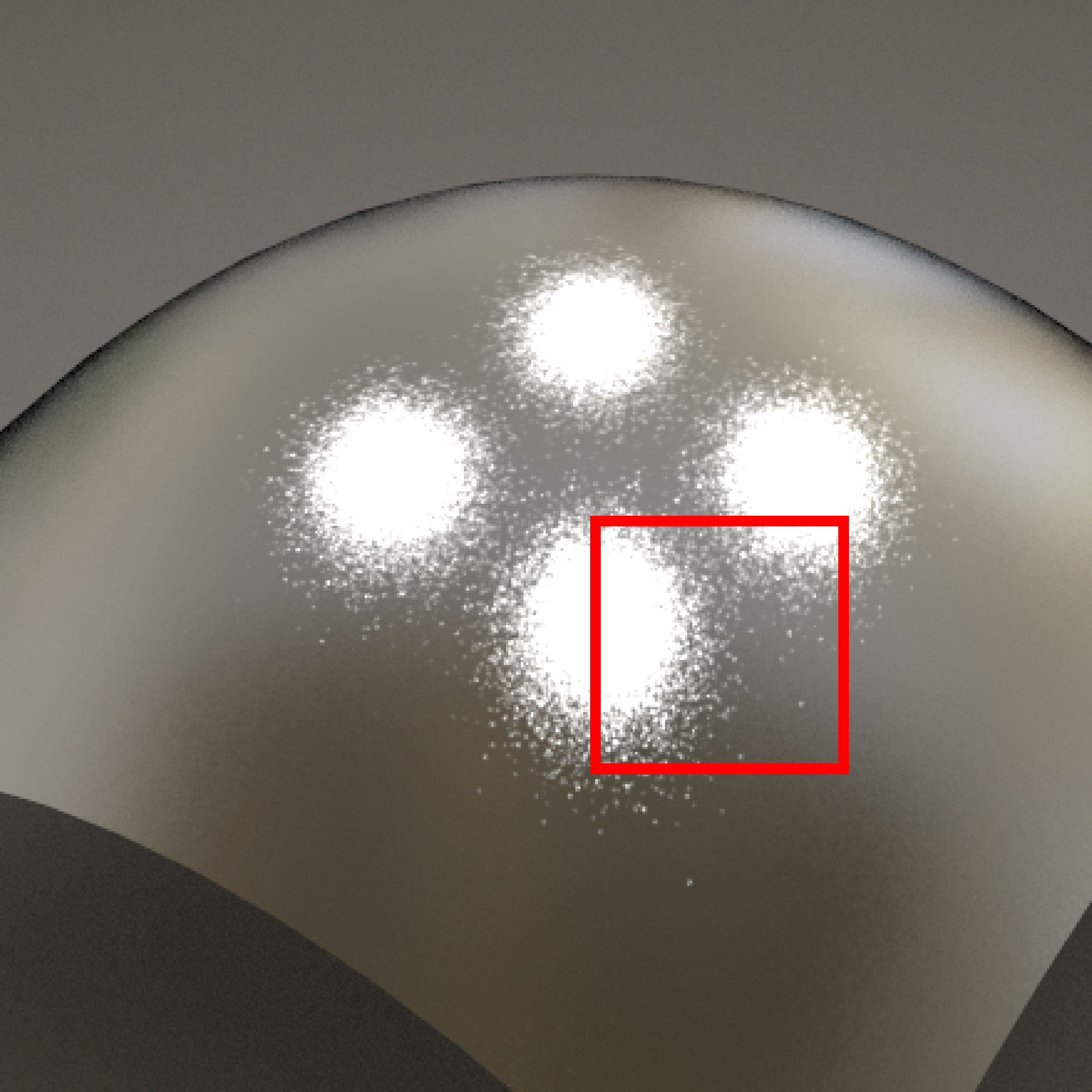}&
    \includegraphics[width=0.16\linewidth]{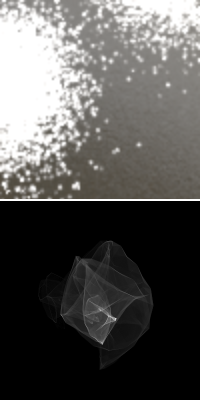}&
    \includegraphics[width=0.16\linewidth]{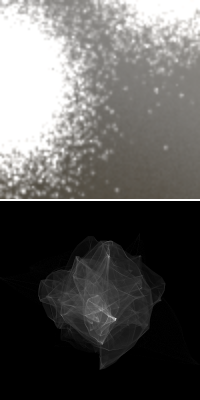}&
    \includegraphics[width=0.16\linewidth]{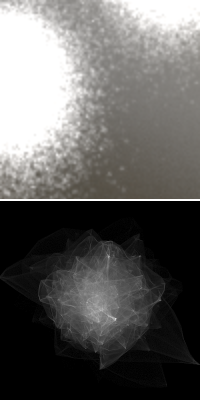}\\[-2pt]
    \includegraphics[width=0.32\linewidth]{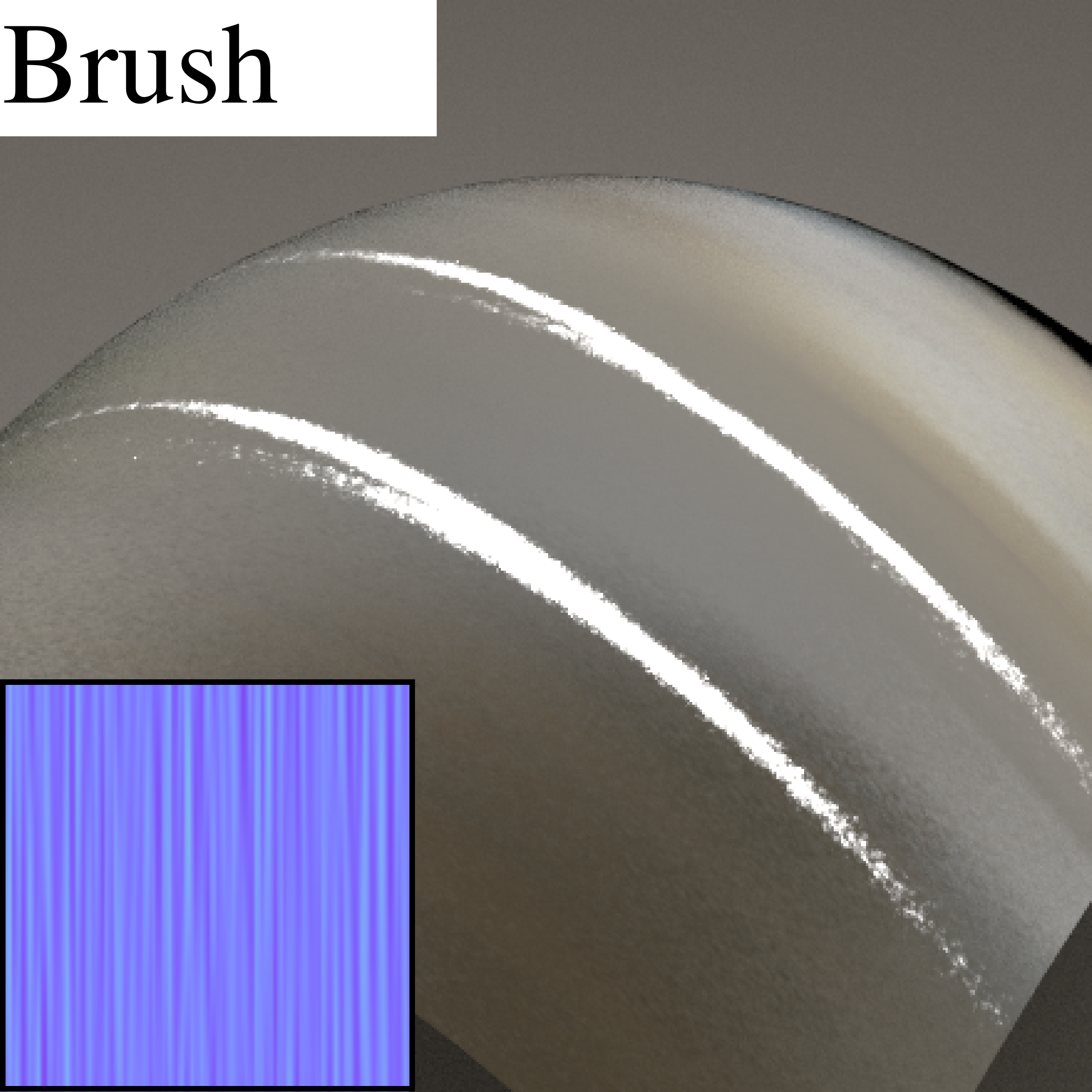}&
    \includegraphics[width=0.32\linewidth]{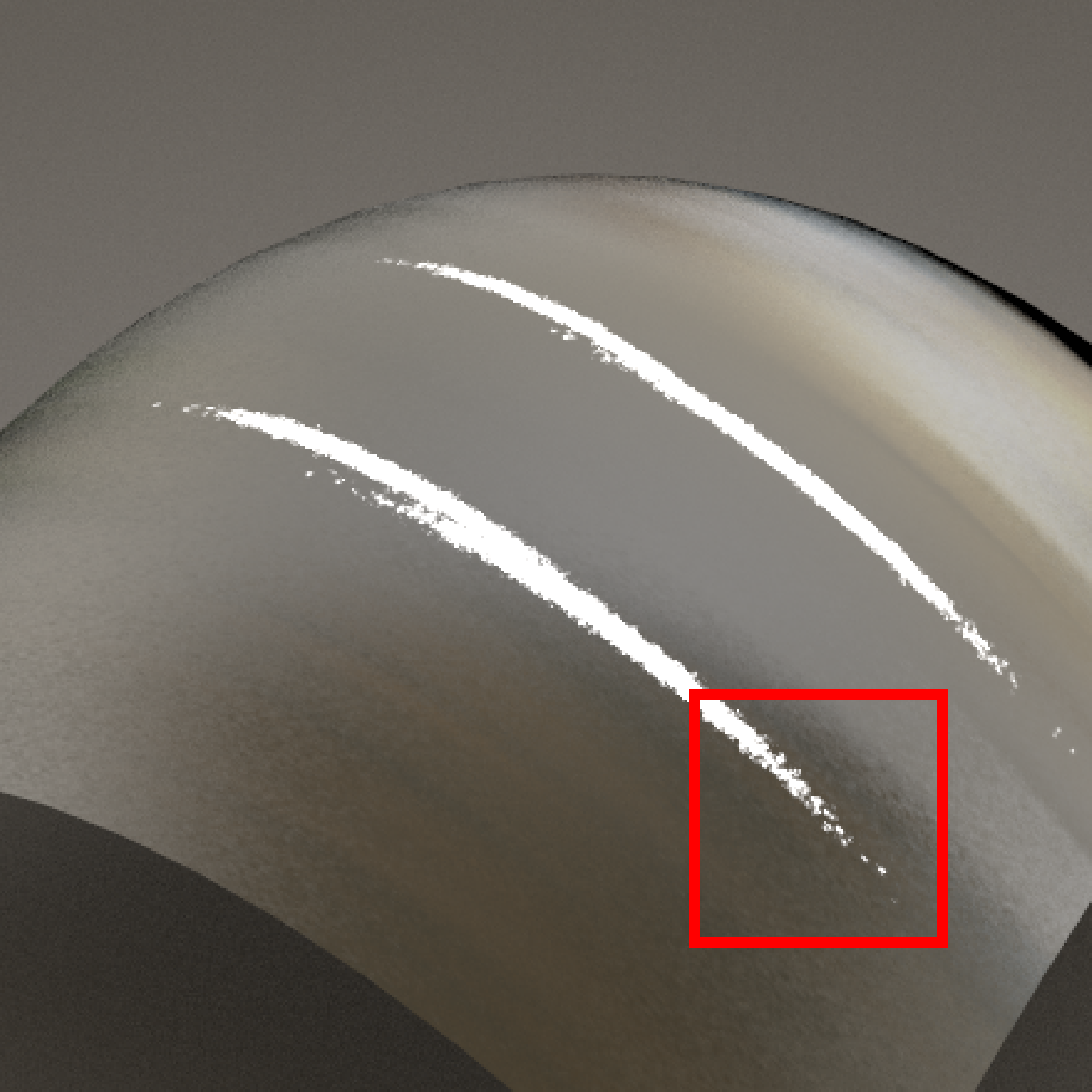}&
    \includegraphics[width=0.16\linewidth]{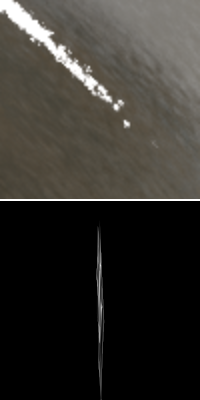}&
    \includegraphics[width=0.16\linewidth]{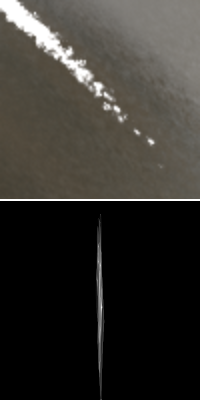}&
    \includegraphics[width=0.16\linewidth]{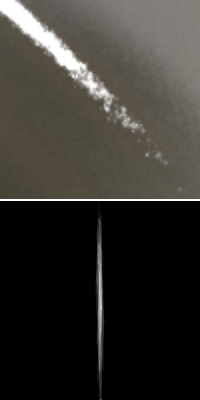}\\[-2pt]
    \includegraphics[width=0.32\linewidth]{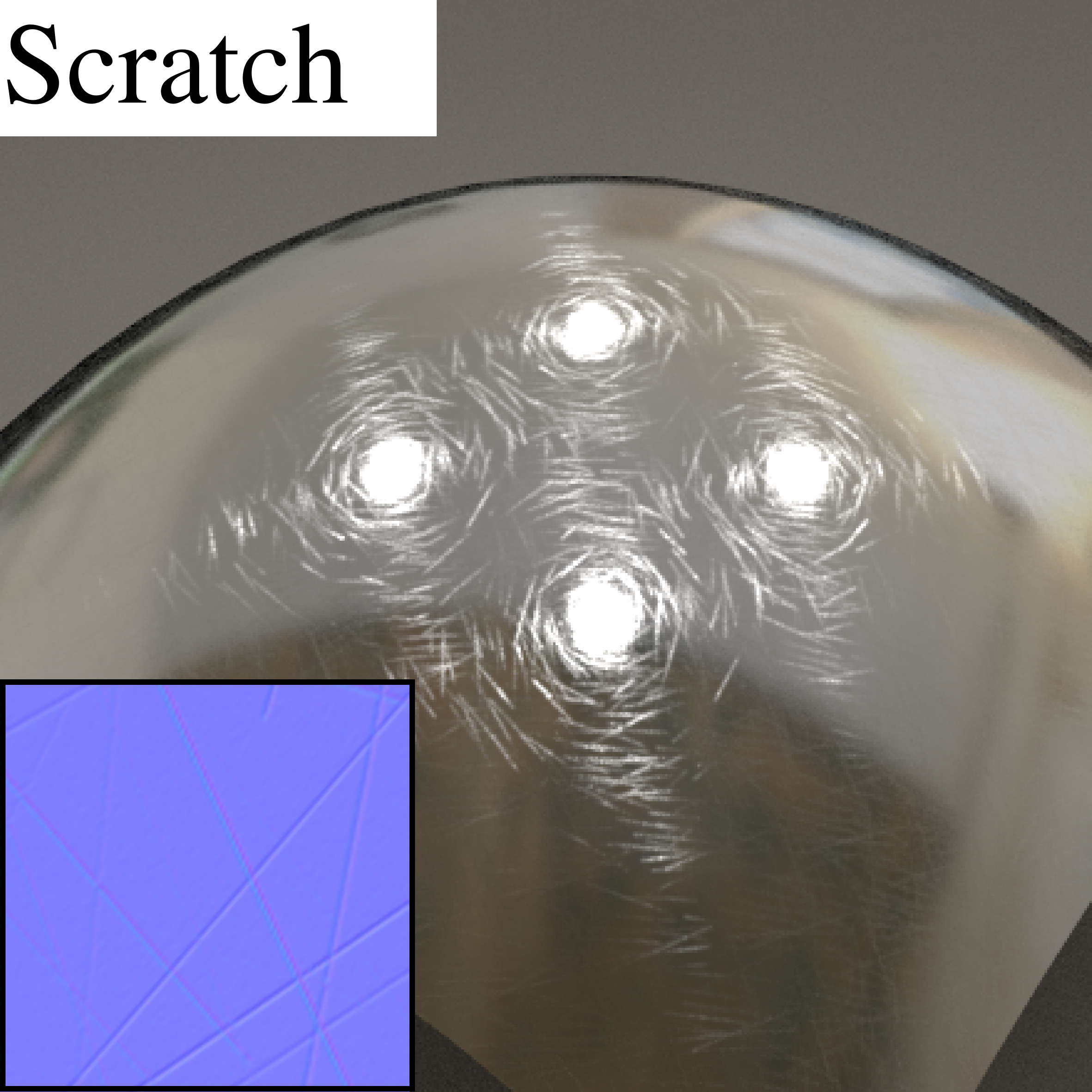}&
    \includegraphics[width=0.32\linewidth]{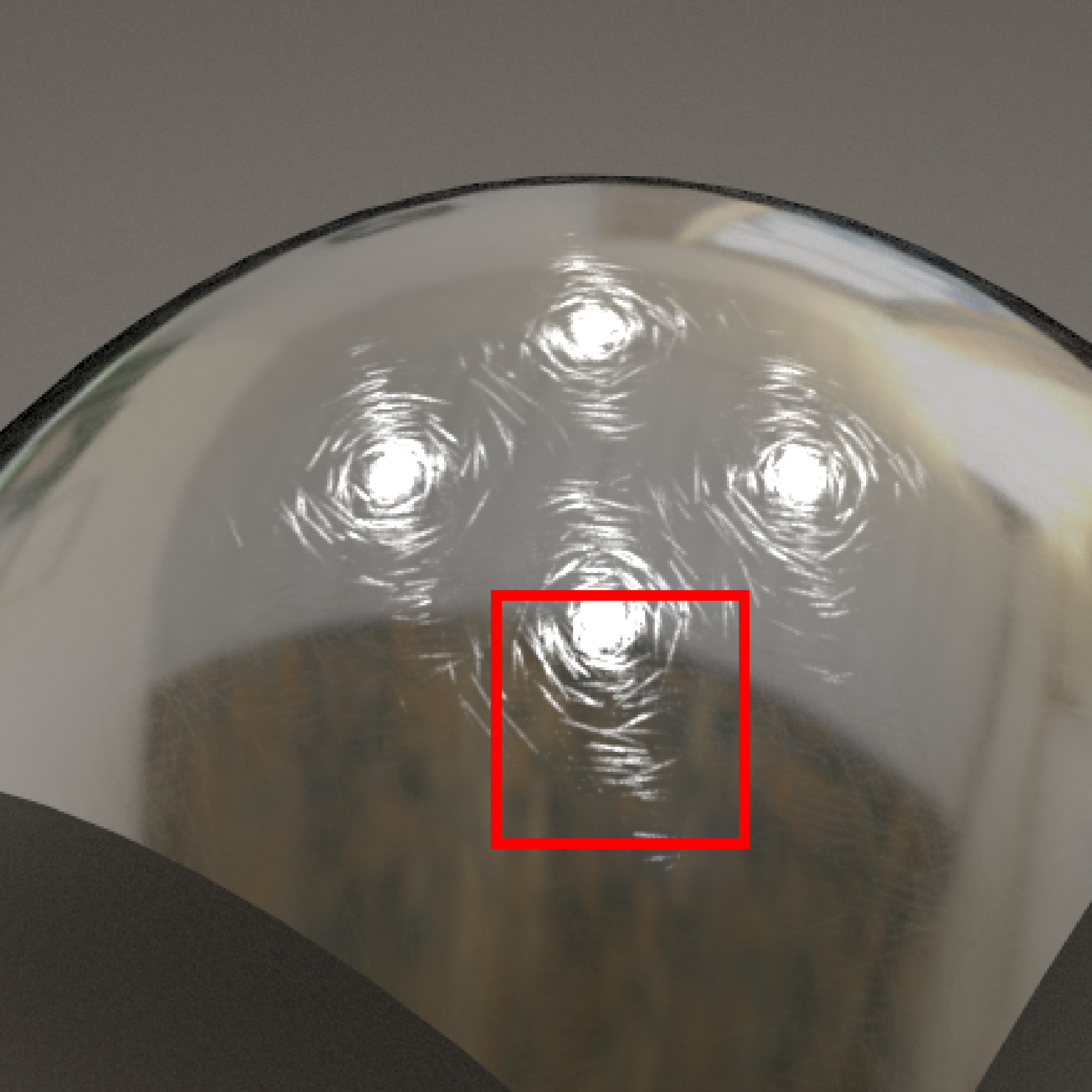}&
    \includegraphics[width=0.16\linewidth]{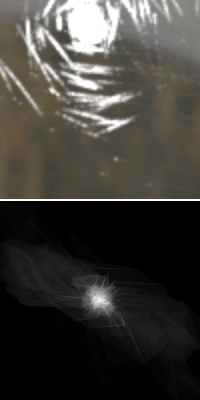}&
    \includegraphics[width=0.16\linewidth]{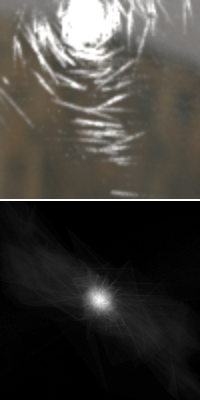}&
    \includegraphics[width=0.16\linewidth]{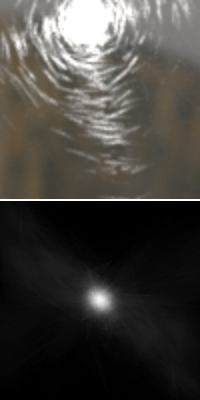}\\[-2pt]
    \citet{yan2016position} & Ours & \multicolumn{3}{c}{Our rendering vs \pndf{}}
    \\
    \end{tabular}
    }
    \caption{
        \textbf{Normal maps used in the experiment and their rendering comparison.}
        Our \pndf{} (column 2) gives similar rendering as its pre-filtered formulation (column 1).
        The insets show the renderings and the NDFs for different footprint scale (texel numbers per unit footprint).
    }
    \label{fig:performance}
\end{figure}

%% file: tables/performance.tex
\begin{table}[t]
    \centering
    \caption{
    \textbf{Quantitative performance comparison of \cref{fig:performance}} shows our method is noticeably faster over different footprint scales.
    }
    \setlength\tabcolsep{2.5pt}
    \resizebox{0.99\linewidth}{!}{
    \begin{tabular}{l |ccc| ccc |ccc |ccc}
    \toprule
    \multicolumn{1}{l}{Method} & 
      \multicolumn{3}{c}{\citet{yan2014rendering}} &
      \multicolumn{3}{c}{\citet{yan2016position}} &
      \multicolumn{3}{c}{Ours no cluster} &
      \multicolumn{3}{c}{Ours}\\
      \multicolumn{1}{l}{Scale} & 
      $64^2$ & $128^2$ & \multicolumn{1}{c}{$256^2$} &
      $64^2$ & $128^2$ & \multicolumn{1}{c}{$256^2$} &
      $64^2$ & $128^2$ & \multicolumn{1}{c}{$256^2$} &
      $64^2$ & $128^2$ & \multicolumn{1}{c}{$256^2$}\\
      \midrule
      \multicolumn{13}{c}{Minutes $\downarrow$}\\
      \midrule
    Isotropic & 
    145 & 491 & 1927 &
    2.29 & 8.03 & 30.0 &
     0.92 & 2.77 & 7.54 &
    \fst{0.37} & \fst{0.72} & \fst{1.20}
    \\
    Brush &
    568 & 1620 & 6712 &
    3.17 & 11.2 & 38.6 &
    1.48 & 4.32 & 11.7 &
    \fst{0.40} & \fst{0.61} & \fst{0.93}
    \\
    Scratch&
    1137 & 4592 & 15724 &
    5.89 & 21.0 & 79.2 &
    4.24 & 12.1 & 36.2 &
    \fst{0.67} & \fst{1.14} & \fst{3.92}
    \\
    \midrule
    \multicolumn{13}{c}{Speed up relative to \citet{yan2016position}$\uparrow$}\\
    \midrule
    Isotropic &
    0.016 & 0.016 & 0.016 & 
    1.00 & 1.00 & 1.00 & 
    2.49 & 2.90 & 3.98 & 
    \fst{6.19} & \fst{11.2} & \fst{25.0} \\
    Brush&
    0.006 & 0.007 & 0.006 & 
    1.00 & 1.00 & 1.00 & 
    2.14 & 2.59 & 3.30 & 
    \fst{7.93} & \fst{18.4} & \fst{41.5} \\
    Scratch&
    0.005 & 0.005 & 0.005 & 
    1.00 & 1.00 & 1.00 & 
    1.39 & 1.74 & 2.19 & 
    \fst{8.79} & \fst{18.4} & \fst{20.2} \\
    \bottomrule
    \end{tabular}
    }
    \label{tab:performance}
\end{table}

%% file: figures/shadow-masking.tex
\begin{figure}[t]
    \centering
    \setlength\tabcolsep{1pt}
    \resizebox{0.99\linewidth}{!}{
    \begin{tabular}{ccc}
         \includegraphics[width=0.34\linewidth]{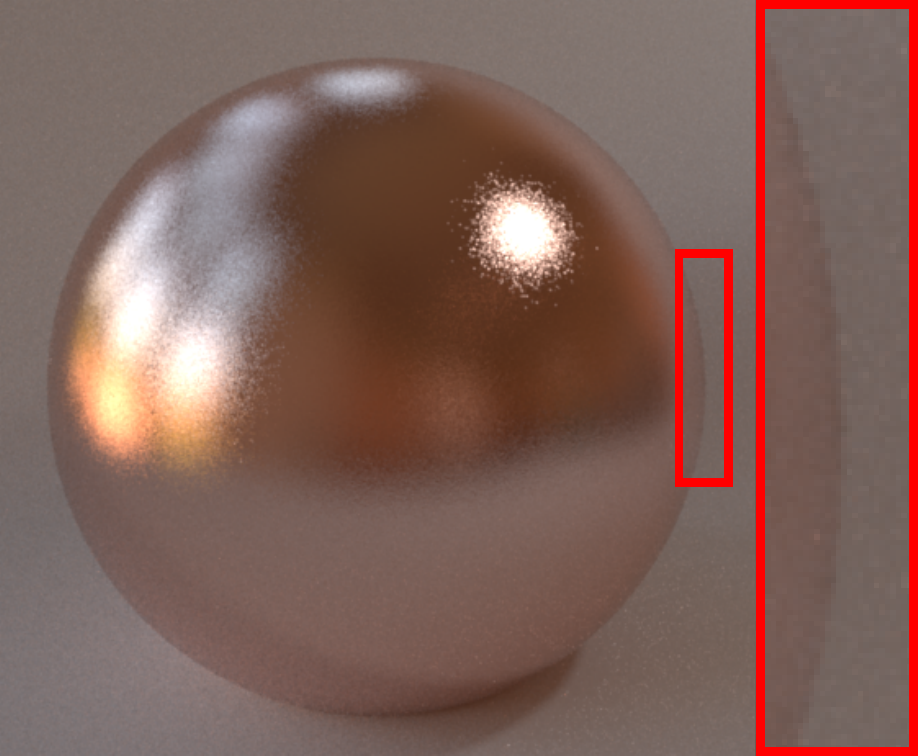}&
         \includegraphics[width=0.34\linewidth]{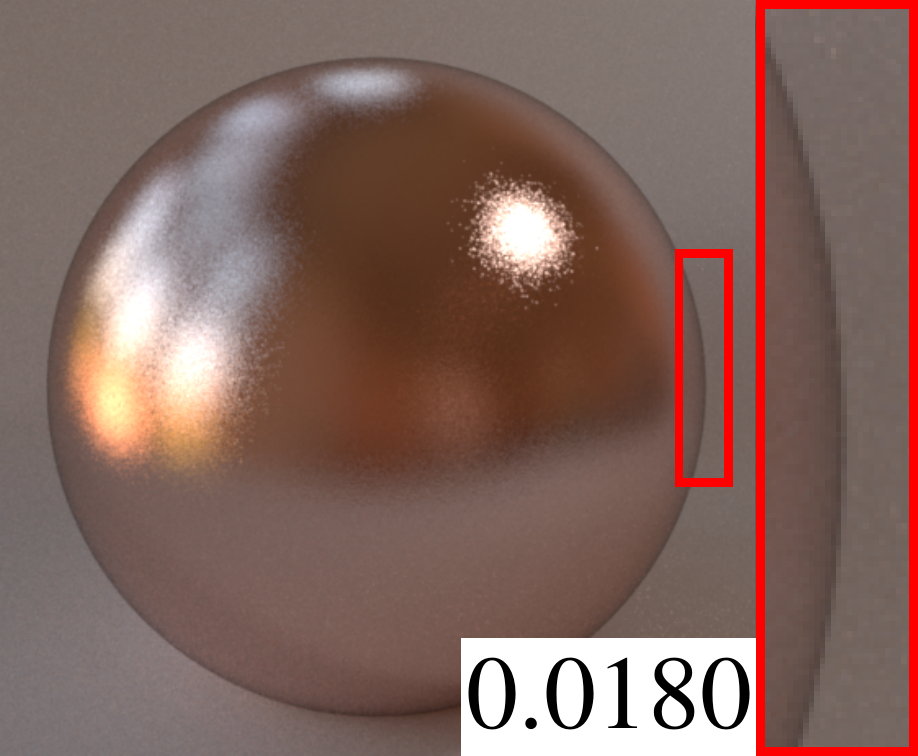}&
         \includegraphics[width=0.34\linewidth]{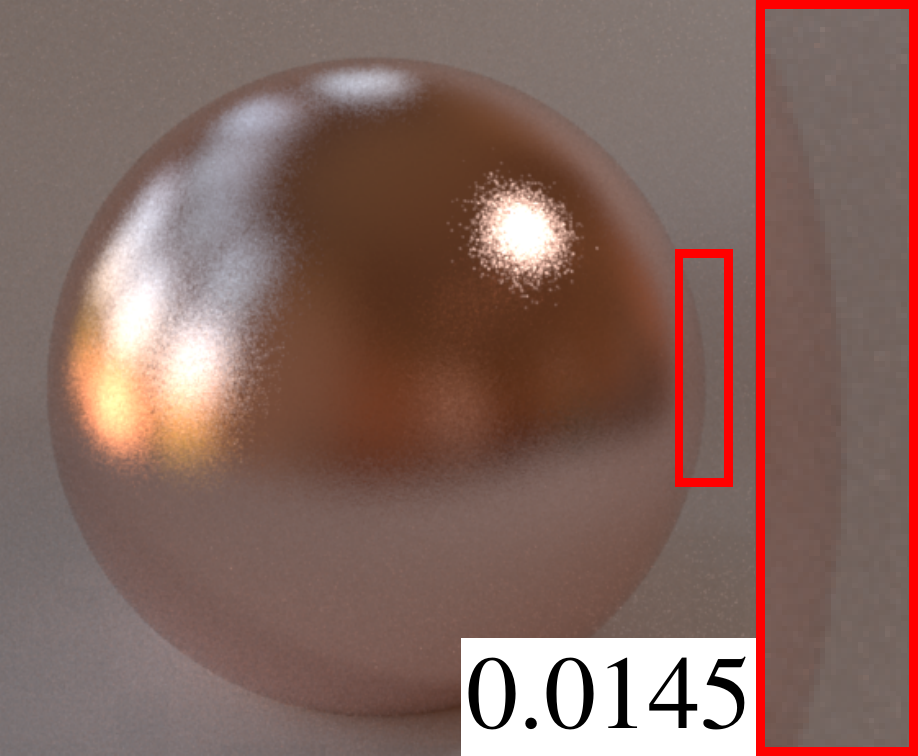}
         \\[-2pt]
         Our analytical (2min) & \citeauthor{yan2016position} (0.38min) & Our GGX (0.43min)\\ 
    \end{tabular}
    }
    \caption{
        \textbf{Shadow-masking comparison on a specular surface} suggests our GGX approximation gives a very close rendering compared to the analytical shadow-masking yet is faster.
        Without correct shadow-masking modeling,
        \citet{yan2016position} produce darker rendering in grazing angles (insets). 
        The numbers on the images show the RMSE.
    }
    \label{fig:shadow-masking}
\end{figure}

%% file: figures/shadow-masking-diffuse.tex
\begin{figure}[t]
    \centering
    \setlength\tabcolsep{0.5pt}
    \resizebox{0.99\linewidth}{!}{
    \begin{tabular}{ccc}
         \multicolumn{3}{l}{
         \hspace{4pt}
         Roughness $\relbar\joinrel\relbar\joinrel\relbar\joinrel\relbar\joinrel\relbar\joinrel\relbar\joinrel\relbar\joinrel\relbar\joinrel\relbar\joinrel\relbar\joinrel\relbar\joinrel\relbar\joinrel\relbar\joinrel\relbar\joinrel\relbar\joinrel\relbar\joinrel\relbar\joinrel\relbar\joinrel\relbar\joinrel\relbar\joinrel\relbar\joinrel\relbar\joinrel\relbar\joinrel\relbar\joinrel\relbar\joinrel\relbar\joinrel\relbar\joinrel\relbar\joinrel\rightarrow$
         }\\
         \includegraphics[width=0.36\linewidth]{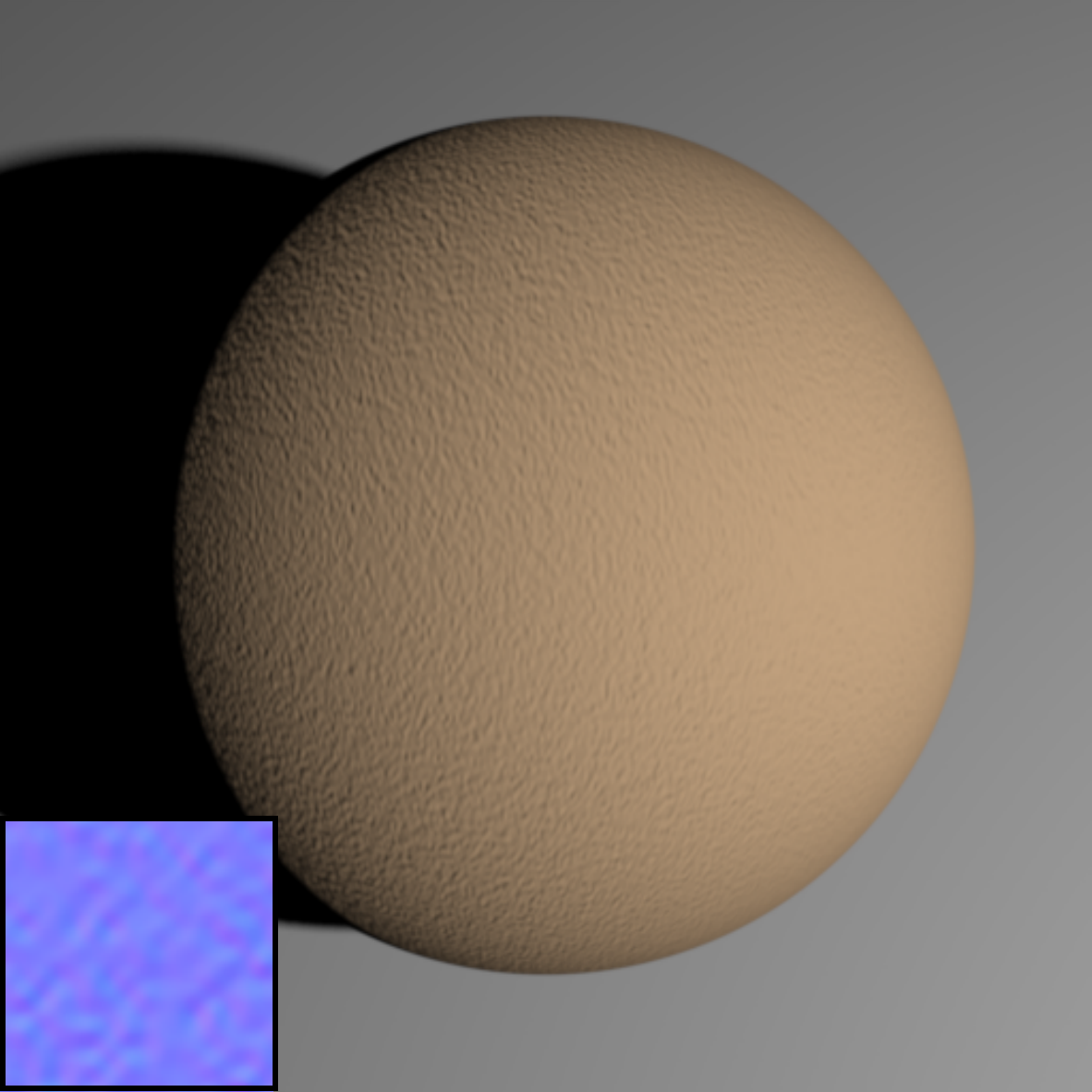}&
         \includegraphics[width=0.36\linewidth]{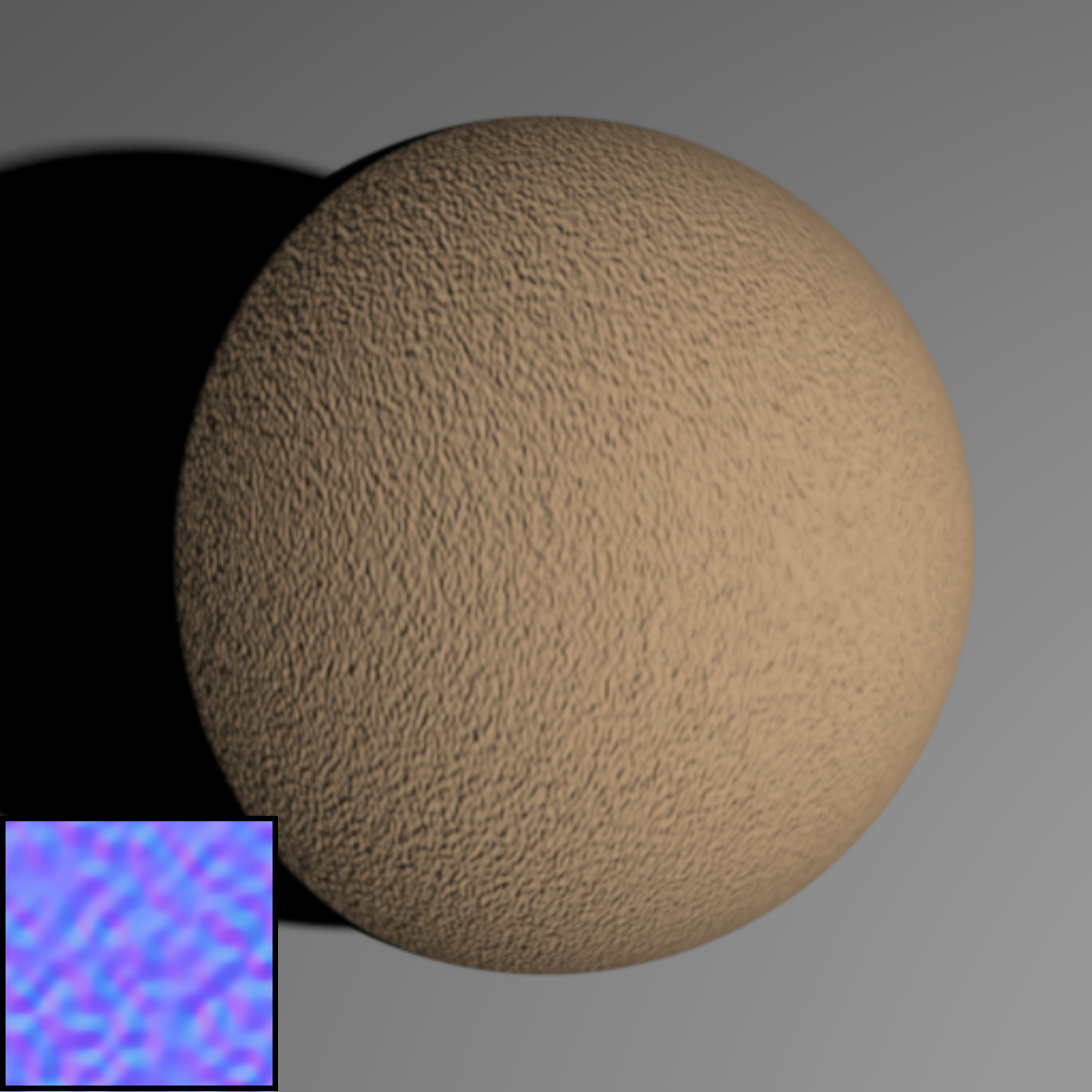}&
         \includegraphics[width=0.36\linewidth]{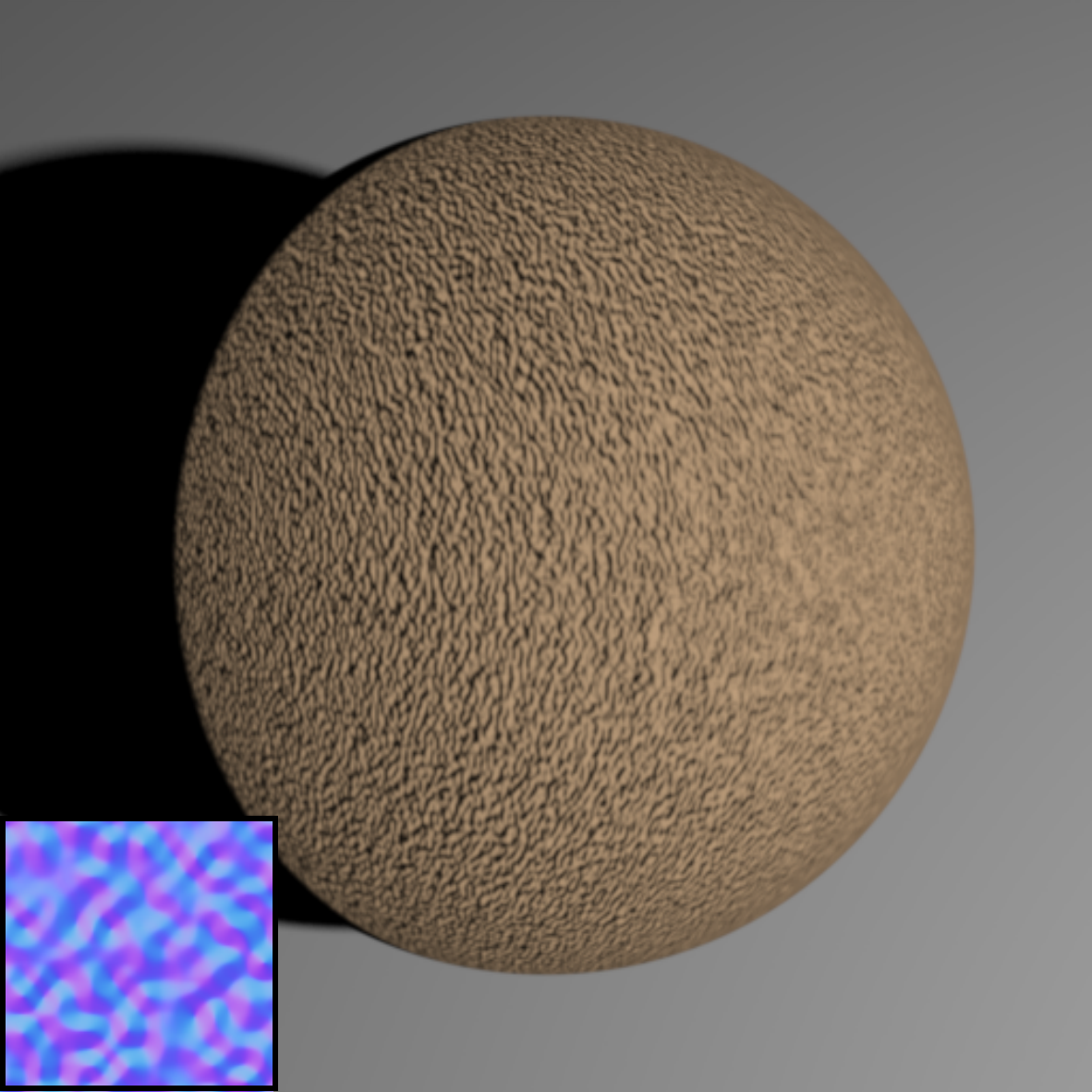}\\[-2pt]
         \multicolumn{3}{c}{Our diffuse}\\
         \includegraphics[width=0.36\linewidth]{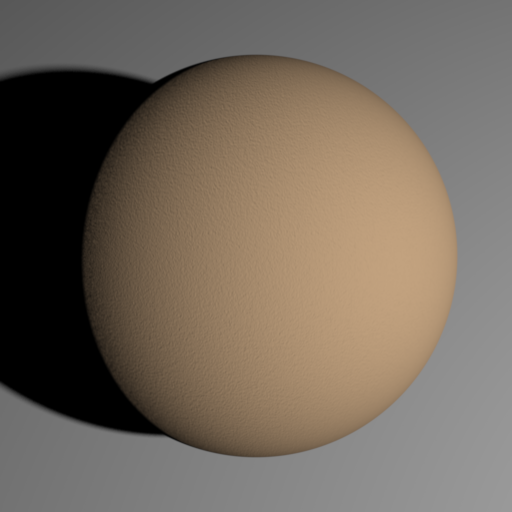}&
         \includegraphics[width=0.36\linewidth]{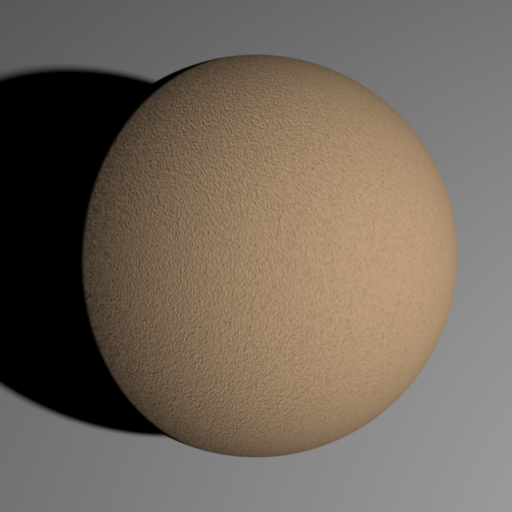}&
         \includegraphics[width=0.36\linewidth]{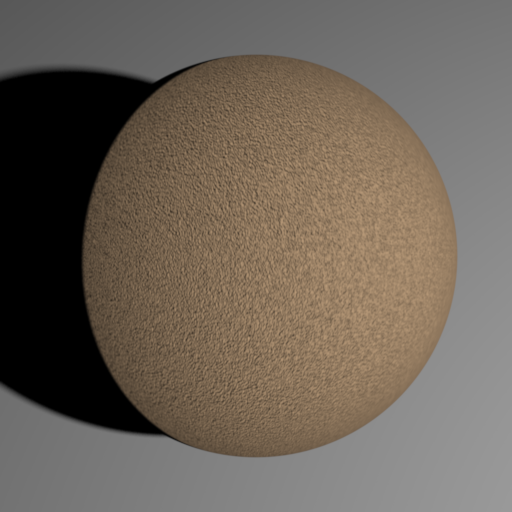}\\[-2pt]
         \multicolumn{3}{c}{Our with larger footprint size}\\
         \includegraphics[width=0.36\linewidth]{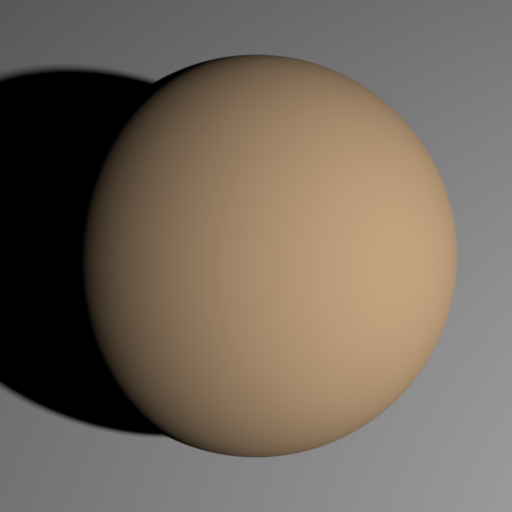}&
         \includegraphics[width=0.36\linewidth]{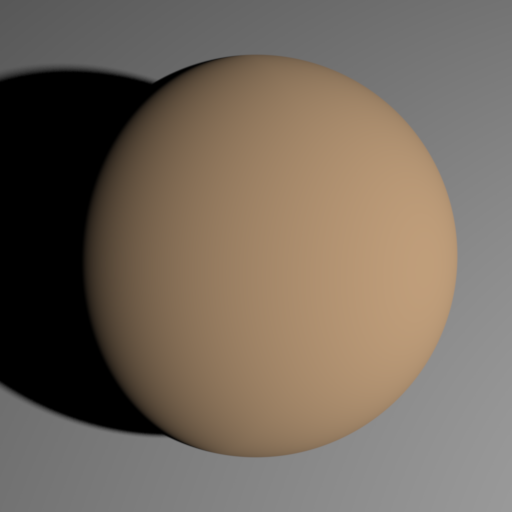}&
         \includegraphics[width=0.36\linewidth]{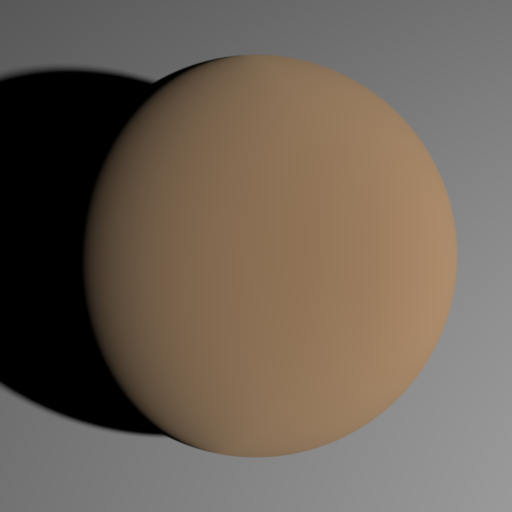}\\[-2pt]
         \multicolumn{3}{c}{Oren-Nayar diffuse}\\
    \end{tabular}
    }
    \caption{
        \textbf{Diffuse appearance developed from our analytical projected area} preserves the detailed appearance and surface variation (1st row) of the underlying normal map (insets).
        With larger query footprint size (2nd row), it resembles the Oren-Nayar BRDF~\cite{oren1994generalization}, where the shading becomes flat for the high roughness surface.
    }
    \label{fig:shadow-masking-diffuse}
\end{figure}

%% file: figures/diffuse.tex
\begin{figure*}[t]
    \centering
    \setlength\tabcolsep{0.5pt}
    \resizebox{0.99\linewidth}{!}{
    \begin{tabular}{ccc}
    Normal-mapped & Oren-Nayar+LEADR Mapping & Ours\\
    \includegraphics[width=0.34\linewidth]{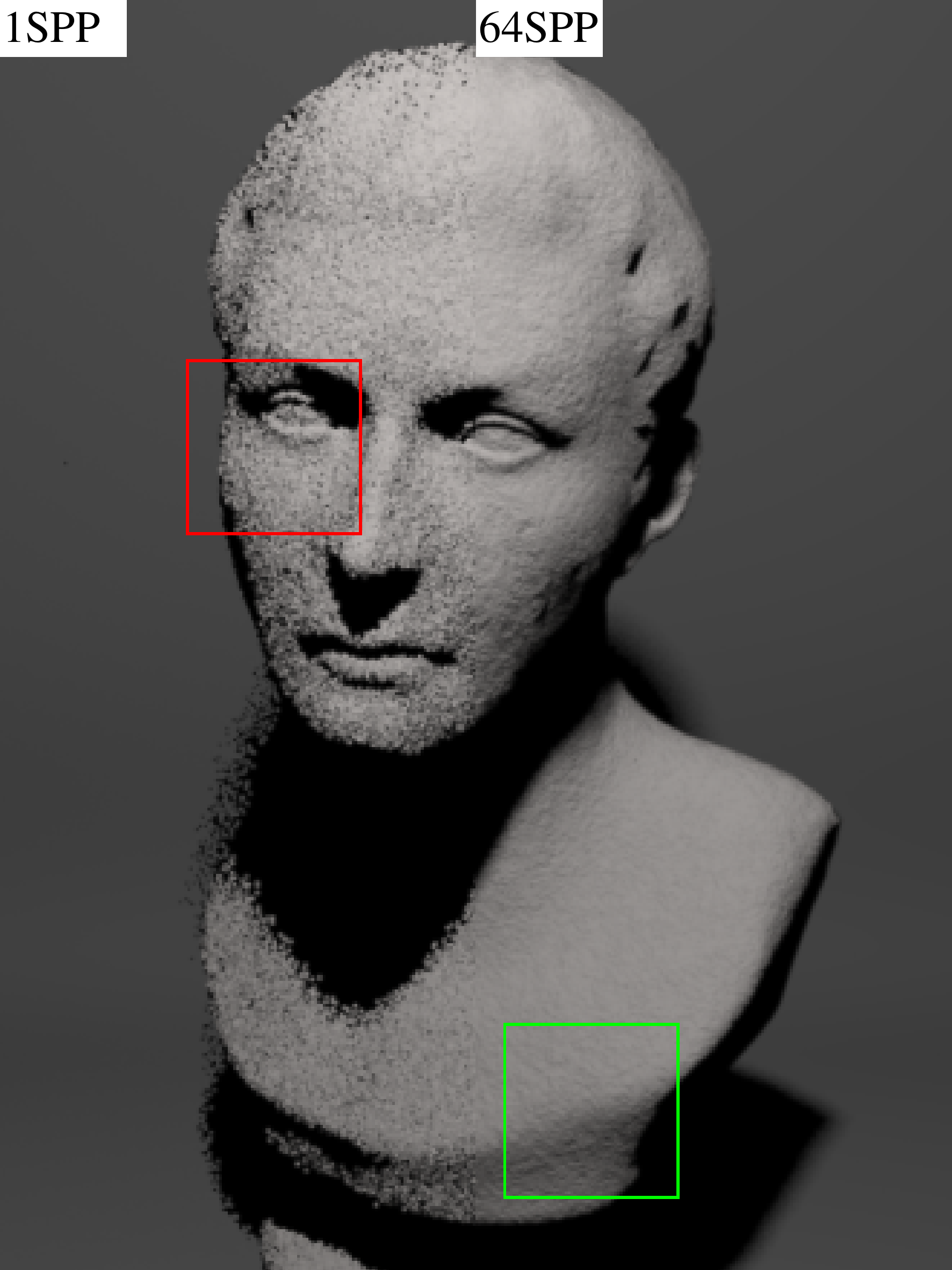}&
    \includegraphics[width=0.34\linewidth]{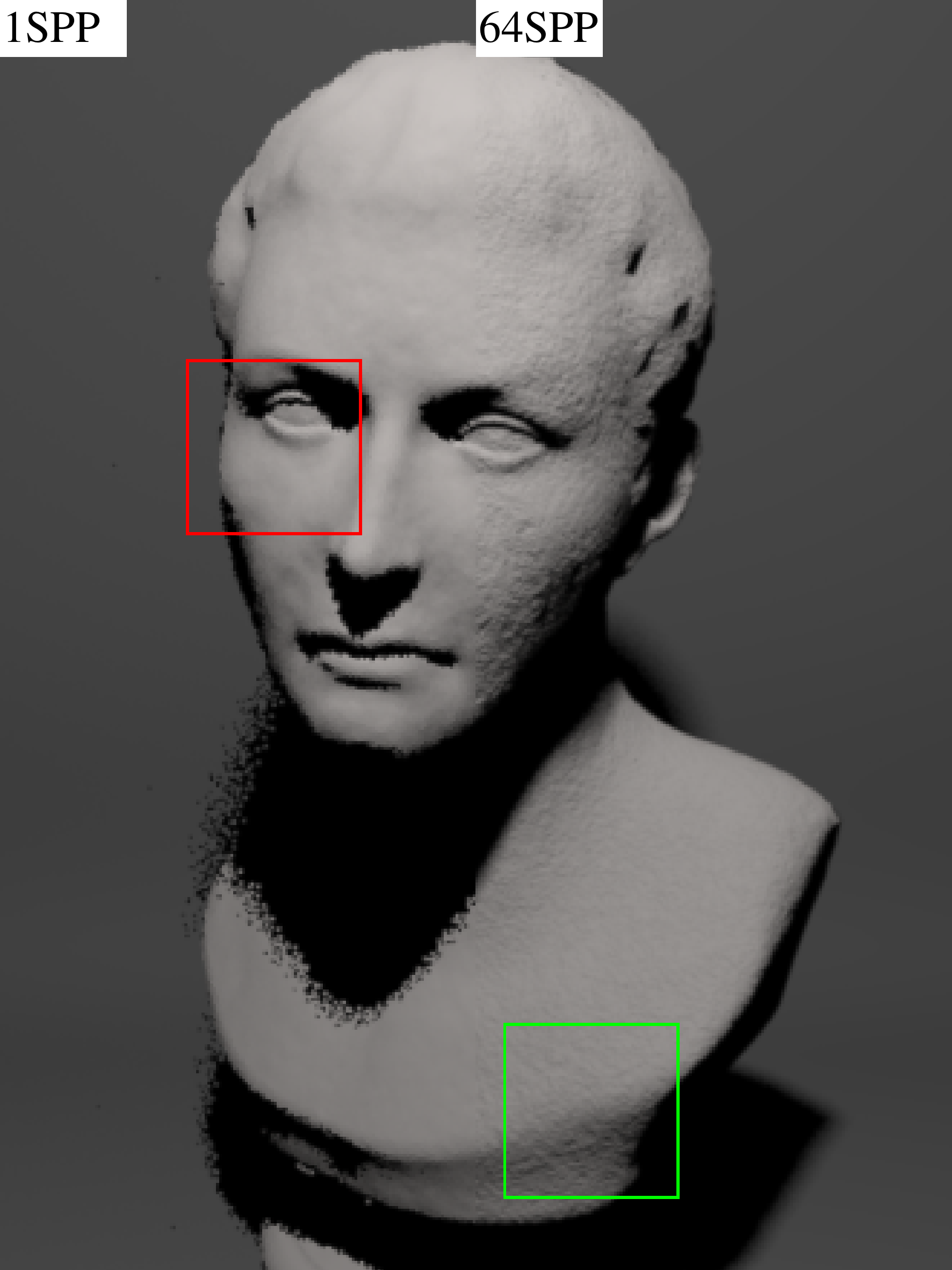}&
    \includegraphics[width=0.34\linewidth]{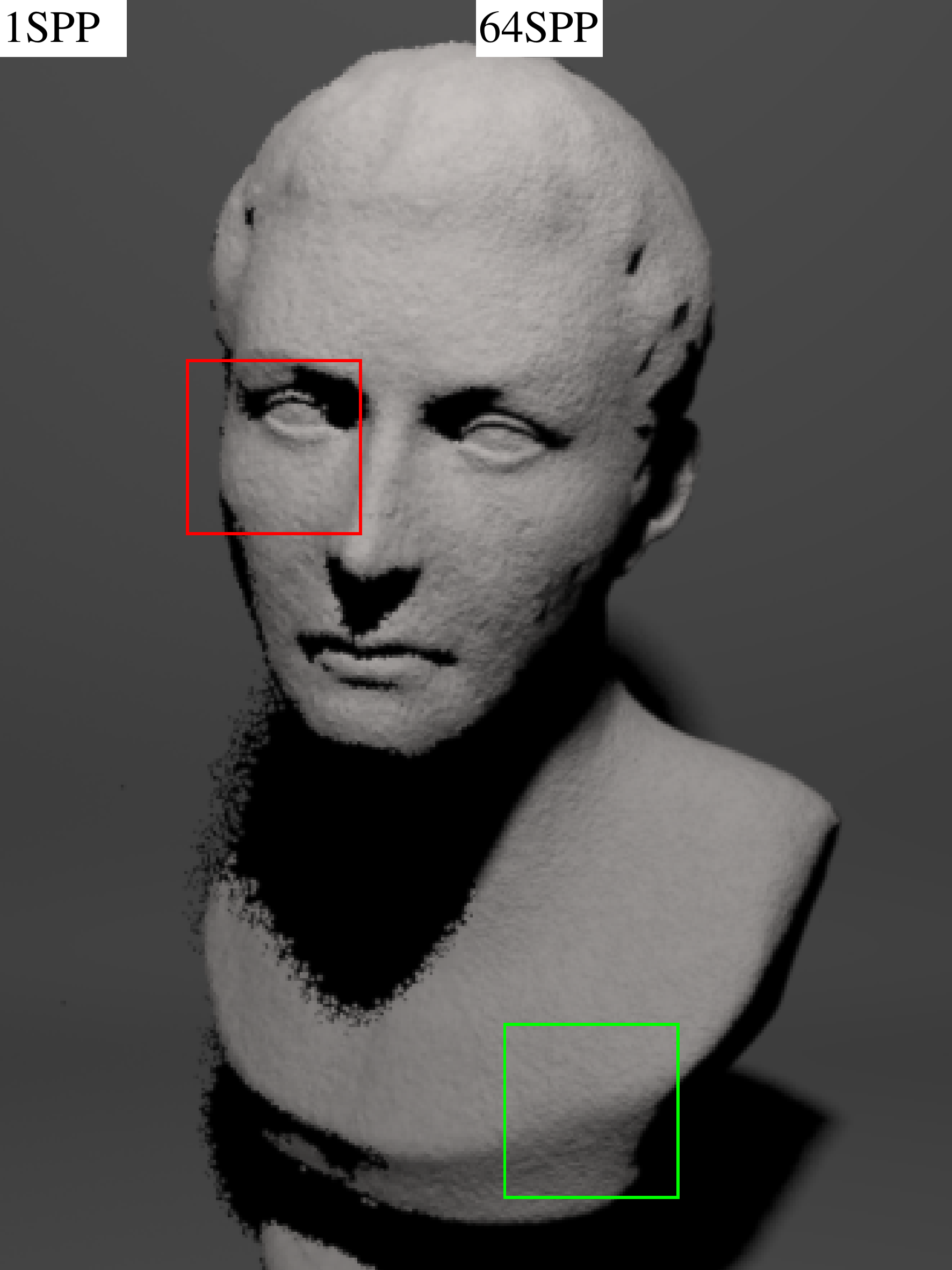}\\[-2pt]
        \includegraphics[width=0.34\linewidth]{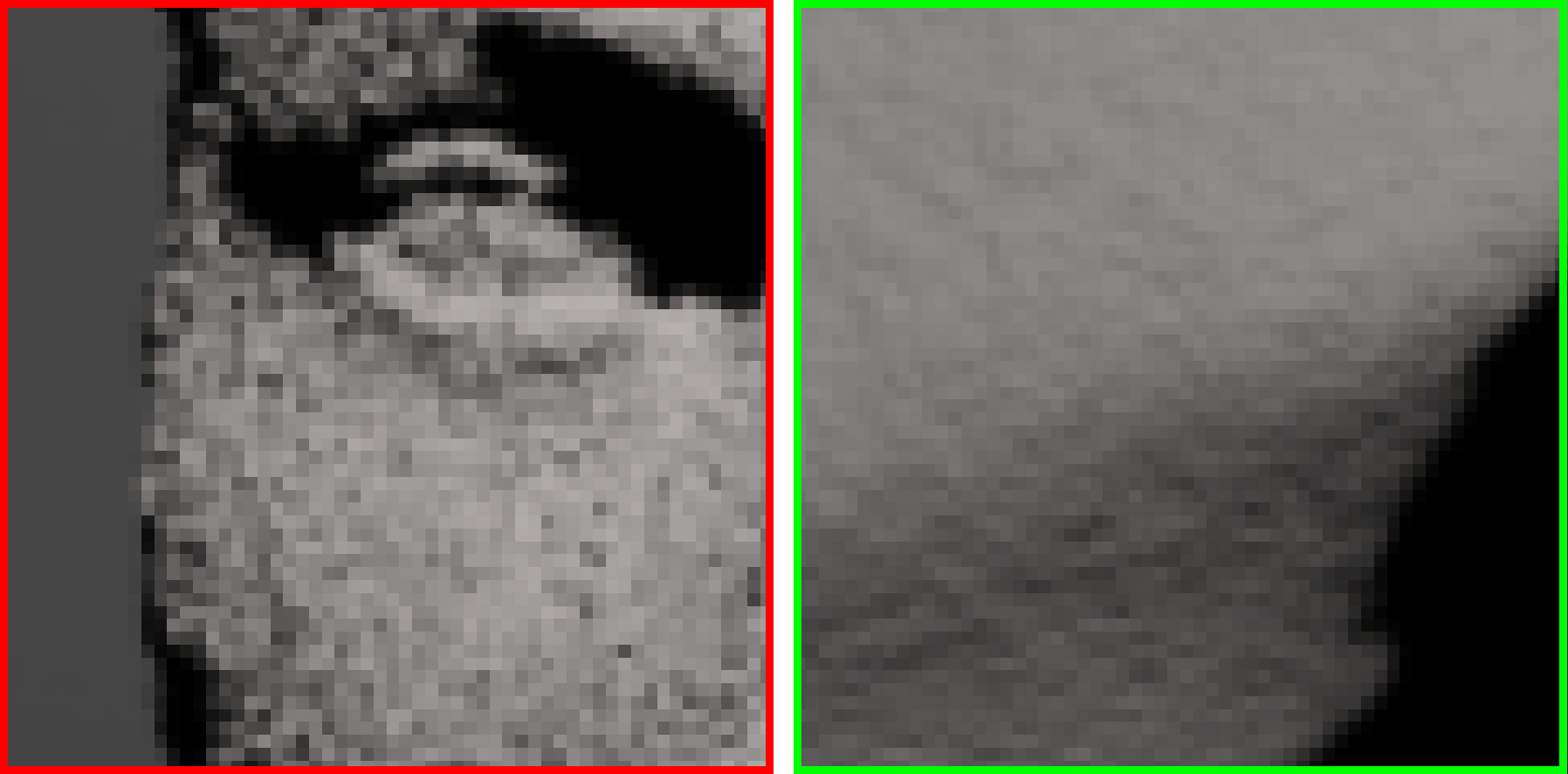}&
    \includegraphics[width=0.34\linewidth]{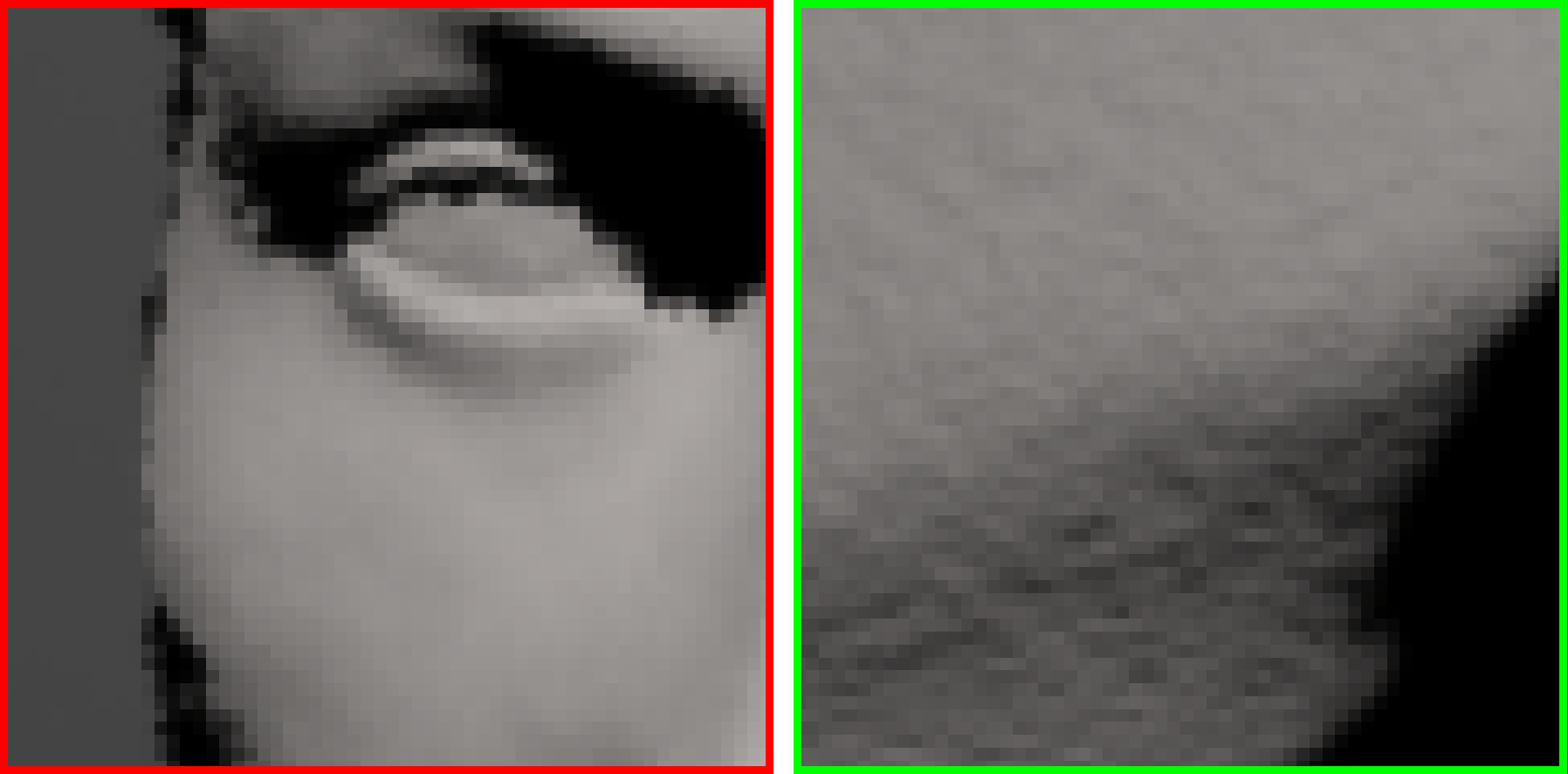}&
    \includegraphics[width=0.34\linewidth]{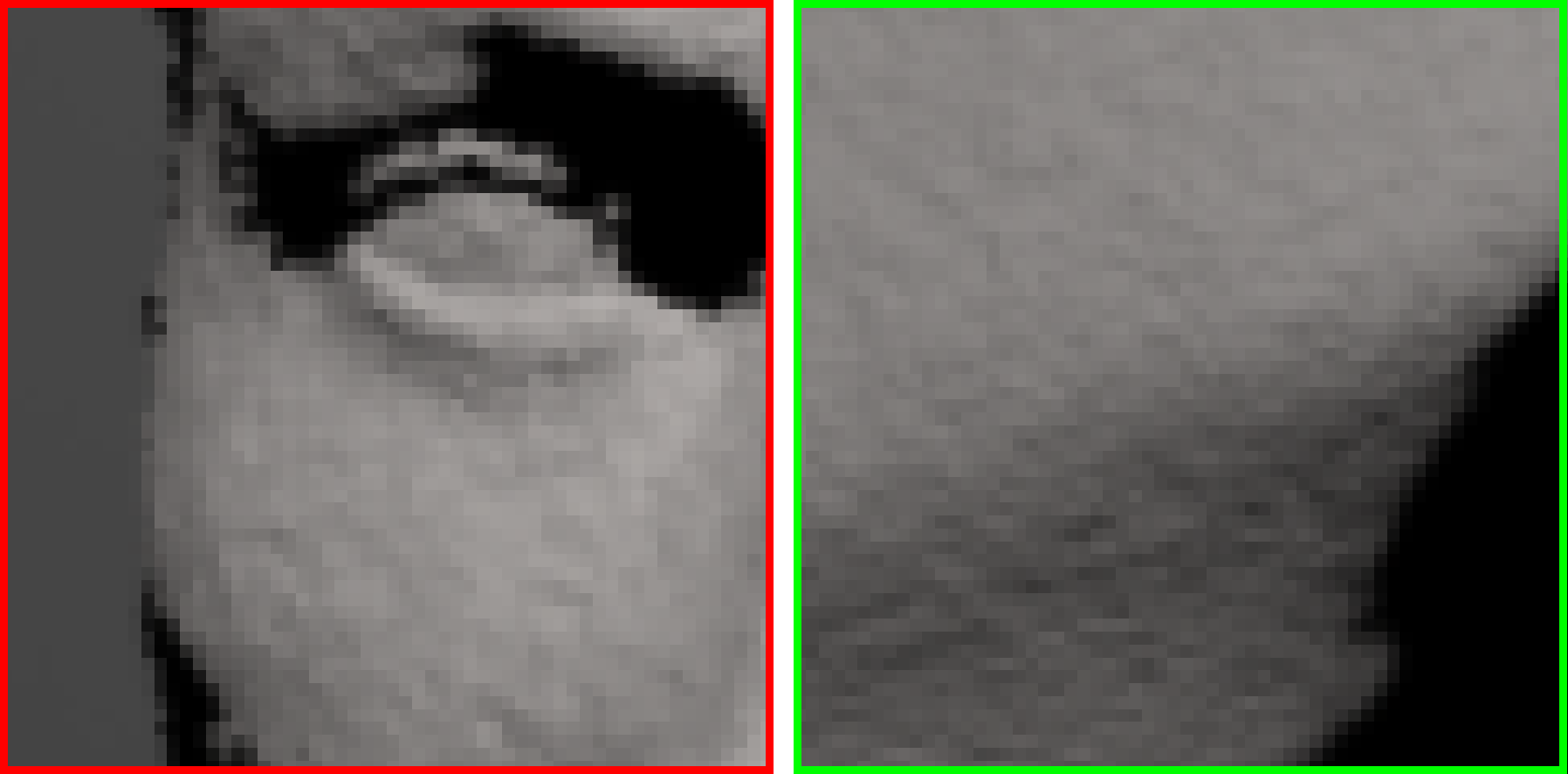}\\
    \end{tabular}
    }
    \caption{
    \textbf{Qualitative comparison between normal-mapped and our aggregated diffuse BRDF.}
    When the surface normal is in small (micro) scale,
    the standard normal mapping method fails to consider each normal texel's contribution within the image pixel, 
    leading to aliasing artifacts for low sampling rate (left).
    Oren-Nayar BRDF~\cite{oren1994generalization} using LEADR-mapping~\cite{dupuy2013linear} (implementation details in supplementary) removes the aliasing but also the normal-map details (middle). 
    Instead, our method analytically integrates all the facets' diffuse reflections within the footprint, 
    so it does not miss important surface reflections even at 1SPP (right). 
    The images are rendered in $720\!\times\!960$ resolution.
    The red and green insets are at 1 and 64 SPP.
    }
    \label{fig:diffuse}
\end{figure*}

%% file: figures/performance-bias.tex
\begin{figure}[t]
    \centering
    \setlength\tabcolsep{0.5pt}
    \resizebox{0.99\linewidth}{!}{
    \begin{tabular}{cccc}
    Reference & $\tau=10^{-4}$ & $\tau=10^{-3}$ & $\tau=10^{-2}$\\[-2pt]
    \includegraphics[width=0.26\linewidth]{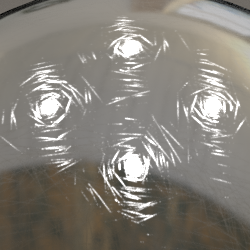}&
    \includegraphics[width=0.26\linewidth]{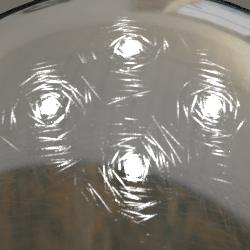}&
    \includegraphics[width=0.26\linewidth]{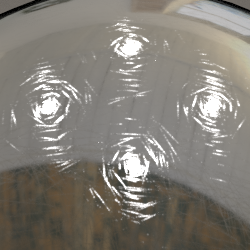}&
    \includegraphics[width=0.26\linewidth]{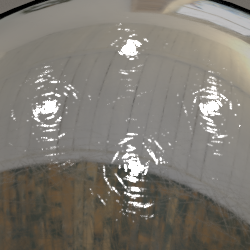}\\
    \end{tabular}
    }
    \caption{
        \textbf{Qualitative ablation of difference clustering threshold.} The glint pattern can be well-preserved when the clustering threshold $\tau$ is selected well (1st and 2nd images).
        For a large $\tau$, the cluster normals no longer match the ground truth,
        causing distorted highlights (4th image).
    }
    \label{fig:performance-bias}
\end{figure}

%% file: tables/performance-bias.tex
\begin{table}[t]
    \centering
    \caption{
    \textbf{Performance-quality ablation} shows increasing the clustering threshold helps reduce the inference speed but hurts the rendering.
    }
    \setlength\tabcolsep{4pt}
    \resizebox{0.8\linewidth}{!}{
    \begin{tabular}{l| c c c c}
        \toprule
        \multicolumn{1}{c}{}& Reference & $\tau\!=\!10^{-4}$ & $\tau\!=\!10^{-3}$ & $\tau\!=\!10^{-2}$\\
        \midrule
        RMSE$\downarrow$ & 0 & 0.0218 & 0.0465 & 0.0818\\
        Minutes$\downarrow$ & 4.24 & 0.67 & 0.49 & 0.26\\ 
        \bottomrule
    \end{tabular}
    }
    \label{tab:performance-bias}
\end{table}

%% file: figures/kernel.tex
\begin{figure}[t]
    \centering
    \setlength\tabcolsep{0.5pt}
    \resizebox{0.99\linewidth}{!}{
    \begin{tabular}{ccc}
         \includegraphics[width=0.36\linewidth]{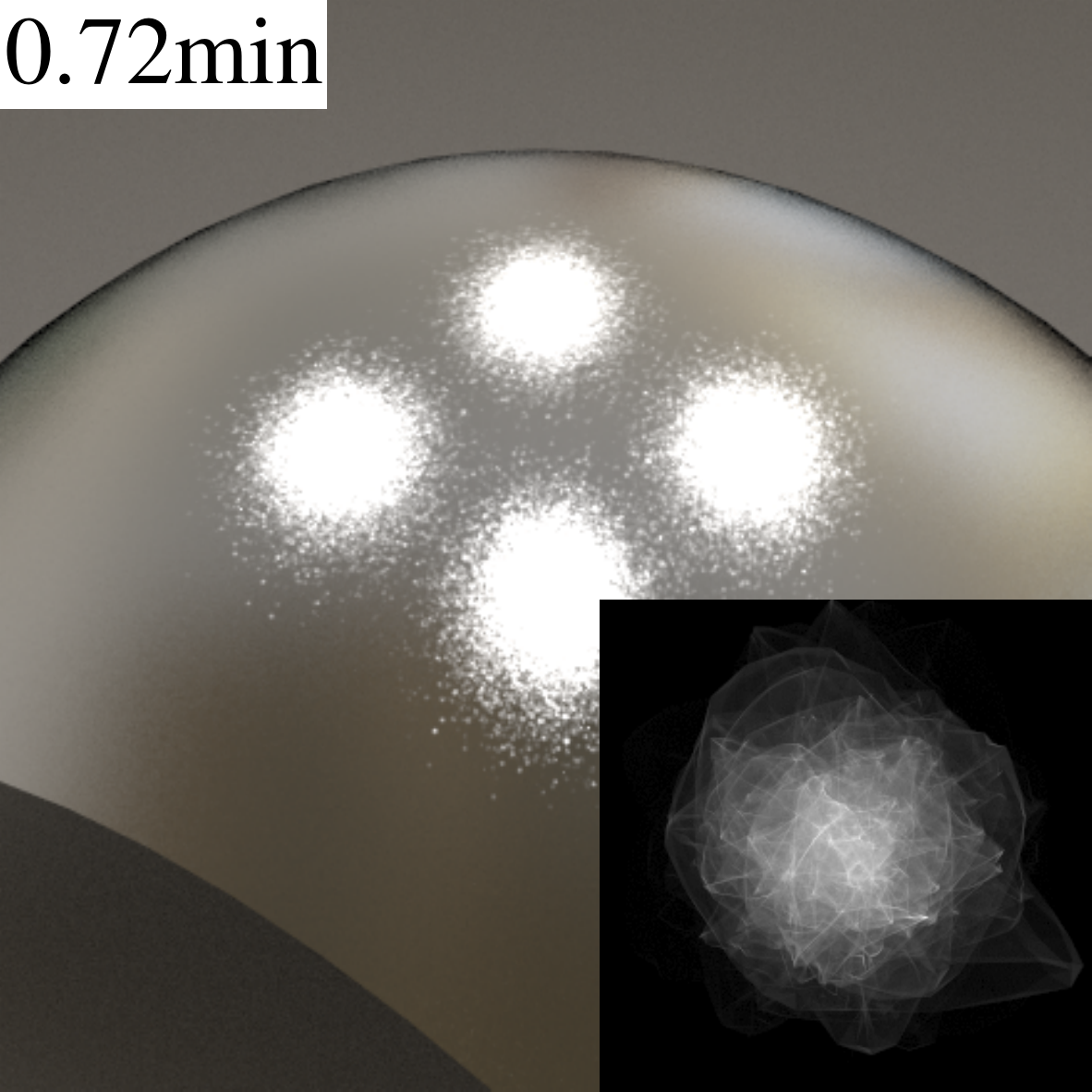}&
         \includegraphics[width=0.36\linewidth]{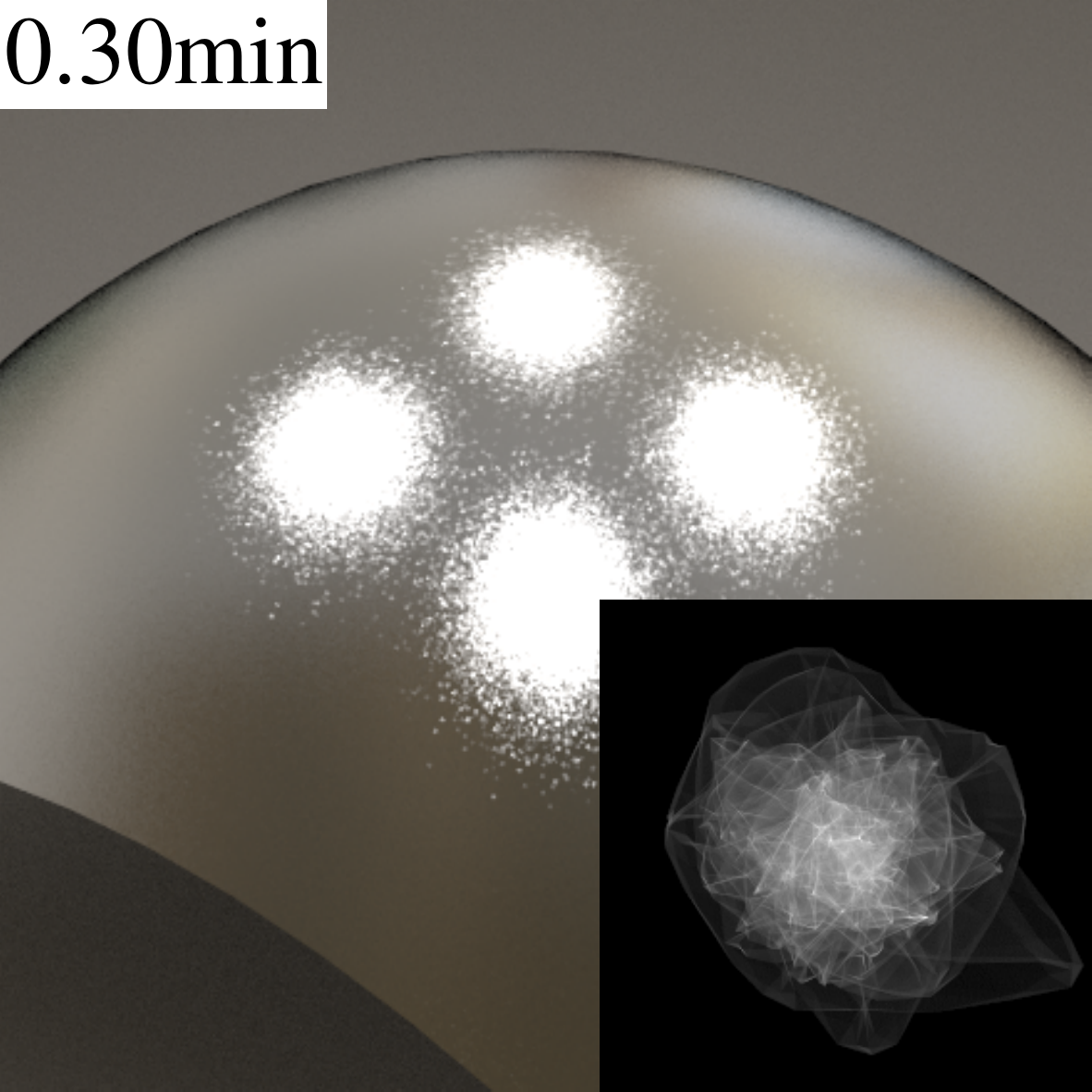}&
         \includegraphics[width=0.36\linewidth]{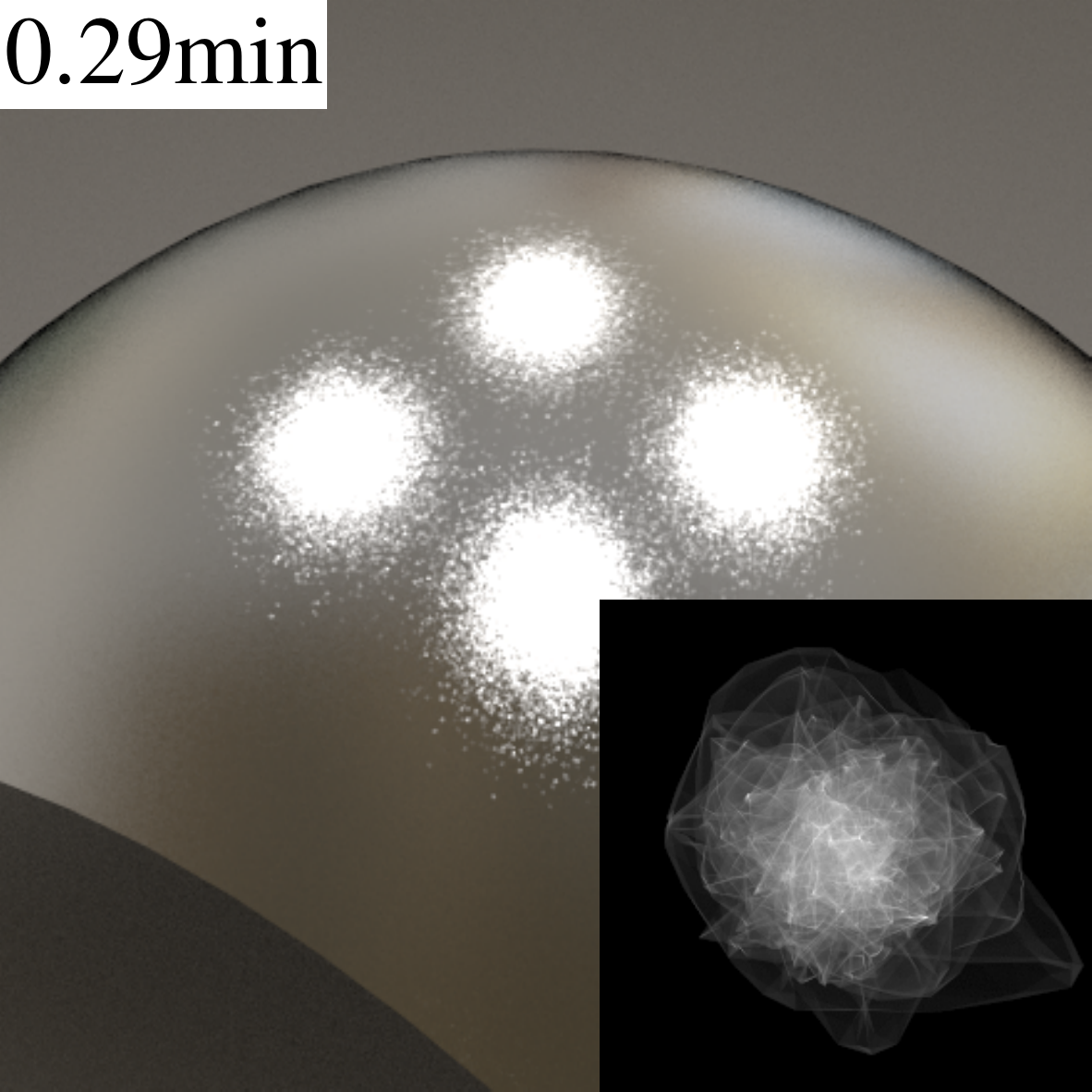}\\[-2pt]
         Gaussian & Disk & Box
    \end{tabular}
    }
    \caption{
        \textbf{Our \pndf{} with different footprint kernels.}
        Both the disk and the box filter give similar \pndf{}s (insets) and renderings compared to the Gaussian filter. However, they have smaller footprint size thus are faster to compute. The numbers on the images show the inference time.
    }
    \label{fig:kernel}
\end{figure}

%% file: 5-conclusion.tex
\section{Conclusion and Future Work}
\label{sec:conclusion}

We presented a manifold-based formulation of the surface NDF by utilizing mesh intersection techniques,
which bring more efficient glint rendering on highly detailed surfaces.
Moreover, we extend the glint BRDF model with an analytical shadow-masking,
as well as introduce a novel approach of filtering detailed diffuse reflections.
In terms of limitations,
our uniform cluster grids are less efficient at modeling high frequency but sparse structures such as the scratched surface (\cref{tab:performance}).
Adapting mesh simplification algorithms~\cite{garland1997surface} to also optimize the grid topology may help.
Our shadow-masking and diffuse BRDF do not consider multiple scattering on the microsurface that can cause energy loss on rough surfaces.
Ideas from manifold exploration~\cite{jakob2012manifold} can potentially be applied to handle these effects.
Also, our method does not take wave optics~\cite{yan2018rendering} into account.

\begin{acks}
This work was supported in part by NSF grant 2212085 and the Ronald L. Graham Chair.
Ramamoorthi acknowledges a part-time appointment at NVIDIA.
We also acknowledge support from ONR grant N00014-23-1-2526, gifts from Adobe, Google, Qualcomm and Rembrand and the UC San Diego Center for Visual Computing.
\end{acks}

%% file: 6-supplemental.tex
\section{Derivations of Cluster Normal Optimization}
Below, we derive $\mathbf{A},\mathbf{B}$ in Eq.~8 of the main paper.
For a cluster in range $[2^li,2^l(i+1)]\times[2^lj,2^l(j+1)]$,
we first translate the cluster and all the triangles within by $(2^li,2^lj)$ to the canonical domain $[0,2^l]\times[0,2^l]$.
Let $\triangle_{ij}^+$ denote the upper triangle with vertices $(i,j),(i+1,j),(i,j+1)$;
$\triangle_{ij}^-$ denote the lower triangle with vertices $(i+1,j),(i+1,j+1),(i,j+1)$;
$\triangle_l^+$ and $\triangle_l^-$ denote the cluster triangles with vertices $(0,0),(2^l,0), (2^l,0)$ and $(2^l,0),(2^l,2^l),(0,2^l)$.
The integral Eq.~7 of the main paper can be written as:
\begin{equation}
\begin{aligned}
    \int \frac{\Vert \mathbf{n}^l(\mathbf{u}/2^l)\!-\!\mathbf{n}(\mathbf{u})\Vert^2}{\vert\det\mathbf{J}(\mathbf{u})\vert} \mathrm{d}\mathbf{u}
    =\\
    \sum_{i,j=0}^{2^l-1} (
    \int\limits_0^1\!\int\limits_0^{1-v}\!
    \frac{\Vert\mathbf{n}^l((u\!+\!i,v\!+\!j)/2^l)\!-\!\mathbf{n}((i\!+\!u,j\!+\!v))\Vert^2}{2\Vert\mathbf{n}(\triangle_{ij}^+)\Vert} \mathrm{d}u\mathrm{d}v\\
    +\int\limits_0^1\!\int\limits_{1-v}^1\!
    \frac{\Vert\mathbf{n}^l((u\!+\!i,v\!+\!j)/2^l)\!-\!\mathbf{n}((i\!+\!u,j\!+\!v))\Vert^2}{2\Vert\mathbf{n}(\triangle_{ij}^-)\Vert} \mathrm{d}u\mathrm{d}v ).
\end{aligned}
\end{equation}%
By applying the normal interpolation function (Eq.~6 of the main paper) to $\mathbf{n}$ and $\mathbf{n}^l$, 
the two integrals above become polynomial integrals of $u,v$ that have closed-form solutions.
Computing their gradients respect to $(\mathbf{n}^l_0,\mathbf{n}^l_1,\mathbf{n}^l_2,\mathbf{n}^l_3)$ using a symbolic solver and letting $a=2^l$, we have:
\input{equations/coefficientA}
\input{equations/coefficientB}%
Here, $\mathbf{n}_0=\mathbf{n}((i,j)),\mathbf{n}_1=\mathbf{n}((i+1,j)),\mathbf{n}_2=\mathbf{n}((i,j+1)),\mathbf{n}_3=\mathbf{n}((i+1,j+1))$,
and the final $\mathbf{A,B}$ can be obtained by summing up all the $\mathbf{A}_{ij}^+,\mathbf{A}_{ij}^-,\mathbf{B}_{ij}^+,\mathbf{B}_{ij}^-$ divided by the corresponding normal triangle areas.

\section{Derivations of Projected-Area Integral}
\paragraph{Integral domain on projected hemisphere.}
Under our canonical setting, it is trivial to see in Fig.~7 of the main paper that the right boundary of $\tilde{\mathbf{m}}\!^\top\!\bm{\omega}\geq0$ is a semi-circle,
and the left boundary is given by $\tilde{\mathbf{m}}\!^\top\!\bm{\omega}=0$. Since $\bm{\omega}_y=0$, we have:
\begin{equation}
\begin{aligned}
\tilde{\mathbf{m}}\!^\top\!\bm{\omega}=\tilde{\mathbf{m}}_x\bm{\omega}_x+\tilde{\mathbf{m}}_z\bm{\omega}_z=0
\Rightarrow
(\tilde{\mathbf{m}}_x\bm{\omega}_x)^2=(\tilde{\mathbf{m}}_z\bm{\omega}_z)^2\\
\Rightarrow
\tilde{\mathbf{m}}^2_x(1-\bm{\omega}_z^2)=(1-\tilde{\mathbf{m}}_x^2-\tilde{\mathbf{m}}_y^2)\bm{\omega}_z^2\\
\Rightarrow 
(\tfrac{\mathbf{m}_x}{\bm{\omega}_z})^2+\mathbf{m}_y^2=1,
\end{aligned}
\end{equation}%
which is an ellipse.
Given a normal triangle with three edges of endpoints $\mathbf{n}^i,\mathbf{n}^{i+1}$ $ (\mathbf{n}^3\!=\!\mathbf{n}^0)$,
we clip these edges to the interior of the semi-ellipse and semi-circle, 
which is done by solving $t$ for the line-ellipse intersection $(\tfrac{\mathbf{n}_x^i(1-t)+\mathbf{n}_x^{i+1}t}{\bm{\omega}_z})^2+(\mathbf{n}_y^i(1-t)+\mathbf{n}_y^{i+1}t)^2=1$ (\cref{alg:clip}).
The clipped endpoints are then connected as either lines or ellipse arcs to form the final integral domain.
\input{equations/clip}

\paragraph{Correctness of the area-line integral conversion.}
By Stokes theorem, the curl of the right-hand-side (RHS) integrand of the main paper Eq.~11 should equal to the left-hand-side (LHS) integrand,
which is true as follow:
\begin{equation}
\begin{aligned}
\nabla_\times\!
\begin{bmatrix}
0\\
\bm{\omega}_z\tilde{\mathbf{m}}_x\!-\!\bm{\omega}_x\tilde{\mathbf{m}}_z\\
0\\
\end{bmatrix}\!\cdot\!
\begin{bmatrix}
    \mathrm{d}\tilde{\mathbf{m}}_{yz}\\
    \mathrm{d}\tilde{\mathbf{m}}_{zx}\\
    \mathrm{d}\tilde{\mathbf{m}}_{xy}
\end{bmatrix}
\!=\!
\tfrac{\partial
(\bm{\omega}_z\tilde{\mathbf{m}}_x\!-\!\bm{\omega}_x\sqrt{1\!-\!\tilde{\mathbf{m}}_{\smash{x}}^{\smash{2}}\!-\!\tilde{\mathbf{m}}_{\smash{y}}^{\smash{2}}})}{\partial\tilde{\mathbf{m}}_x}\mathrm{d}\tilde{\mathbf{m}}_{xy}\\
=(\bm{\omega}_z+
\bm{\omega}_x\tfrac{\tilde{\mathbf{m}}_x}{\sqrt{1-\tilde{\mathbf{m}}_{\smash{x}}^{\smash{2}}-\tilde{\mathbf{m}}_{\smash{y}}^{\smash{2}}}})\mathrm{d}\mathbf{m}
=(\bm{\omega}_x\tfrac{\tilde{\mathbf{m}}_x}{\tilde{\mathbf{m}}_z}+\bm{\omega}_z)\mathrm{d}\mathbf{m}.
\end{aligned}
\end{equation}

\paragraph{Line integral of the ellipse arcs.}
On the ellipse, we have $\mathbf{m}_x=-\bm{\omega}_z\sqrt{1-\mathbf{m}_{\smash{y}}^2}$,
and the line integral becomes:
\begin{equation}
\begin{aligned}
\oint_{\mathbf{n}^i}^{\mathbf{n}^{i+1}}\!
(\bm{\omega}_z\tilde{\mathbf{m}}_x-\bm{\omega}_x\tilde{\mathbf{m}}_z)\mathrm{d}\tilde{\mathbf{m}}_y\\
=\int_{\mathbf{n}^i}^{\mathbf{n}^{i+1}}\!
\left(-\bm{\omega}_z^2\sqrt{1-\mathbf{m}_{\smash{y}}^{\smash{2}}}
-(1-\bm{\omega}_{\smash{z}}^{\smash{2}})
\sqrt{1-\mathbf{m}_{\smash{y}}^{\smash{2}}}\right)\mathrm{d}\mathbf{m}_y\\
=\int_{\mathbf{n}^i}^{\mathbf{n}^{i+1}}\!
-\sqrt{1-\mathbf{m}_{\smash{y}}^{\smash{2}}}\mathrm{d}\mathbf{m}_y
=\tfrac{1}{2}\left[\arcsin{p}+p\sqrt{1-p^{\smash{2}}}\right]_{\mathbf{n}^{i+1}_y}^{\mathbf{n}^{i}_y}.
\end{aligned}
\end{equation}

\input{figures/line_integral}
\paragraph{Line integral of the line segments.}
On a line segment, 
the integral of $\bm{\omega}_z\tilde{\mathbf{m}}_x$ is:
\begin{equation}
\begin{aligned}
\oint_{\mathbf{n}^i}^{\mathbf{n}^{i+1}}\!
\bm{\omega}_z\tilde{\mathbf{m}}_x \mathrm{d}\tilde{\mathbf{m}}_y
=\bm{\omega}_z\int_0^1\!
(\mathbf{n}^i_x(1-t)+\mathbf{n}^{i+1}_xt)(\mathbf{n}_y^{i+1}\!-\!\mathbf{n}_y^i)\mathrm{d}t\\
=
\tfrac{\bm{\omega}_z}{2}(\mathbf{n}^{i+1}_y-\mathbf{n}^{i}_y)(\mathbf{n}^{i+1}_x\!+\!\mathbf{n}^i_x),
\end{aligned}
\end{equation}
and we show a geometric derivation of $\bm{\omega}_x\tilde{\mathbf{m}}_z$'s integration in \cref{fig:line_integral}.
The line segment unprojected onto the hemisphere gives a circle arc,
where $\mathbf{m}_z$ is its ordinate and the line is in the direction of the abscissa.
Therefore, integrating $\tilde{\mathbf{m}}_zd\tilde{\mathbf{m}}_y$ is equivalent to projecting the arc's underlying area to the $\tilde{\mathbf{m}}_y$ axis (\cref{fig:line_integral} top).
Let $\mathbf{d}=\tfrac{\mathbf{n}^{i+1}-\mathbf{n}^i}{\Vert\mathbf{n}^{i+1}-\mathbf{n}^i\Vert_2}$ denotes the tangent of the line.
The line's distance to the origin is $s=\vert\mathbf{d}_y\mathbf{n}_x^i-\mathbf{d}_x\mathbf{n}_y^i\vert$, so the circle's radius is $r=\sqrt{1-s^{\smash{2}}}$ (\cref{fig:line_integral} bottom).
Projecting $\mathbf{n}^i,\mathbf{n}^{i+1}$ to the tangent direction,
we get their abscissas on the circle plane $p^i=\mathbf{d}\!^\top\!\mathbf{n}^i,p^{i+1}=\mathbf{d}\!^\top\!\mathbf{n}^{i+1}$,
and the underlying arc area is:
\begin{equation}
\begin{aligned}
    \int_{p^i}^{p^{i+1}}\sqrt{r^{2}-p^{2}}\mathrm{d}p
=\int_{p^i/r}^{p^{i+1}/r}r^2\sqrt{1-(p/r)^{2}}\mathrm{d}(p/r)  
\\    
=\tfrac{1}{2}r^2\left[\arcsin{p}+p\sqrt{1\!-\!p^{\smash{2}}}\right]^{\mathbf{d}\!^\top\!\mathbf{n}^{i+1}/r}_{\mathbf{d}\!^\top\!\mathbf{n}^i/r}.
\end{aligned}
\end{equation}
The cosine term between the line and the $\tilde{\mathbf{m}}_y$ axis is $\mathbf{d}_x$,
thus, the line integral of $\bm{\omega}_x\tilde{\mathbf{m}}_z$ is:
\begin{equation}
    \oint_{\mathbf{n}^i}^{\mathbf{n}^{i+1}}\!
\bm{\omega}_x\tilde{\mathbf{m}}_z \mathrm{d}\tilde{\mathbf{m}}_y
=
\tfrac{\bm{\omega}_x}{2}r^2\mathbf{d}_y\left[\arcsin{p}+p\sqrt{1\!-\!p^{\smash{2}}}\right]^{\mathbf{d}\!^\top\!\mathbf{n}^{i+1}/r}_{\mathbf{d}\!^\top\!\mathbf{n}^i/r}.
\end{equation}

\section{Experiment details}
\paragraph{Entry level of the acceleration structures.}
Given a normal map of resolution $N^2$, queried location $\mathbf{x}$, and footprint $\mathbf{r}$,
we start the traversal of the acceleration structures at level $l=\lceil\max(\log\mathbf{r}_x,\log\mathbf{r}_y)\rceil$ of coordinate $\lfloor\mathbf{x}/2^l\rfloor$.
Here, $l\!=\!0$ corresponds to the level of leaf nodes. 

\paragraph{Oren-Nayar BRDF with LEADR mapping.}
In Fig.~16 of the paper,
the Oren-Nayar~\cite{oren1994generalization} BRDF baseline is evaluated in mip-mapped normal shading frame using mip-mapped slope variance as in the LEADR mapping~\cite{dupuy2013linear}.
We mip-map $\mathbf{n}$ then get $\tilde{\mathbf{n}}_z=\sqrt{1-\lVert\mathbf{n}\rVert^2}$,
and the mip-mapped variance is derived through the mip-mapped slope $\tfrac{\lVert\tilde{\mathbf{m}}_{xy}\rVert_2}{\tilde{\mathbf{m}}_z}$ and slope's second-order moment.
Oren-Nayar BRDFs use Beckmann NDF so fit perfectly with the LEADR mapping.

\paragraph{Diffuse appearance renderings in the paper Fig.~11.}
We use a footprint scale of $16^2$ for the first row and $32^2$ for the second row with corresponding rendering time 15s and 32s.
The Oren-Nayar model in the last row takes 3s to render.

\input{figures/supp_normalmapped}%
\input{figures/supp_comparison}%
\input{tables/performance_supp}%
\input{tables/timing}%

\section{Additional results.}

\paragraph{Comparison with normal-mapped ground truth.}
We show the glint renderings of the `ground truth' normal-mapped specular surface (mirror reflection) in \cref{fig:supp-normalmapped}.
It can be seen that the standard normal mapping produces aliasing artifacts with a low SPP, suggesting the necessity of the \pndf modeling.
Meanwhile, glint patterns from our formulation are closer to the ground truth than \citet{yan2016position} that use intrinsic roughness, 

\paragraph{Additional comparison with \citet{yan2016position}.}
Figure~\ref{fig:supp-comparison} shows qualitative comparison with \citeauthor{yan2016position} under different footprint scales on scenes in Fig.~9 of
the paper.
We additionally show renderings using the flake normal map with coating,
whose timing is provided in \cref{tab:performance_supp}.
Because the flake normal map is nearly piecewise constant that can be efficiently pruned by the bounding box hierarchy,
both \citet{yan2016position} and our evaluation without the clustering (ours no cluster) demonstrate faster inference speed compared to their performances on other normal maps.

\paragraph{Timing of different stages.}
Table~\ref{tab:timing} shows the timing of different rendering stages for the scenes in Fig.~14 of the paper.
For indirect bounces, Mitsuba~\cite{mitsuba} sets their ray differentials to zero that corresponds to querying the \pndf{} with a small footprint,
thus, their computations are fast.
While the BRDF sampling by the paper Eq.~(6) should be fast,
the PDF (\pndf) evaluation in the BRDF sampling stage is more expensive than that in the emitter sampling stage.
This is because their sampled normals usually have high NDF responses,
such that the evaluation of the paper Eq.~(5) may contain more intersections that cannot be pruned out.

%% file: equations/coefficientA.tex
\begin{equation}
\small
\begin{aligned}
\underset{(\mathbf{n}^l_0,\mathbf{n}^l_1,\mathbf{n}^l_2,\mathbf{n}^l_3)}{\nabla}
\int \frac{\Vert \mathbf{n}^l(\mathbf{u}/2^l)\!-\!\mathbf{n}(\mathbf{u})\Vert^2}{\vert\det\mathbf{J}(\mathbf{u})\vert} \mathrm{d}\mathbf{u}=
\!\sum_{i,j=0}^{2^l-1}(\tfrac{\mathbf{B}^+_{ij}}{2\Vert\mathbf{n}(\triangle_{ij}^+)\Vert}+
\tfrac{\mathbf{B}^-_{ij}}{2\Vert\mathbf{n}(\triangle_{ij}^-)\Vert})
\\
+\sum_{i,j=0}^{2^l-1}
    (\tfrac{\mathbf{A}^+_{ij}}{2\Vert\mathbf{n}(\triangle_{ij}^+)\Vert}
    +\tfrac{\mathbf{A}^-_{ij}}{2\Vert\mathbf{n}(\triangle_{ij}^-)\Vert})(\mathbf{n}^l_0,\mathbf{n}^l_1,\mathbf{n}^l_2,\mathbf{n}^l_3)\!^\top\!.
\end{aligned}
\end{equation}

\begin{equation}
\small
\begin{aligned}
\mathbf{A}_{ij}^+=
\scalebox{0.95}{$
\left[
\begingroup 
\setlength\arraycolsep{2pt}
\begin{matrix}
\frac{6 a^{2} - 4 a \left(3 i + 3 j + 2\right) + 6 i^{2} + 12 i j + 8 i + 6 j^{2} + 8 j + 3}{6 a^{2}} & \frac{\frac{a \left(3 i + 1\right)}{3} - i^{2} - i j - i - \frac{j}{3} - \frac{1}{4}}{a^{2}}\\\frac{\frac{a \left(3 i + 1\right)}{3} - i^{2} - i j - i - \frac{j}{3} - \frac{1}{4}}{a^{2}} & \frac{6 i^{2} + 4 i + 1}{6 a^{2}}\\\frac{\frac{a \left(3 j + 1\right)}{3} - i j - \frac{i}{3} - j^{2} - j - \frac{1}{4}}{a^{2}} & \frac{i j + \frac{i}{3} + \frac{j}{3} + \frac{1}{12}}{a^{2}}\\0 & 0\end{matrix}
\endgroup
\right.
$}
\\
\scalebox{0.95}{$
\left.
\begingroup 
\setlength\arraycolsep{2pt}
\begin{matrix}\frac{\frac{a \left(3 j + 1\right)}{3} - i j - \frac{i}{3} - j^{2} - j - \frac{1}{4}}{a^{2}} & 0\\\frac{i j + \frac{i}{3} + \frac{j}{3} + \frac{1}{12}}{a^{2}} & 0\\\frac{6 j^{2} + 4 j + 1}{6 a^{2}} & 0\\0 & 0\end{matrix}
\endgroup
\right]
$}
\\
\mathbf{B}_{ij}^+=
\scalebox{0.95}{$
\left[
\begingroup 
\setlength\arraycolsep{2pt}
\begin{matrix}\frac{2 \mathbf{n_0} \left(- 2 a + 2 i + 2 j + 1\right) + \mathbf{n_1} \left(- 4 a + 4 i + 4 j + 3\right) + \mathbf{n_2} \left(- 4 a + 4 i + 4 j + 3\right)}{12 a}\\\frac{- \mathbf{n_0} \cdot \left(4 i + 1\right) - 2 \mathbf{n_1} \cdot \left(2 i + 1\right) - \mathbf{n_2} \cdot \left(4 i + 1\right)}{12 a}\\\frac{- \mathbf{n_0} \cdot \left(4 j + 1\right) - \mathbf{n_1} \cdot \left(4 j + 1\right) - 2 \mathbf{n_2} \cdot \left(2 j + 1\right)}{12 a}\\0\end{matrix}
\endgroup
\right]
$}
\\
\text{for} \quad \triangle_{ij}^+ \in \triangle_{l}^+
\end{aligned}
\end{equation}

\begin{equation}
\small
\begin{aligned}
\mathbf{A}_{ij}^+=
\scalebox{0.95}{$
\left[
\begingroup 
\setlength\arraycolsep{2pt}
\begin{matrix}0 & 0 & 0\\0 & \frac{6 a^{2} - 12 a j - 4 a + 6 j^{2} + 4 j + 1}{6 a^{2}} & \frac{a^{2} - \frac{a \left(3 i + 3 j + 2\right)}{3} + i j + \frac{i}{3} + \frac{j}{3} + \frac{1}{12}}{a^{2}}\\0 & \frac{a^{2} - \frac{a \left(3 i + 3 j + 2\right)}{3} + i j + \frac{i}{3} + \frac{j}{3} + \frac{1}{12}}{a^{2}} & \frac{6 a^{2} - 12 a i - 4 a + 6 i^{2} + 4 i + 1}{6 a^{2}}\\0 & \frac{- a^{2} + a \left(i + 2 j + 1\right) - i j - \frac{i}{3} - j^{2} - j - \frac{1}{4}}{a^{2}} & \frac{- a^{2} + a \left(2 i + j + 1\right) - i^{2} - i j - i - \frac{j}{3} - \frac{1}{4}}{a^{2}}\end{matrix}
\endgroup
\right.
$}
\\
\scalebox{0.95}{$
\left.
\begingroup 
\setlength\arraycolsep{2pt}
\begin{matrix}0\\\frac{- a^{2} + a \left(i + 2 j + 1\right) - i j - \frac{i}{3} - j^{2} - j - \frac{1}{4}}{a^{2}}\\\frac{- a^{2} + a \left(2 i + j + 1\right) - i^{2} - i j - i - \frac{j}{3} - \frac{1}{4}}{a^{2}}\\\frac{6 a^{2} - 4 a \left(3 i + 3 j + 2\right) + 6 i^{2} + 12 i j + 8 i + 6 j^{2} + 8 j + 3}{6 a^{2}}\end{matrix}
\endgroup
\right]
$}
\\
\mathbf{B}_{ij}^+=
\scalebox{0.95}{$
\left[
\begingroup 
\setlength\arraycolsep{2pt}
\begin{matrix}0\\\frac{\mathbf{n_0} \left(- 4 a + 4 j + 1\right) + \mathbf{n_1} \left(- 4 a + 4 j + 1\right) + 2 \mathbf{n_2} \left(- 2 a + 2 j + 1\right)}{12 a}\\\frac{\mathbf{n_0} \left(- 4 a + 4 i + 1\right) + 2 \mathbf{n_1} \left(- 2 a + 2 i + 1\right) + \mathbf{n_2} \left(- 4 a + 4 i + 1\right)}{12 a}\\\frac{- 2 \mathbf{n_0} \left(- 2 a + 2 i + 2 j + 1\right) - \mathbf{n_1} \left(- 4 a + 4 i + 4 j + 3\right) - \mathbf{n_2} \left(- 4 a + 4 i + 4 j + 3\right)}{12 a}
\end{matrix}
\endgroup
\right]
$}
\\
\text{for} \quad \triangle_{ij}^+ \in \triangle_{l}^-
\end{aligned}
\end{equation}

%% file: equations/coefficientB.tex
\begin{equation}
\small
\begin{aligned}
\mathbf{A}_{ij}^-=
\scalebox{0.95}{$
\left[
\begingroup 
\setlength\arraycolsep{2pt}
\begin{matrix}\frac{6 a^{2} - 4 a \left(3 i + 3 j + 4\right) + 6 i^{2} + 12 i j + 16 i + 6 j^{2} + 16 j + 11}{6 a^{2}} & \frac{\frac{a \left(3 i + 2\right)}{3} - i^{2} - i j - 2 i - \frac{2 j}{3} - \frac{11}{12}}{a^{2}}\\\frac{\frac{a \left(3 i + 2\right)}{3} - i^{2} - i j - 2 i - \frac{2 j}{3} - \frac{11}{12}}{a^{2}} & \frac{6 i^{2} + 8 i + 3}{6 a^{2}}\\\frac{\frac{a \left(3 j + 2\right)}{3} - i j - \frac{2 i}{3} - j^{2} - 2 j - \frac{11}{12}}{a^{2}} & \frac{12 i j + 8 i + 8 j + 5}{12 a^{2}}\\0 & 0\end{matrix}
\endgroup
\right.
$}
\\
\scalebox{0.95}{$
\left.
\begingroup 
\setlength\arraycolsep{2pt}
\begin{matrix}\frac{\frac{a \left(3 j + 2\right)}{3} - i j - \frac{2 i}{3} - j^{2} - 2 j - \frac{11}{12}}{a^{2}} & 0\\\frac{12 i j + 8 i + 8 j + 5}{12 a^{2}} & 0\\\frac{6 j^{2} + 8 j + 3}{6 a^{2}} & 0\\0 & 0\end{matrix}
\endgroup
\right]
$}
\\
\mathbf{B}_{ij}^-=
\scalebox{0.95}{$
\left[
\begingroup 
\setlength\arraycolsep{2pt}
\begin{matrix}\frac{\mathbf{n}_1 \left(- 4 a + 4 i + 4 j + 5\right) + \mathbf{n}_2 \left(- 4 a + 4 i + 4 j + 5\right) + 2 \mathbf{n}_3 \left(- 2 a + 2 i + 2 j + 3\right)}{12 a}\\\frac{- \mathbf{n}_1 \cdot \left(4 i + 3\right) - 2 \mathbf{n}_2 \cdot \left(2 i + 1\right) - \mathbf{n}_3 \cdot \left(4 i + 3\right)}{12 a}\\\frac{- 2 \mathbf{n}_1 \cdot \left(2 j + 1\right) - \mathbf{n}_2 \cdot \left(4 j + 3\right) - \mathbf{n}_3 \cdot \left(4 j + 3\right)}{12 a}\\0\end{matrix}
\endgroup
\right]
$}
\\
\text{for} \quad \triangle_{ij}^- \in \triangle_{l}^+
\end{aligned}
\end{equation}

\begin{equation}
\small
\begin{aligned}
\mathbf{A}_{ij}^-=
\scalebox{0.9}{$
\left[
\begingroup 
\setlength\arraycolsep{2pt}
\begin{matrix}0 & 0 & 0\\0 & \frac{6 a^{2} - 12 a j - 8 a + 6 j^{2} + 8 j + 3}{6 a^{2}} & \frac{12 a^{2} - 4 a \left(3 i + 3 j + 4\right) + 12 i j + 8 i + 8 j + 5}{12 a^{2}}\\0 & \frac{12 a^{2} - 4 a \left(3 i + 3 j + 4\right) + 12 i j + 8 i + 8 j + 5}{12 a^{2}} & \frac{6 a^{2} - 12 a i - 8 a + 6 i^{2} + 8 i + 3}{6 a^{2}}\\0 & \frac{- a^{2} + a \left(i + 2 j + 2\right) - i j - \frac{2 i}{3} - j^{2} - 2 j - \frac{11}{12}}{a^{2}} & \frac{- a^{2} + a \left(2 i + j + 2\right) - i^{2} - i j - 2 i - \frac{2 j}{3} - \frac{11}{12}}{a^{2}}\end{matrix}
\endgroup
\right.
$}
\\
\scalebox{0.9}{$
\left.
\begingroup 
\setlength\arraycolsep{2pt}
\begin{matrix}0\\\frac{- a^{2} + a \left(i + 2 j + 2\right) - i j - \frac{2 i}{3} - j^{2} - 2 j - \frac{11}{12}}{a^{2}}\\\frac{- a^{2} + a \left(2 i + j + 2\right) - i^{2} - i j - 2 i - \frac{2 j}{3} - \frac{11}{12}}{a^{2}}\\\frac{6 a^{2} - 4 a \left(3 i + 3 j + 4\right) + 6 i^{2} + 12 i j + 16 i + 6 j^{2} + 16 j + 11}{6 a^{2}}\end{matrix}
\endgroup
\right]
$}
\\
\mathbf{B}_{ij}^+=
\scalebox{0.9}{$
\left[
\begingroup 
\setlength\arraycolsep{2pt}
\begin{matrix}0\\\frac{2 \mathbf{n}_1 \left(- 2 a + 2 j + 1\right) + \mathbf{n}_2 \left(- 4 a + 4 j + 3\right) + \mathbf{n}_3 \left(- 4 a + 4 j + 3\right)}{12 a}\\\frac{\mathbf{n}_1 \left(- 4 a + 4 i + 3\right) + 2 \mathbf{n}_2 \left(- 2 a + 2 i + 1\right) + \mathbf{n}_3 \left(- 4 a + 4 i + 3\right)}{12 a}\\\frac{- \mathbf{n}_1 \left(- 4 a + 4 i + 4 j + 5\right) - \mathbf{n}_2 \left(- 4 a + 4 i + 4 j + 5\right) - 2 \mathbf{n}_3 \left(- 2 a + 2 i + 2 j + 3\right)}{12 a}\end{matrix}
\endgroup
\right]
$}
\\
\text{for} \quad \triangle_{ij}^- \in \triangle_{l}^-.
\end{aligned}
\end{equation}

%% file: equations/clip.tex
\begin{algorithm}[t]
\SetAlgoNoLine
\KwIn{Endpoints $\mathbf{n}^i,\mathbf{n}^{i+1}$ of a triangle edge and $\bm{\omega}_z$.}
\KwOut{The clipped endpoints $\mathbf{n}^i_\text{out},\mathbf{n}^{i+1}_\text{out}$.}
$\mathbf{d}=\mathbf{n}^{i+1}-\mathbf{n}^i$\;
$a=\bm{\omega}_z^2\mathbf{d}_y^2+\mathbf{d}_x^2$;
$b=2(\bm{\omega}_z^2\mathbf{n}^{i}_y\mathbf{d}_y+\mathbf{n}^i_x\mathbf{d}_x)$\;
$c=\bm{\omega}_z^2((\mathbf{n}^{i}_y)^2-1)+(\mathbf{n}^i_x)^2$;
$c'=\bm{\omega}_z^2((\mathbf{n}^{i+1}_y)^2-1)+(\mathbf{n}^{i+1}_x)^2$\;
$\Delta=b^2-4ac$\;
$clip_0=c>0 \ and \ \mathbf{n}^i_x<0$\;
$clip_1=c'>0 \ and \ \mathbf{n}^{i+1}_x<0$\;
$\mathbf{n}^i_\text{out}=\mathbf{n}^i;\mathbf{n}^{i+1}_\text{out}=\mathbf{n}^{i+1}$\;
\If{$clip_0$ or $clip_1$
}{
\If{$\Delta<0$} {
drop the edge\;
\Return
}
$t_0=\tfrac{-b-\sqrt{\Delta}}{2a}$;$t_1=\tfrac{-b+\sqrt{\Delta}}{2a}$\;
\If{$clip_0$} {
$\mathbf{n}^{i}_\text{out}=\mathbf{n}^i(1-t_0)+\mathbf{n}^{i+1}t_0$\;
}
\If{$clip_1$} {
$\mathbf{n}^{i+1}_\text{out}=\mathbf{n}^i(1-t_1)+\mathbf{n}^{i+1}t_1$\;
}
\If{($clip_0$ and $clip_1$) and ($t_0<0$ or $t_0>1$ or $t_1<0$ or $t_1>1$)} {
drop the edge\;
\Return
}
}
\caption{Clip triangle edge}
\label{alg:clip}
\end{algorithm}

%% file: figures/line_integral.tex
\begin{figure}[t]
    \centering
    \setlength\tabcolsep{0.1pt}
    \begin{tabular}{cc}    
    \includegraphics[width=0.8\linewidth]{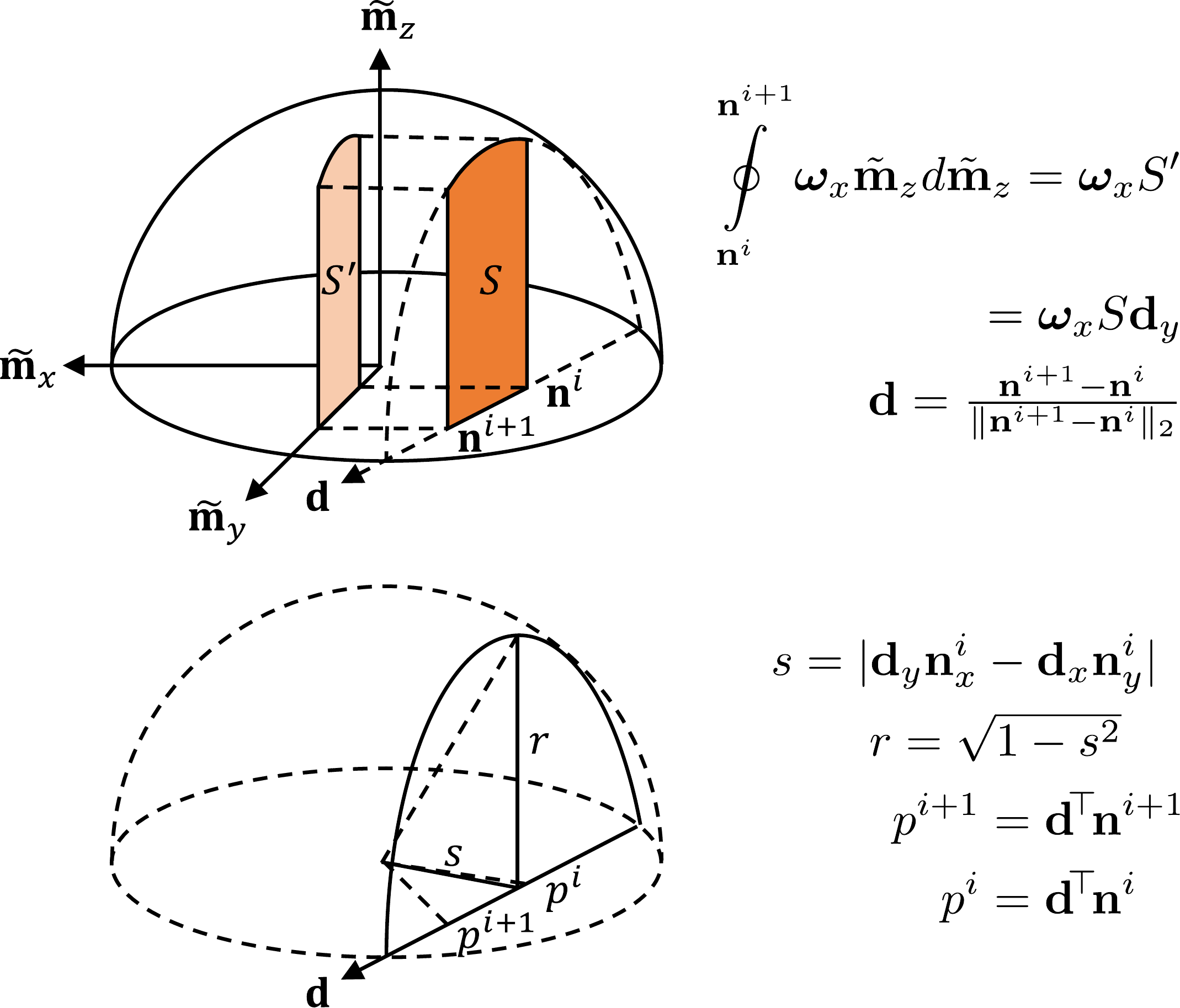}
    \end{tabular}
    \caption{
    \textbf{Geometric derivation of $\bm{\omega}_x\tilde{\mathbf{m}}_z$'s integration.}
    Top: the line integral corresponds to the area $S'$, which is the projection of circle arc's underlying area $S$ on $\mathbf{m}_y$ axis.
    Bottom: derivation of the circle's radius $r$ and the endpoints' abscissas $p^i,p^{i+1}$ on the circle plane.
    }
    \label{fig:line_integral}
\end{figure}

%% file: figures/supp_normalmapped.tex
\begin{figure*}
    \centering
    \setlength\tabcolsep{0.5pt}
    \resizebox{0.99\linewidth}{!}{
    \begin{tabular}{cccc}
         Normal-mapped (16384SPP) & Normal-mapped (128SPP) & \citet{yan2016position} (128SPP) & Ours (128SPP)\\
         \includegraphics[width=0.26\linewidth]{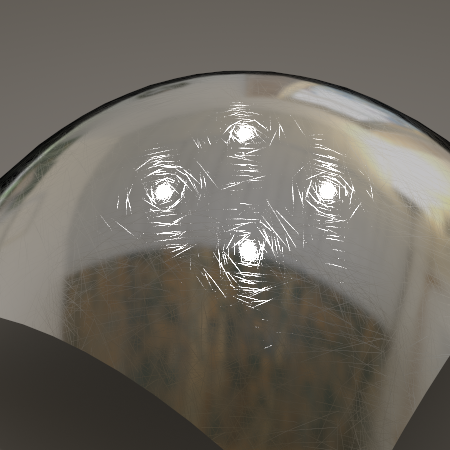}&
         \includegraphics[width=0.26\linewidth]{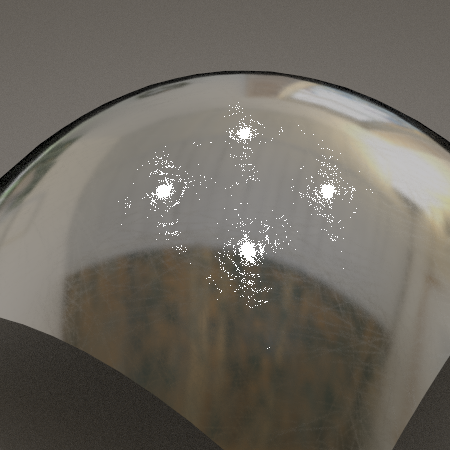}&
         \includegraphics[width=0.26\linewidth]{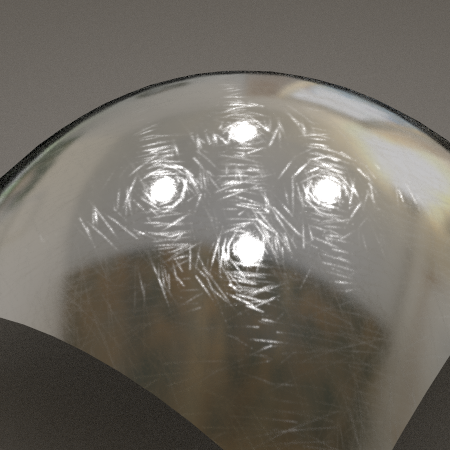}&
         \includegraphics[width=0.26\linewidth]{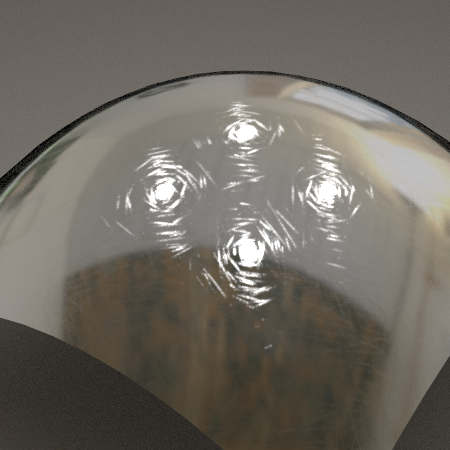}\\
    \end{tabular}
    }
    \caption{\textbf{Qualitative comparison with normal-mapped ground truth rendering.} The standard normal mapping requires a very large SPP to capture the glint pattern (1st and 2nd images). Our approach is a more accurate approximation of this normal-mapped ground truth than \citeauthor{yan2016position}.}
    \label{fig:supp-normalmapped}
\end{figure*}

%% file: figures/supp_comparison.tex
\begin{figure*}
    \centering
    \setlength\tabcolsep{0.5pt}
    \resizebox{0.99\linewidth}{!}{
    \begin{tabular}{ccc@{\hskip4pt}ccc}
         $64^2$ & $128^2$ & $256^2$ & $64^2$ & $128^2$ & $256^2$\\
         \includegraphics[width=0.16\linewidth]{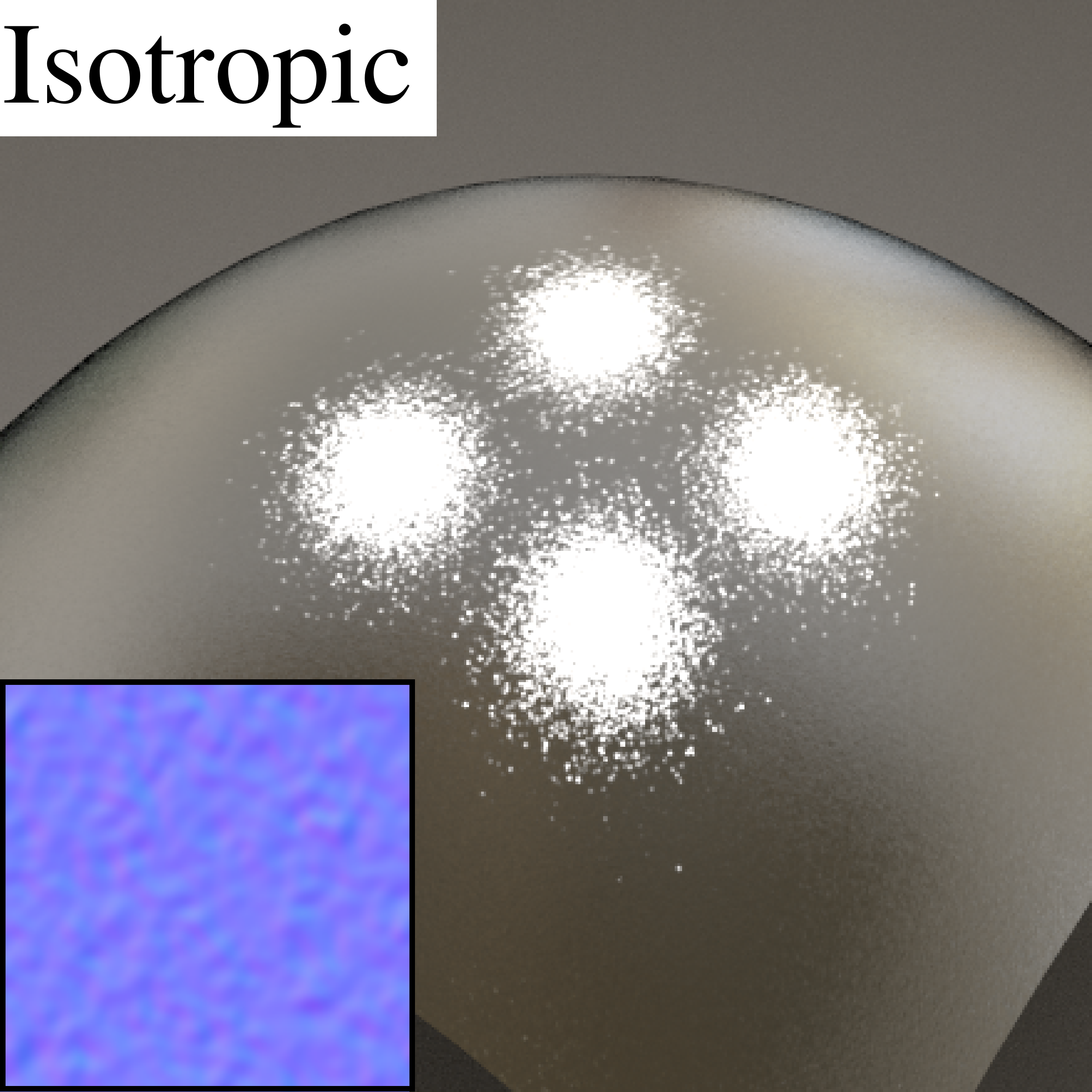}&
         \includegraphics[width=0.16\linewidth]{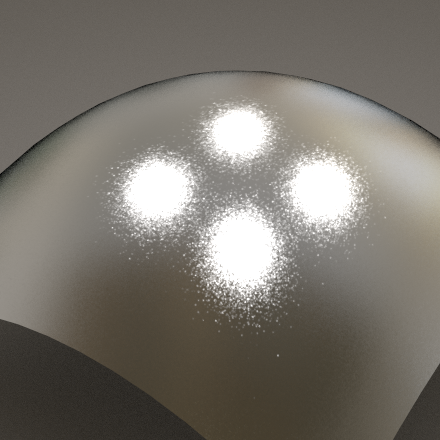}&
         \includegraphics[width=0.16\linewidth]{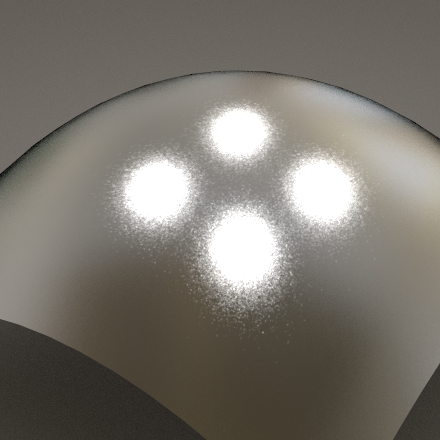}&
         \includegraphics[width=0.16\linewidth]{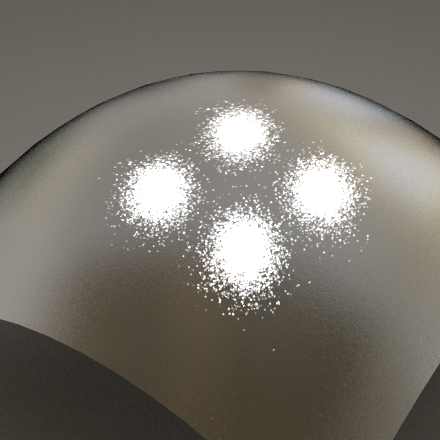}&
         \includegraphics[width=0.16\linewidth]{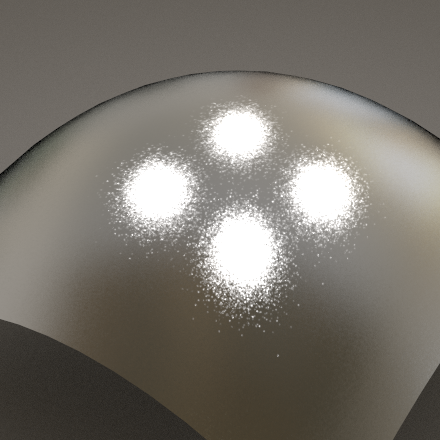}&
         \includegraphics[width=0.16\linewidth]{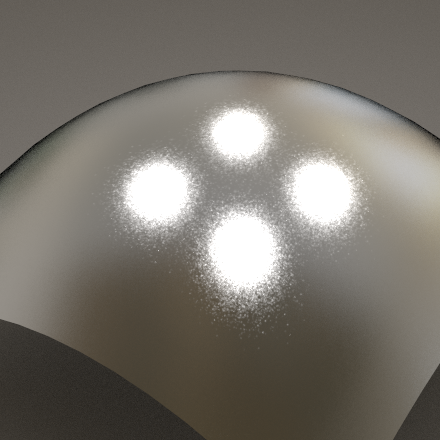}\\[-2pt]

         \includegraphics[width=0.16\linewidth]{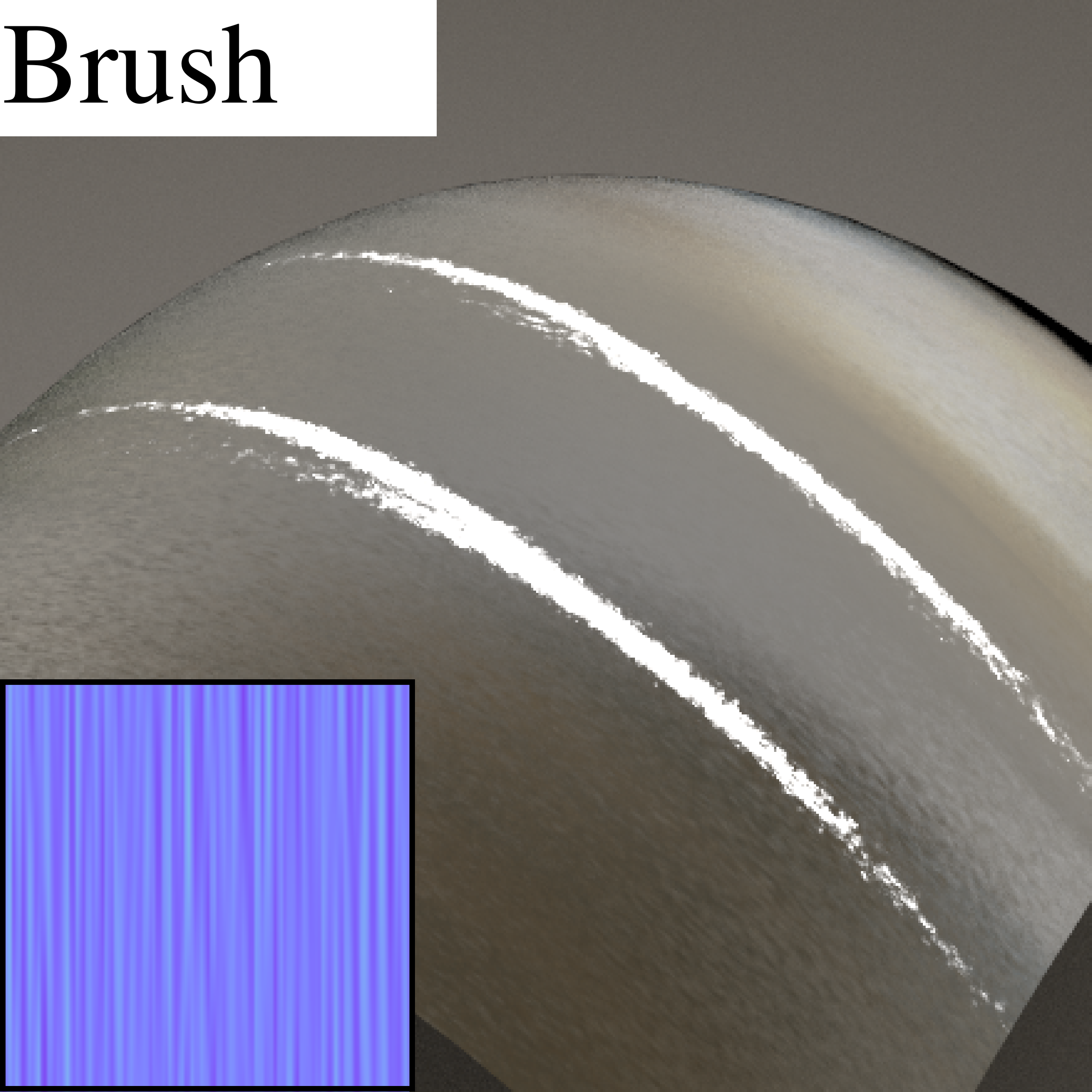}&
         \includegraphics[width=0.16\linewidth]{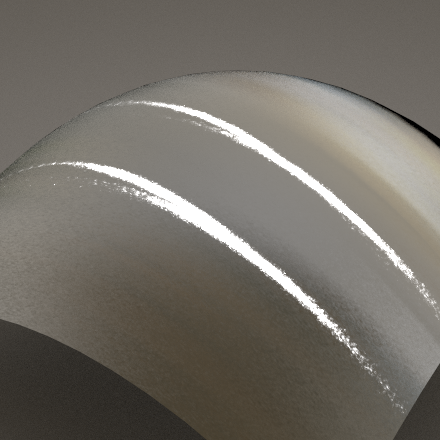}&
         \includegraphics[width=0.16\linewidth]{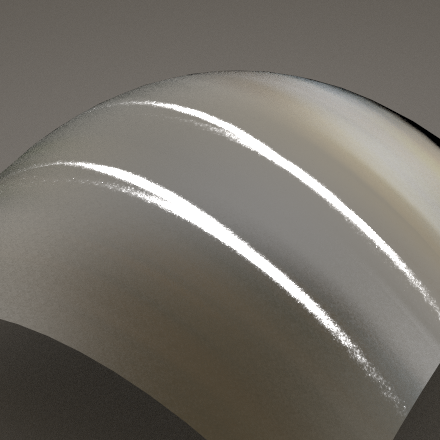}&
         \includegraphics[width=0.16\linewidth]{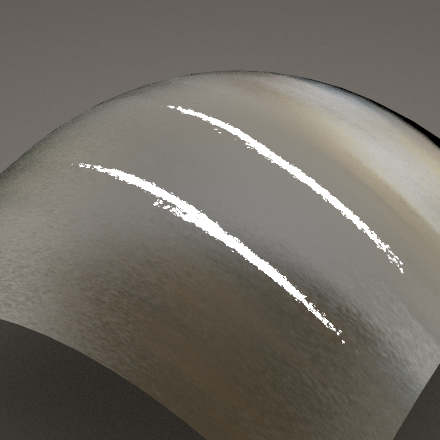}&
         \includegraphics[width=0.16\linewidth]{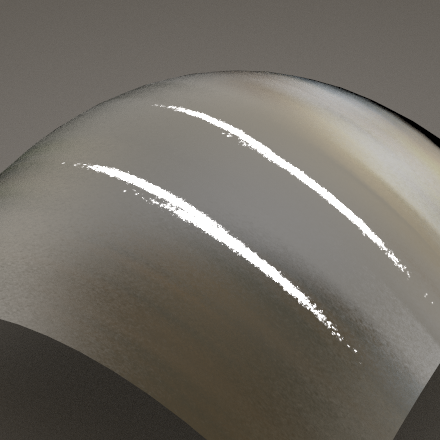}&
         \includegraphics[width=0.16\linewidth]{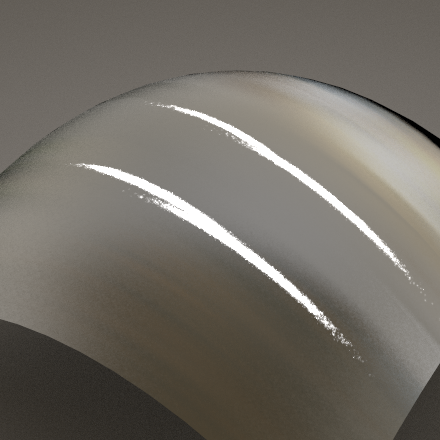}\\[-2pt]

         \includegraphics[width=0.16\linewidth]{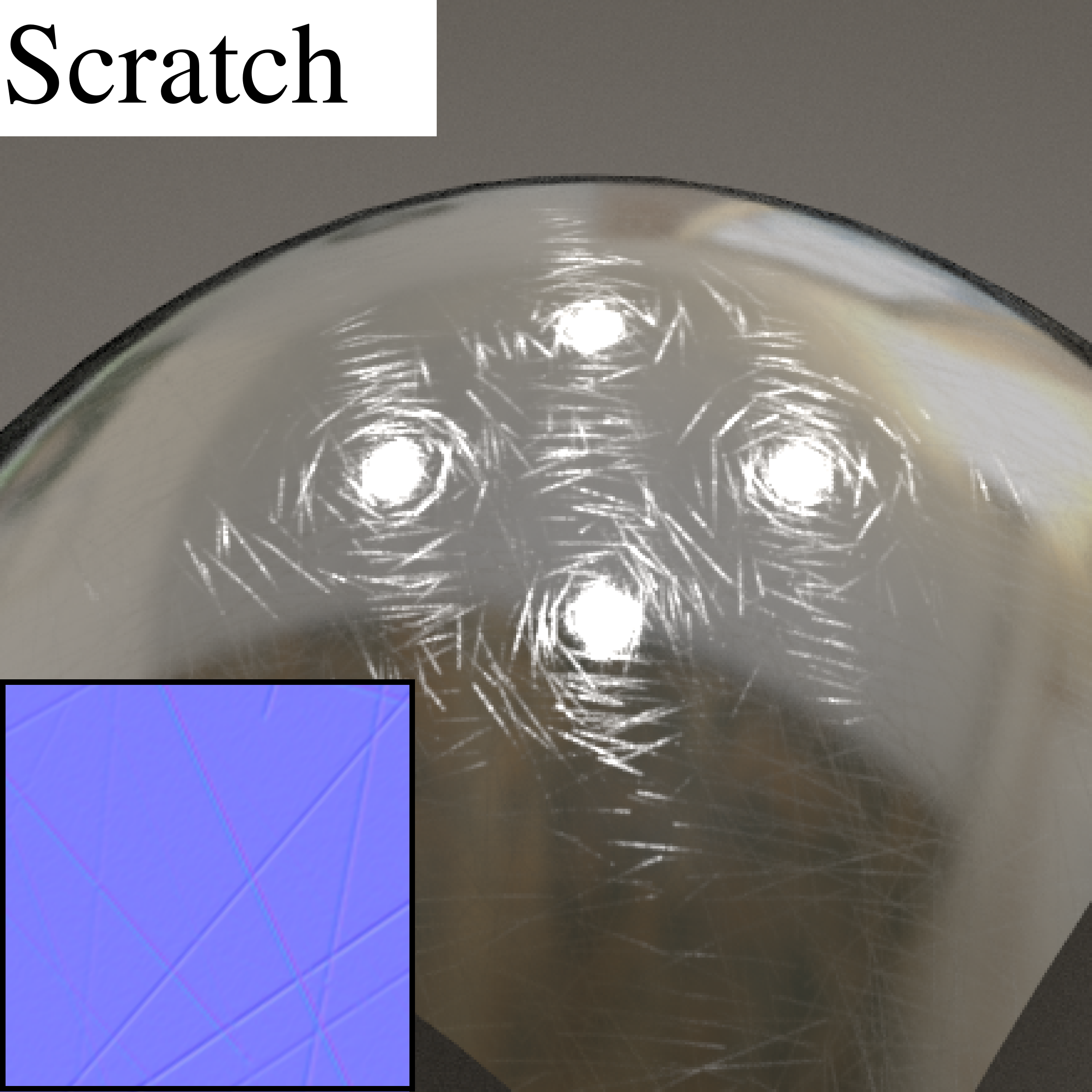}&
         \includegraphics[width=0.16\linewidth]{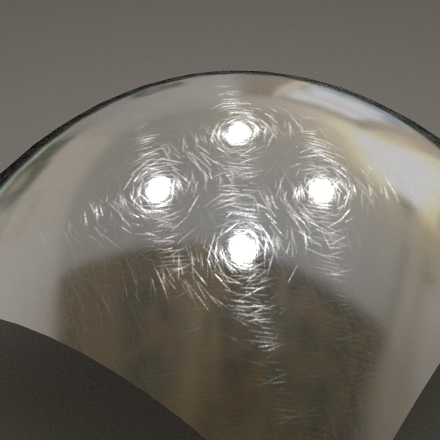}&
         \includegraphics[width=0.16\linewidth]{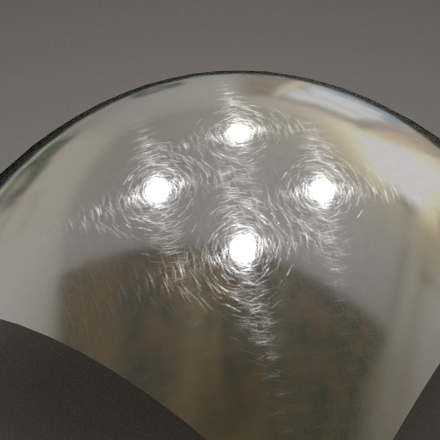}&
         \includegraphics[width=0.16\linewidth]{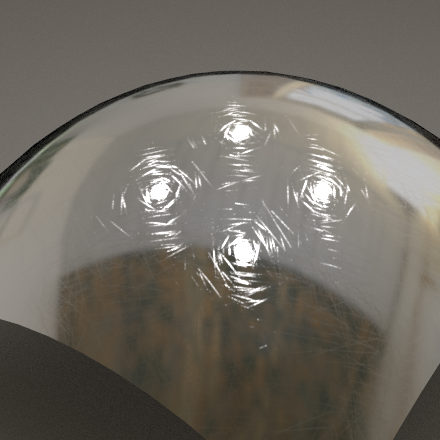}&
         \includegraphics[width=0.16\linewidth]{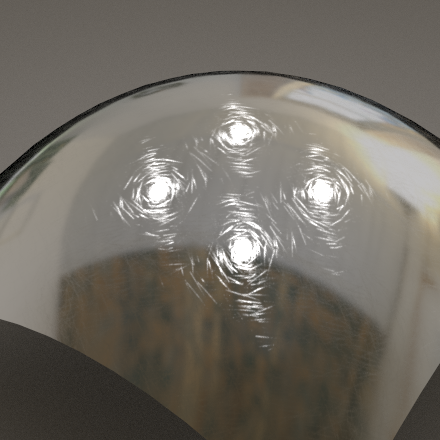}&
         \includegraphics[width=0.16\linewidth]{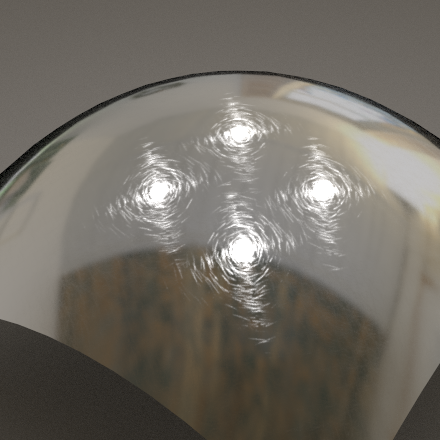}\\[-2pt]
         
         \includegraphics[width=0.16\linewidth]{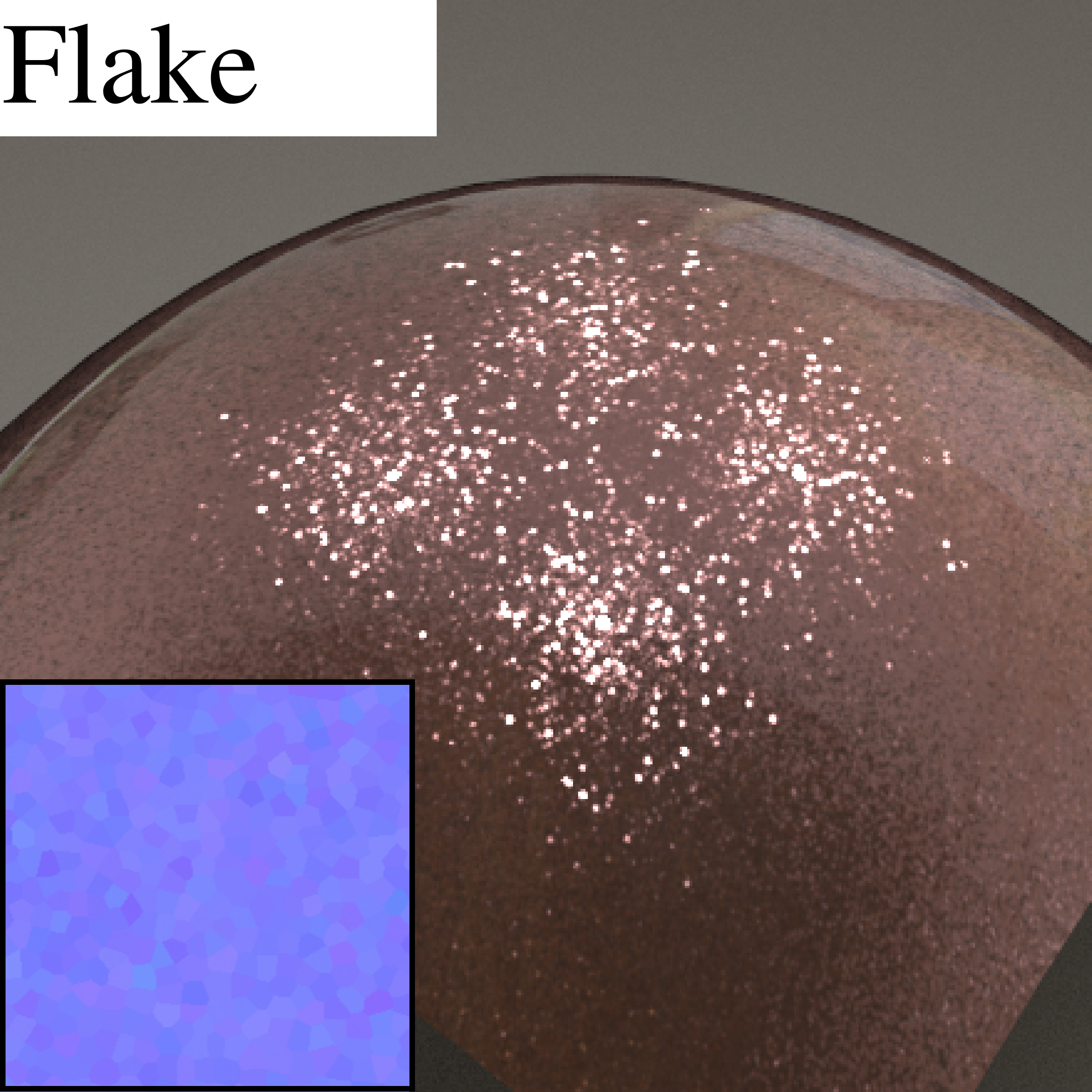}&
         \includegraphics[width=0.16\linewidth]{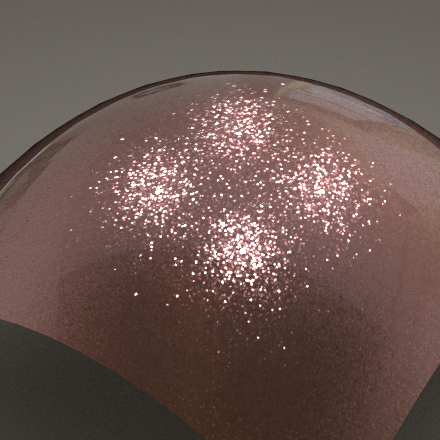}&
         \includegraphics[width=0.16\linewidth]{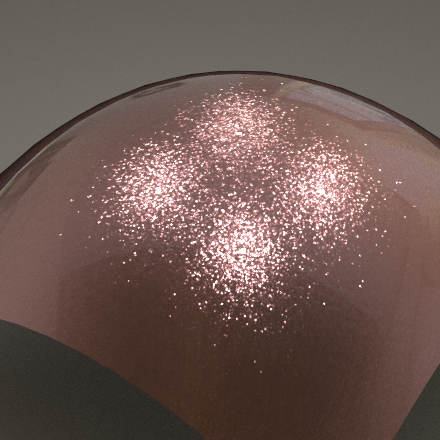}&
         \includegraphics[width=0.16\linewidth]{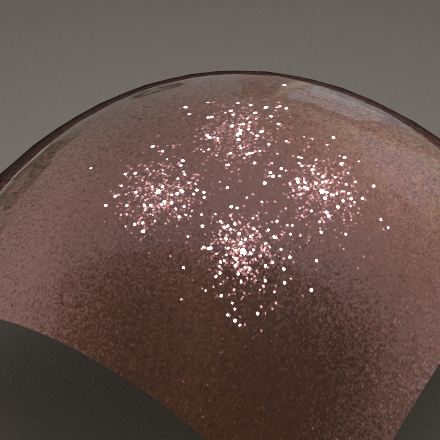}&
         \includegraphics[width=0.16\linewidth]{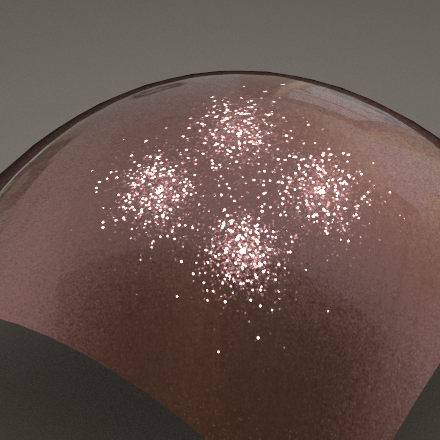}&
         \includegraphics[width=0.16\linewidth]{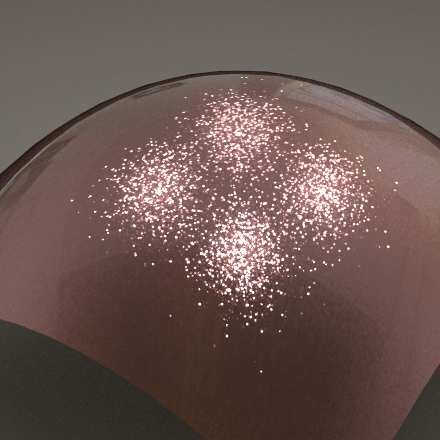}\\[-2pt]
         \multicolumn{3}{c}{\citet{yan2016position}} & \multicolumn{3}{c}{Ours}\\
    \end{tabular}
    }
    \caption{\textbf{Qualitative comparison with \citet{yan2016position} on each normal map.}
    We use a conductor BRDF for renderings with an additional coating layer applied to the flake normal map (4th row).
    }
    \label{fig:supp-comparison}
\end{figure*}

%% file: tables/performance_supp.tex
\begin{table}[t]
    \centering
    \caption{
    \textbf{Performance comparison on the flake normal map in minutes.}
    \citet{yan2014rendering} takes days to render so is not measured here.
    }
    \setlength\tabcolsep{5pt}
    \resizebox{0.99\linewidth}{!}{
    \begin{tabular}{l | ccc |ccc |ccc}
    \toprule
    \multicolumn{1}{l}{Method} & 
      \multicolumn{3}{c}{\citet{yan2016position}} &
      \multicolumn{3}{c}{Ours no cluster} &
      \multicolumn{3}{c}{Ours}\\
      \multicolumn{1}{l}{Scale} &
      $64^2$ & $128^2$ & \multicolumn{1}{c}{$256^2$} &
      $64^2$ & $128^2$ & \multicolumn{1}{c}{$256^2$} &
      $64^2$ & $128^2$ & \multicolumn{1}{c}{$256^2$}\\
      \midrule
    Flake & 
    0.94 & 2.88 & 10.72 &
    0.59 & 1.26 & 3.12 &
    \fst{0.25} & \fst{0.36} & \fst{0.55}\\
    \bottomrule
    \end{tabular}
    }
    \label{tab:performance_supp}
\end{table}

%% file: tables/timing.tex
\begin{table}[]
    \centering
    \caption{
    \textbf{Timing of different rendering stages in minutes.}
    The direct illumination computation dominates the inference, 
    where the \pndf{} evaluation during the BRDF sampling is the most time-consuming part.
    }
    \setlength\tabcolsep{4pt}
    \resizebox{0.99\linewidth}{!}{
    \begin{tabular}{l cc cc}
         \toprule
         Scene (paper Fig.~14) & Direct BRDF & Direct emitter& Indirect & Total\\
         \midrule
         Wrench& 
         0.31 & 0.14 & 0.19 & 0.63\\
         Kettle&
         0.21 & 0.12 & 0.15 & 0.48\\
         Plate \& Cutlery&
         0.72 & 0.22 & 0.10 & 1.04\\
         \bottomrule
    \end{tabular}
    }
    \label{tab:timing}
\end{table}